Observatório Nacional

Ministério da Ciência e Tecnologia

# Observações solares:
# Estudo das variações do diâmetro e suas correlações

Eugênio Reis Neto

**Tese de mestrado**

Orientador: Dr. Alexandre Humberto Andrei

*À memória de meu pai*

*Eugênio*

# Agradecimentos

Ao Dr. Alexandre Humberto Andrei, pela dedicação, apoio, paciência, compreensão dispensados ao longo deste trabalho e, principalmente pela amizade e bom humor.

A Dra. Jucira Lousada Penna, por todo seu entusiasmo, experiência e carinho.

Ao Dr. Eugeni Jilinski pela amizade, interesse e auxílio indispensáveis.

Ao Dr. Victor de Amorim d'Ávila, por todo o incentivo, afeição e ajuda.

A todos os amigos do terceiro andar do Observatório Nacional.

Ao CNPq e FAPERJ, pelas bolsas de estudo que possibilitaram a realização deste trabalho.

Aos caríssimos amigos e colegas da pós-graduação do Observatório Nacional, pela amizade, apoio e por todos os bons e divertidos momentos.

A todos os meus amigos de fora do Observatório Nacional, que sempre me apoiaram.

A minha mãe, pela sua eterna compreensão e amor.

# Resumo

Este trabalho teve como objetivo analisar os dados da campanha de monitoramento do diâmetro solar de 1998 até 2000. O instrumento utilizado foi um astrolábio Danjon adaptado para observações solares, localizado no campus do Observatório Nacional. Durante este intervalo foram realizadas 10807 medidas individuais do diâmetro solar, divididas entre observações a Leste a Oeste do meridiano e homogeneamente distribuídas.

Um tratamento das imagens, levando-se em conta a *corrente de escuro* e o *flat field* da câmara foi testada, fazendo uso das rotinas IRAF. Um estudo das influências sistemáticas que as condições observacionais têm sobre o resultado final da redução foi realizado. A temperatura média na hora da observação mostrou-se como o parâmetro mais influente sobre o resultado final, além da variação da temperatura durante a sessão, do fator de Fried e do desvio padrão do ajuste da parábola refletida que apresentaram um complexo e menor grau de influência.

As correções introduzidas são da ordem de centésimos de segundo de arco, o que é dez vezes menor que o erro associado a uma única observação. O valor encontrado para o semidiâmetro médio deste período foi de 959",107±0",006.

Através do algoritmo CLEAN, os termos periódicos das variações do valor do semidiâmetro foram obtidos, com destaque para o período de 515 dias, de maior amplitude. A utilização de um segundo algoritmo, o DCDFT, serviu de contraprova para os períodos encontrados.

Uma dependência do semidiâmetro com a sua heliolatitude foi encontrada. A diferença entre o raio equatorial e polar pôde ser calculada (Δr = 0",013±0",004) e o valor do momento de quadrupólo do Sol inferido: $| J_2 | = (3{,}61 \pm 2{,}90) \times 10^{-6}$.

Para melhorar as investigações de variações do diâmetro e suas relações com a constante solar, eventos do Sol e resposta climática, é necessária precisão da ordem de milésimo de segundos de arco, em escala de tempo de poucos dias. Para isto, um novo instrumento encontra-se em desenvolvimento no Observatório Nacional-MCT.

# Abstract

This work has aimed to analyze the 1998 to 2000 campaign of solar diameter surveying. The employed instrument was a Danjon astrolabe, at the Observatório Nacional campus, and specially modified for the solar observations. During the time lapse, 10807 independent measurements of the solar diameter were made, Eastwards and Westwards from the local meridian and evenly distributed.

An image treatment has been devised to account for the camera dark current and flat field, using IRAF routines. In other stance, a study was made of the systematic bearing of the observational conditions upon the reduction final outcome. The mean temperature at the observation is shown as the most influential parameter upon the final result. Next to it, follows the temperature variation, the Fried's factor, and the standard deviation of the reflected parabola, all these presenting a minor and complex degree of influence.

The derived corrections are of the order of hundredths of arc seconds, thus being tenfold smaller than the typical error of one observation. The mean semi diameter for the time lapse is found as 959",107±0",006.

Through the use of a CLEAN algorithm the periodic terms of the semi diameter were obtained. The largest amplitude is attached to that of 515 days. By using a second type of algorithm, namely the DCDFT, the found periods stand additional proof.

A dependency of the semi diameter on the observed heliolatitude is verified. The difference between the equatorial and polar radii was calculated as $\Delta r$ = 0",013±0",004, and the Solar quadrupole moment was inferred as $|J_2| = (3,61 \pm 2,90) \times 10^{-6}$.

In order to further improve on the diameter variations surveying, and on the relationships to the solar constant, solar events and the climate response, the required accuracy is at the level of the thousandth of arc seconds, in the time scale of few days. To achieve such level, a state-of-the-art instrument is being developed at the Observatório Nacional-MCT.

# Sumário

# Índice de figuras

*Figura* *página*

*Figura* *página*

*Figura* *página*

# Índice de tabelas

*Tabela* *página*

# Capítulo 1

## Introdução

A observação do Sol no Observatório Nacional utilizando-se um Astrolábio Danjon começou em 1978, primeiramente orientada para a determinação dos parâmetros orbitais da Terra e do sistema de referência astrométrico (Penna *et al.* 1996, 1997). Uma posterior modificação no instrumento, equipando-o com um prisma refletor de ângulo variável e uma câmera CCD, permitiu que, a partir de 1997, se iniciasse um programa de monitoramento do diâmetro solar em cooperação com o CERGA/Observatoire de la Côte d'Azur (Andrei *et al.* 1996). Em quatro anos o número de medidas já era de 12.546 medidas individuais do diâmetro solar (Jilinski *et al.*, 1998, 1999; Puliaev *et al.*, 2000).

O valor do diâmetro do Sol chamou a atenção de diversas figuras históricas. Utilizando um instrumento, descrito por Hiparcos (século II a.C.), que permitia a comparação do diâmetro aparente do Sol e da Lua com um pequeno disco móvel sobre uma haste, Arquimedes (287-212 a.C.) calculou que o valor do diâmetro do Sol estaria entre 27' e 33'. Aristarco (≈ 310-230 a.C.) atribuiu o valor de 30' e Ptolomeu (≈ 87-151 d.C.) o valor de 31'20", sendo este último monitorado por um ano, sem apresentar variações importantes.

Em 1591, Tycho Brahe (1546-1601), projetando a imagem do Sol num anteparo, realizou 11 medidas do seu diâmetro. Johannes Kepler (1571-1630) selecionou e reduziu duas destas observações e concluiu que o valor mínimo do diâmetro solar seria de 30'30" (Sigismondi e Fraschetti, 2001).

Apesar das dificuldades inerentes a estas observações, Auwers, em 1891, mediu o raio solar com um heliômetro (instrumento que mede a separação angular de limbos opostos do Sol) obtendo um valor médio de 959",53.

Devido à importância das considerações que uma possível variação do seu raio traria para a física do Sol (e para os modelos de evolução estelar), um grande esforço tem sido feito nos últimos anos na direção do desenvolvimento de novos instrumentos e técnicas para a astrometria solar (Laclare e Guy, 1991; Wittmann, 1997; Penna *et al.*, 1997; Andrei *et al.*, 2001).

Dados históricos apontam para um raio do Sol maior durante o período conhecido como Mínimo de Maunder, que coincidiu com períodos de extremo frio na Europa (Ribes, 1991).

Diversos autores têm tentado extrair informações relevantes analisando dados antigos, cobrindo até algumas centenas de anos, e observações contemporâneas. Porém medidas de solo são sensíveis a várias distorções instrumentais e atmosféricas, dificultando a interpretação das aparentes mudanças e originando resultados contraditórios. As variações observadas podem ser de caráter periódico (Gavryusev, 1994) ou não periódicos (Andrei *et al.*, 2001).

Um exemplo destas dificuldades está na questão a respeito do Sol apresentar uma variação secular no valor do seu diâmetro. Em 1979, Eddy e Boornazian descobriram uma tendência de decréscimo do raio solar de 1,1 segundo de arco/século, baseando-se nas medidas com o círculo meridiano do Observatório Real de Greenwich, no intervalo 1836-1953 (Sveshnikov, 2002). A magnitude de uma tal variação, sua pressuposta constância por mais de um século e, finalmente, a verificação de mudanças instrumentais não consideradas, deram origem, na década de 80, a intensa controvérsia sobre os resultados experimentais na área. Entretanto, diversos trabalhos subseqüentes, destacando-se a re-análise de Gilliland (1981), que propõe um ciclo de variação de 76 anos, re-introduzindo este tipo de medição na fronteira mais avançada da pesquisa solar. Toulmonde (1997) comparou 30 séries de observações, obtidas em diversas épocas ao longo dos últimos três séculos, juntamente com aproximadamente 900 medidas contemporâneas. Seus resultados revelaram a necessidade de correções instrumentais, notadamente o efeito da difração. Uma vez feitas as correções, uma série homogênea foi obtida se estendendo por mais de 300 anos, não revelando nenhuma variação secular sensível do diâmetro solar. Golbasi *et al.* (2001), na análise das observações solares de 1999 e 2000, conduzidas no astrolábio de Antalya, Turquia, mostrou que os resultados apresentavam uma tendência de aumento de aproximadamente 5 *mas*/ano para o semidiâmetro observado. A mesma tendência, só que um pouco maior (8,1±0,9 *mas*/ano), também está presente nos resíduos das medidas do raio solar do MDI-SOHO, em três anos de observação (Emílio *et al.*, 2000).

Uma variação do raio solar com o ciclo de 11 anos é um fato inquestionável para observações históricas e modernas (Sveshnikov, 2002). Porém, a correlação encontrada entre a atividade solar e o diâmetro observado também é um exemplo de desacordo. Noël (1997) e Basu (1998) propõem uma correlação positiva, enquanto Delache (1988) e Laclare *et al.* (1996) relatam uma correlação negativa. O raio solar aparente observado entre 1998 e 2000, com o astrolábio de Santiago parece também variar em fase com a atividade solar (Noël, 2001).

Uma possível explicação para estes resultados conflitantes sobre a fase das variações do diâmetro observado e a atividade solar (ou seu mais acessível índice: o número de manchas solares), pode estar na técnica utilizada para a observação. Ulrich e Bertello (1995), por exemplo, observaram o diâmetro do Sol no comprimento de onda do ferro neutro ($\lambda$ = 525nm), sendo o raio aparente do Sol definido como a distância média entre o centro do disco e o ponto onde a intensidade cai para 25% do valor registrado no centro do disco. Através deste método, eles relatam uma correlação positiva entre a atividade solar e o diâmetro observado. Se o Sol é observado através de uma luz monocromática, ou por uma banda espectral estreita, a posição verdadeira do limbo depende do comprimento de onda, numa relação de quanto maior $\lambda$, maior o diâmetro observado (Neckel, 1995). Neste ponto, vale a pena lembrar que, para a física solar, eventuais variações do semidiâmetro solar são mais interessantes do que o seu real valor (Ribes *et al.* 1991; Toulmonde 1997).

Observações do Sol em rádio (48GHz) feitas durante o período de 1991-1993, no Rádio-Observatório de Itapetinga, São Paulo, também se mostram em fase com a atividade solar (Costa *et al.* 1999).

Na definição de um diâmetro para o Sol, a curva de luminosidade do limbo solar é o parâmetro ideal por ser o mais diretamente observável. O diâmetro do Sol é definido então pela a distância máxima entre os pontos de inflexão daquela curva, sendo sensível a qualquer variação deste parâmetro. Este é o argumento é utilizado por grupos que questionam se as variações observadas são ou não reais. Wittman *et al.* (1991) analisaram o semidiâmetro solar proveniente de 1122 observações feitas com o método de *drift scan*, entre julho e outubro de 1990, utilizando dois telescópios opticamente idênticos localizados em sítios diferentes (Izaña, Tenerife e Locarno, Suíça), e concluíram que as flutuações observadas no valor do semidiâmetro (±0",3) provinham da turbulência atmosférica.

Ribes *et al.* (1991) objetam que as variações observadas no raio solar nestes três séculos são na verdade variações da função de obscurecimento do bordo e não variações verdadeiras do raio. Argumentam que, como as observações visuais e fotoelétricas apresentam variações de amplitudes bem diferentes, os períodos presentes nas séries poderiam originar-se de um processo que muda as propriedades da função de obscurecimento do limbo ou um processo atmosférico que sistematicamente muda as características das imagens. Brown e Christensen-Dalsgaard (1998) analisaram seis anos de medidas (1981-1987) do *Solar Diameter Monitor* e também compartilham da opinião de que as variações anteriormente relatadas do diâmetro solar com a atividade solar sejam, de fato, um reflexo de variações na inclinação da função de escurecimento do bordo.

Um mecanismo para a mudança das propriedades da curva de luminosidade do limbo solar e o deslocamento do ponto de inflexão da mesma, alterando o valor do diâmetro observado, pode ser encontrado no trabalho de Kuhn *et al.* (1985) que, analisando os dados de 1983 do *Princeton Solar Distortion Telescope*, detectaram a existência de uma dependência do brilho do limbo solar com a heliolatitude, com diferenças em torno de 0,6±0,1 K. Os valores destas grandezas se apresentam em contra fase.

Na análise dos dados astrométricos e fotométricos obtidos pelo experimento com o *Michelson Doppler Imager* (MDI) durante março de 1996 e março de 1997 Kuhn *et al.* (1998) reafirmaram que a forma do Sol e sua temperatura variam com a latitude, de uma maneira complexa.

Uma vez que o raio e a temperatura do limbo solar derivam de experimentos diferentes e independentes, a anticorrelação encontrada entre estas duas grandezas indica que de alguma maneira elas estão relacionadas (Noël, 1998).

Para Pecker (1994) parte da variação observada do raio solar é devida à deformação da figura do Sol ligada, por um lado, à chamada "zona real" (entre 20º e 30º de heliolatitude), onde a existência de manchas diminui o brilho observado e conseqüentemente o valor do raio medido e, por outro lado, a zonas ativas (70º - 80º) onde a presença de fáculas tem o efeito inverso.

Diferenças de temperatura na superfície do Sol já haviam sido detectadas por Altrock e Canfield (1972), que realizaram *scans* fotoelétricos do limbo equatorial e meridional do Sol, em junho de 1971, numa banda espectral estreita centrada em 0,77 Å.

Foram encontradas diferenças de temperatura de ordem de poucos graus, de forma que atribuíram o achatamento da figura do Sol encontrado por Dicke e Goldenberg (1967) a estas diferenças. Esta mesma opinião é compartilhada por Ingersoll e Spiegel (1971) e Durnay e Werner (1971).

A existência e a magnitude de um achatamento do Sol é ainda matéria de discussão (Rozelot e Bois, 1998), mas é um fato esperado, uma vez que se trata de um objeto em rotação em torno do seu eixo. Comprovada a forma oblata do Sol, isto levaria a conseqüências para a dinâmica do sistema solar e para aplicações da relatividade geral na mecânica celeste (Dicke *et al.*, 1985), como a explicação do valor de 4 segundos de arco/ano a mais no movimento do periélio de Mercúrio (Dicke e Goldenberg, 1973).

Vários experimentos foram realizados para se medir não só o achatamento, mas também os possíveis modos de vibração do Sol.

Rozelot e Rösch (1997) deduziram uma variação entre 6 e 19 *mas* e período compatível com os 11 anos do ciclo solar, para o achatamento da figura do Sol analisando 15 anos de observações do raio solar do Observatório de Calern e as observações de 1993 e 1994 do Observatório de Pic du Midi.

Recentemente, a heliossismologia veio a contribuir na busca de flutuações do raio solar (Dziembowiski, 2000) com conclusões que a forma da figura do Sol, chamada de helióide, não é estática, não é esférica ou puramente oblata, mas apresenta uma complexa forma pulsante (Rozelot e Rösch, 1997), podendo o próprio achatamento ser variável no tempo.

Pijpers (1998), utilizando os dados das observações heliossismológicas provenientes do *Solar Heliospheric Satellite* (SoHO) e do *Global Oscillations Network Group* (GONG), determinou o momento de quadrupólo do Sol ($J_2$) como tendo o valor de $(2{,}18\pm0{,}06)\times10^{-7}$. Este valor é consistente com os resultado de Paterno, Sofia e DiMauro (1996) que usaram medidas diretas do achatamento do disco solar para inferir o valor do momento de quadrupólo.

Godier e Rozelot (1999) computaram teoricamente o valor do momento de quadrupólo do Sol utilizando modelos contemporâneos de massa e densidade, o modelo de rotação derivados da heliossismologia e levando em conta as diferentes rotações por camada. O valor encontrado para esta grandeza foi de $|J_2| = 1{,}60 \times 10^{-7}$.

Na próxima década, pelo menos duas missões espaciais estão programadas para medir o raio solar, a irradiância e os modos-g de vibração: SPHERIS (*Solar Physics*

*Explorer for Radius, Irradiance and Shape*), um projeto de responsabilidade da NASA e o PICARD, um microsatélite da CNES (*Centre National d'Etudes Spatiales*). SPHERIS e PICARD irão monitorar pequenas mudanças no raio, achatamento e mudanças de formato de ordens superiores com grau de precisão jamais alcançado em solo. O conhecimento da evolução temporal destas grandezas permitirá um melhor entendimento dos mecanismos físicos envolvidos no transporte de energia do interior do Sol e uma expressiva contribuição no estudo das relações entre estas mudanças e seus efeitos significativos sobre o clima na Terra (Sofia *et al.* 1979; Pap *et al.*, 2001).

Os dados destes satélites servirão também para calibrar as medidas de solo e contribuindo para refinar os modelos atmosféricos.

A coleta de dados de solo continuará a ser importante, não só pela sua altíssima relação custo/benefício, quando comparado a projetos espaciais, mas pelo fato, conforme será demonstrado neste trabalho, que apesar de obviamente uma simples medida não ter a precisão necessária, o valor médio de uma longa série de observações cuidadosamente controlada e posteriormente tratada (para evitar viés instrumentais, atmosféricas e do observador), pode alcançar a acurácia necessária para se obter resultados de qualidade, como o valor do achatamento solar e a dependência do raio com a heliolatitude, com precisão comparável entre os valores obtidos por grupos diversos, e contribuir para o aprimoramento dos modelos físicos envolvidos.

# Capítulo 2

## Observação e Redução

As medidas do diâmetro solar tem requerimentos instrumentais específicos de difícil reconciliação: estabilidade em curtos períodos (da ordem do dia), repetibilidade em longos períodos (da ordem do ano) e acurácia (da ordem de $10^{-4}$). Os diferentes requerimentos encontraram resposta adequada na metrologia do Astrolábio Danjon. Este foi construído para ao mesmo tempo ser um instrumento de precisão astrométrica e um instrumento de campo. Os resultados do projeto MERIT (*Monitor Earth Rotation and Intercompare the Techniques of observation and analysis*), o qual envolvia diferentes métodos para a definição do referencial terrestre, comprovaram aquelas qualidades.

O nível de acurácia é definido pelas flutuações rápidas da constante solar (0,01%), que devem ser filtradas dos dados obtidos. Analogamente, as metas de estabilidade e repetibilidade devem ser atingidas em comparação com o intervalo de variação das grandezas locais, ambientais e observacionais capazes de influenciar as medidas. Por exemplo, a variação média da temperatura durante uma sessão, para toda a campanha, foi de 6,7°C ($\sigma$ = 1,9 °C).

A medida obtida, então, deve ter precisão suficiente para que se possa eliminar as fontes de ruído e assim monitorar variações do diâmetro de origem unicamente solar.

Uma dificuldade adicional é trazida pela circunstância das observações solares não admitirem calibração instrumental externa independente.

O Sol, paradoxalmente, tendo em vista o nível do sinal, único em astronomia, é um dos objetos de mais difícil observação. Assim, conforme será descrito no restante do capítulo, é necessário obter robustez estatística tanto do instrumento, como do método de observação e tratamento de dados.

## 2.1 O Instrumento

É utilizado um astrolábio Danjon adaptado para observações solares. Possui um filtro de densidade neutra que reduz aproximadamente $10^4$ a intensidade da luz. As imagens são obtidas com uma câmera CCD COHU 4710 com as seguintes características:

- Dimensão da matriz do CCD: 6,4x4,8 mm
- Número de *pixels*: 699(H)x576(V)
- Dimensão de cada pixel: 9,2(H)x8,4(V)µm
- Sensibilidade: 0,04 lux (com controle automático de ganho)
- Tempo integração (obturação eletrônica da imagem): 20ms

O instrumento está localizado no campus do Observatório Nacional ($\phi = -22^o 53'42'',50 \quad \Lambda = +2^h 52^m 53^s,479 \quad h = 33m$), e é equipado com um prisma refletor de ângulo variável, definindo a distância zenital instrumental, a qual varia entre 25º e 55º, o que permite o monitoramento do diâmetro solar durante todo o ano.

Um conjunto de dois filtros passa banda estão instalados diante da câmera CCD. A banda passante de 1680 Å foi considerada como sendo o intervalo onde 50% da luz é transmitida, que vai de 5230 Å até 6910 Å. O comprimento de onda onde o índice de transmissão é máximo (≈75%) é 5635 Å, considerado como o comprimento de onda efetivo de observação.

A figura a seguir traz o teste de transmissão realizado no conjunto destes filtros.

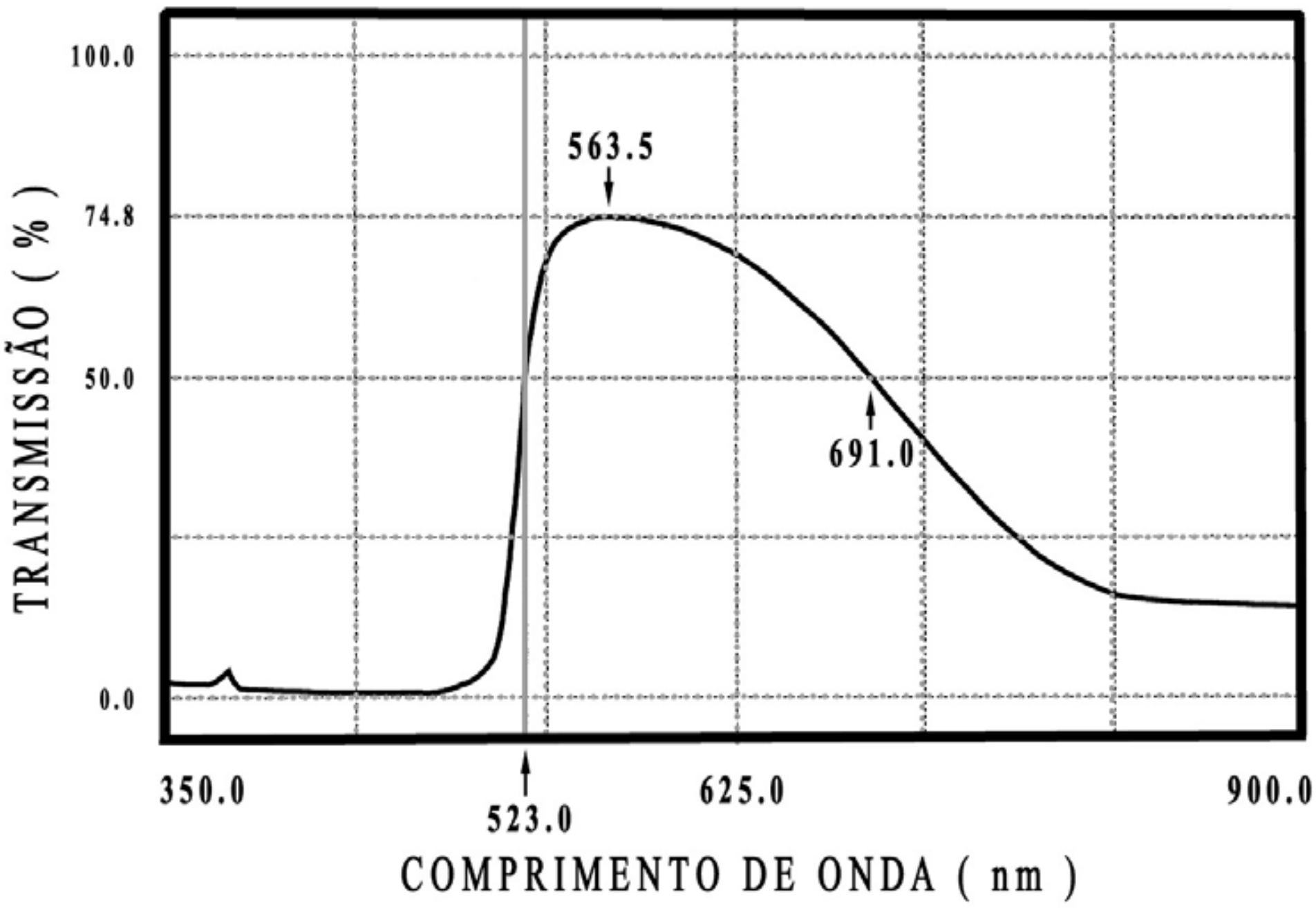

Figura 2.1. Transmissão do conjunto de filtros da câmera CCD do astrolábio solar entre 3500 Å e 9000 Å. O máximo de transmissão (≈75%) ocorre para $\lambda$ = 5635 Å e o intervalo onde 50% da luz é transmitida vai de 5230 Å até 6910 Å.

## 2.2 Metodologia das observações

O sistema de aquisição captura imagens, fornecendo em tempo real a visualização destas. A redução das imagens, conforme discutido acima, privilegia a robustez do processo. Por exemplo, a escala e orientação do CCD são redeterminadas para cada observação. Tanto a aquisição como a redução das imagens são inteiramente impessoais. Com a finalidade de simplificar a comparação de resultados e padronizar os desenvolvimentos, são utilizados os mesmos métodos no Rio de Janeiro e em Calern, baseados naquele desenvolvido no Observatório de Paris (Sinceac, 1998).

A observação consiste em 46 *frames*, cada um contendo uma dupla imagem da porção do disco solar que se aproxima do almucantar definido pelo ângulo do prisma, sendo uma imagem direta e outra refletida numa superfície de mercúrio.

Um *frame* é formado por 256 linhas e 512 colunas e o comprimento vertical do pixel corresponde a 0",56, desta forma, aproximadamente 5% da circunferência do disco solar é imageada. A distância zenital é fixada pelo prisma refletor, associado à superfície de mercúrio. A medida refere-se ao trânsito do bordo solar, superior e inferior, através da distância zenital instrumental.

O acionamento do sistema de aquisição é feito por decisão do observador.

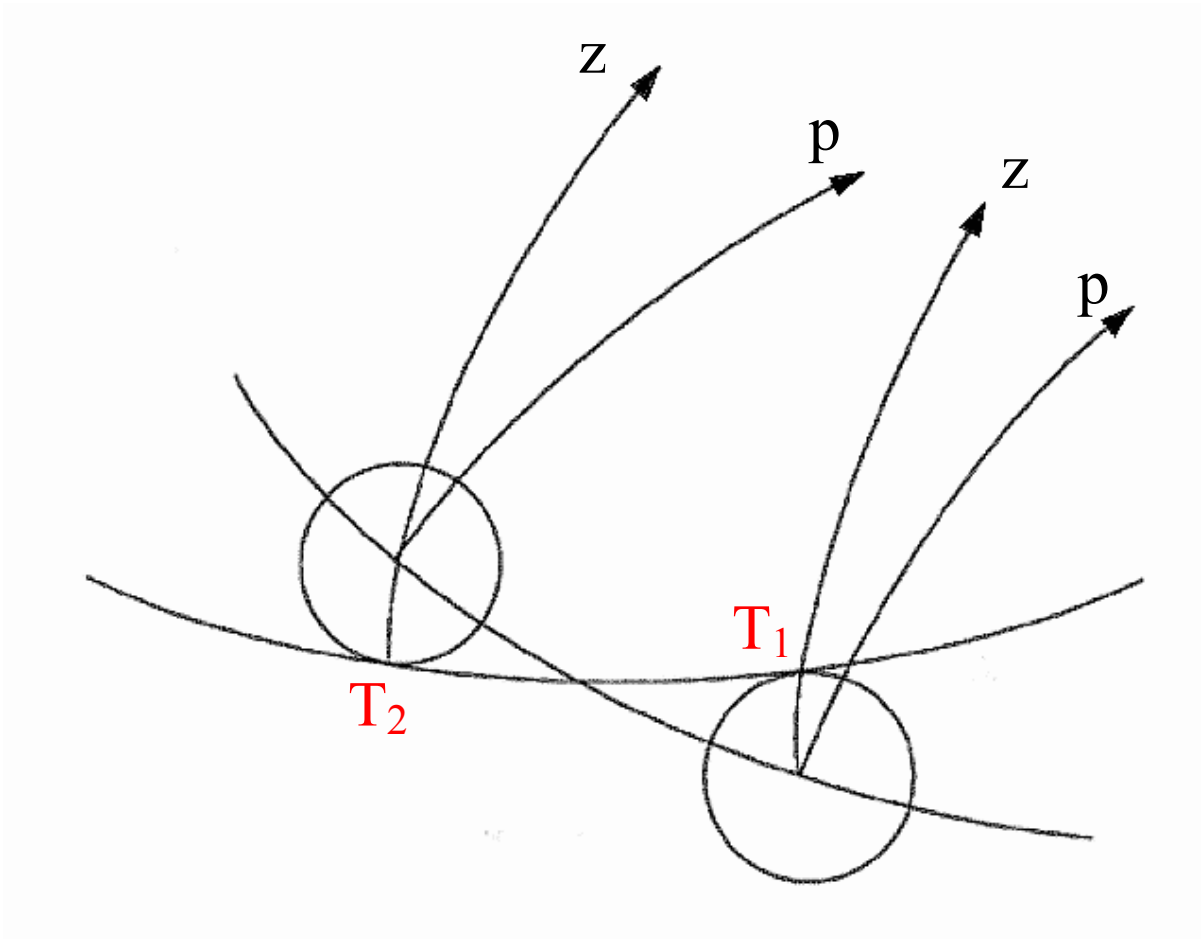

Figura 2.2. Esquema da passagem do Sol através de um círculo de distância zenital *z* definido pelo instrumento. $T_1$ e $T_2$ representam pontos da tangência.

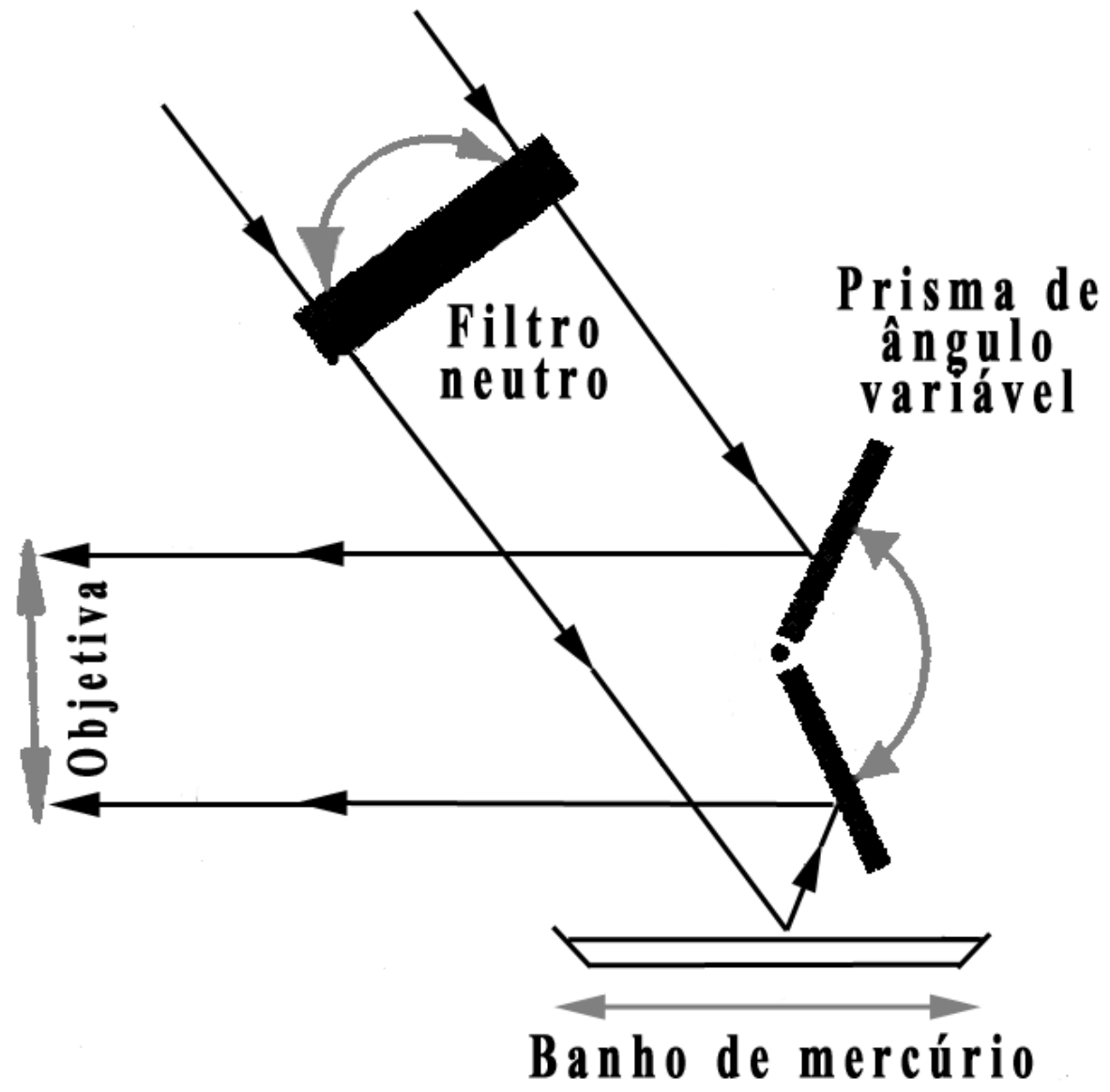

Figura 2.3. Desenho do sistema do prisma refletor de ângulo variável e do espelho de mercúrio fornecendo a dupla imagem para a objetiva. O prisma define a distância zenital de observação e o banho de mercúrio materializa o plano horizontal.

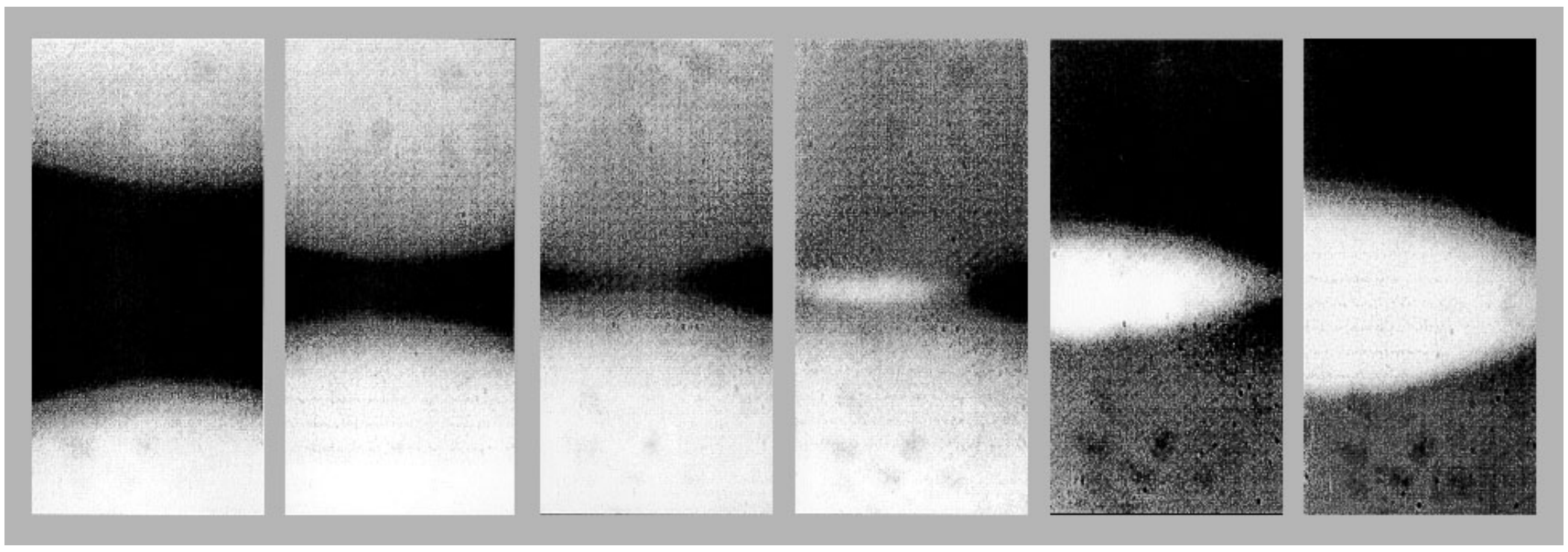

Figura 2.4. *Frames* não consecutivos de uma mesma passagem.
As imagens direta (disco inferior) e refletida (disco superior) se cruzam após o trânsito. Durante a superposição das imagens, o ajuste automático do ganho da câmera atua para evitar a saturação dos *pixels*.

# 2.3 Redução dos dados

O bordo solar é definido como sendo o ponto de inflexão da função de luminosidade I(x,y) ao longo das linhas do CCD. Os pontos de inflexão são encontrados através de uma derivação numérica, retendo-se o a coordenada $x_i$, onde valor máximo da primeira derivada de $I(x_i, y_i)$ ocorre, para cada linha $y_i$ do CCD (para compensar a turbulência atmosférica, o baricentro da figura que representa a primeira derivada da função de luminosidade define a coordenada $x_i$).

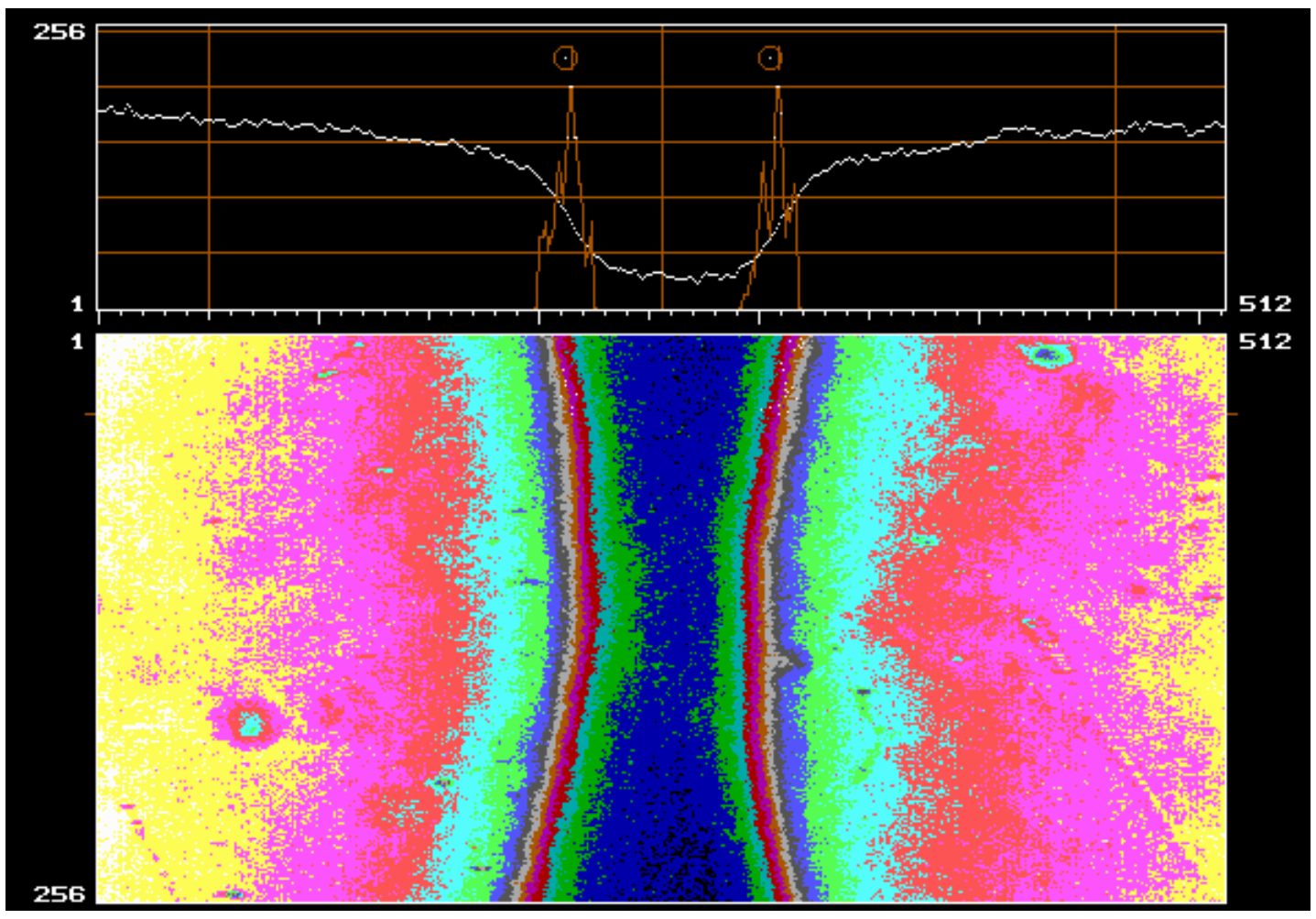

Figura 2.5. Redução de um *frame* da imagem astrolábio. Na parte superior da imagem vê-se a função de luminosidade da linha 36. Os pontos de inflexão de ambos os bordos podem ser identificados na curva superposta.

Do conjunto destes pontos, duas parábolas são ajustadas representando o bordo solar direto e refletido. O ajuste de parábolas, ao invés de arcos de círculo, minimiza defeitos ópticos, responde melhor à forma retangular dos *pixels*, e contempla o movimento do disco solar durante os 20ms de integração. Os pontos correspondentes às coordenadas dos vértices destas parábolas são gravados como função do tempo $\left(X_j^d, Y_j^d, X_j^r, Y_j^r, t\right)$, onde $X_j$ e $Y_j$ são a abscissa e a ordenada do vértice do *frame j*, os índices *d* e *r* referem-se às imagens direta e refletida e *t* é o correspondente Tempo Universal.

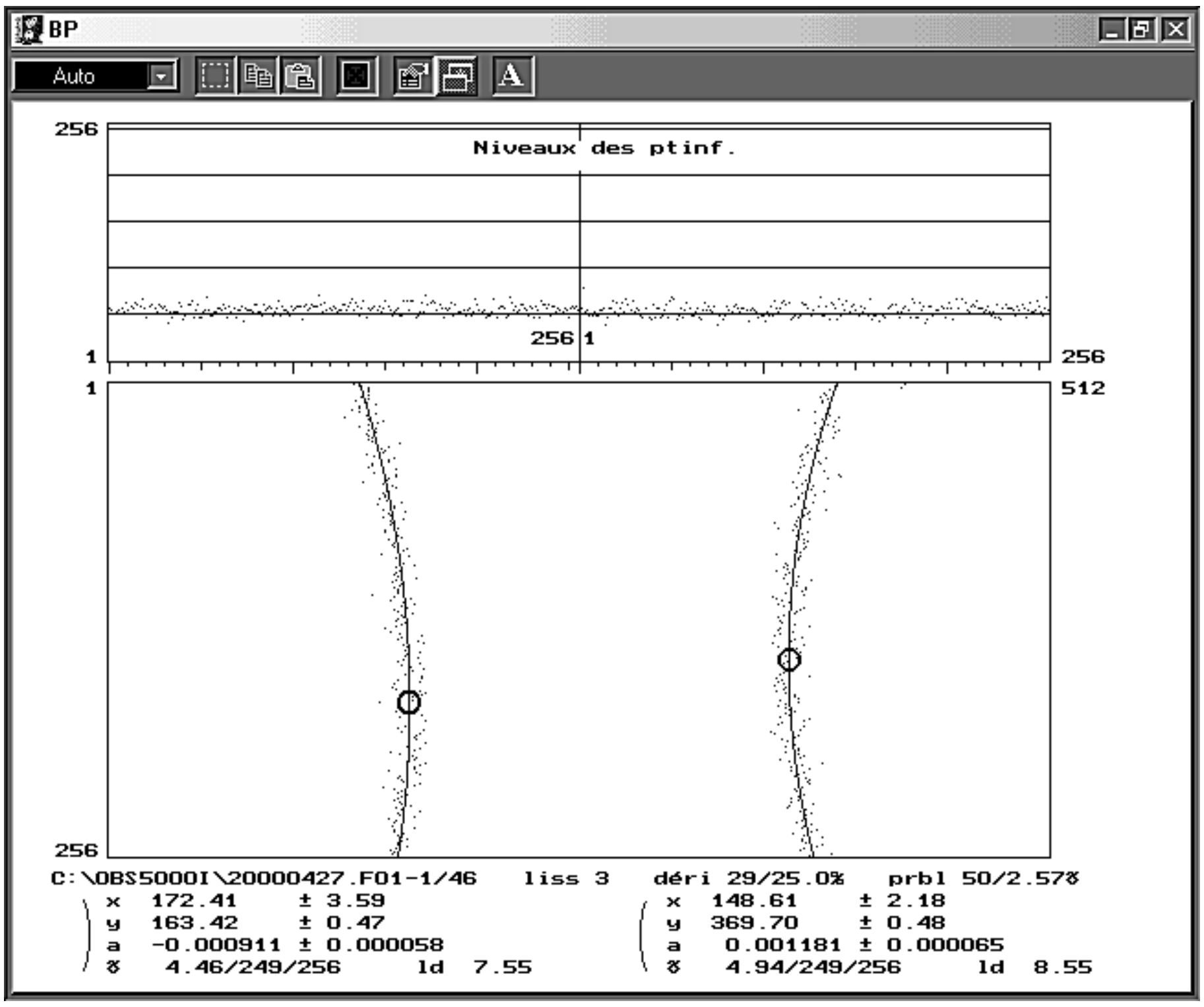

Figura 2.6. Os pontos de inflexão encontrados em cada linha, assim como as parábolas ajustadas. Os vértices das parábolas estão marcados por dois pequenos círculos.

Duas retas são ajustadas, por mínimos quadrados a estas novas duas séries de pontos, correspondentes aos pares $(Y_j, t)$, minimizando os efeitos da turbulência atmosférica e compensando o fato de o programa de redução não conseguir ajustar as parábolas correspondentes aos bordos no momento em que as imagens começam a se superpor. A localização do ponto de interseção destas duas retas dá o instante do trânsito do bordo observado, através da distância zenital do instrumento.

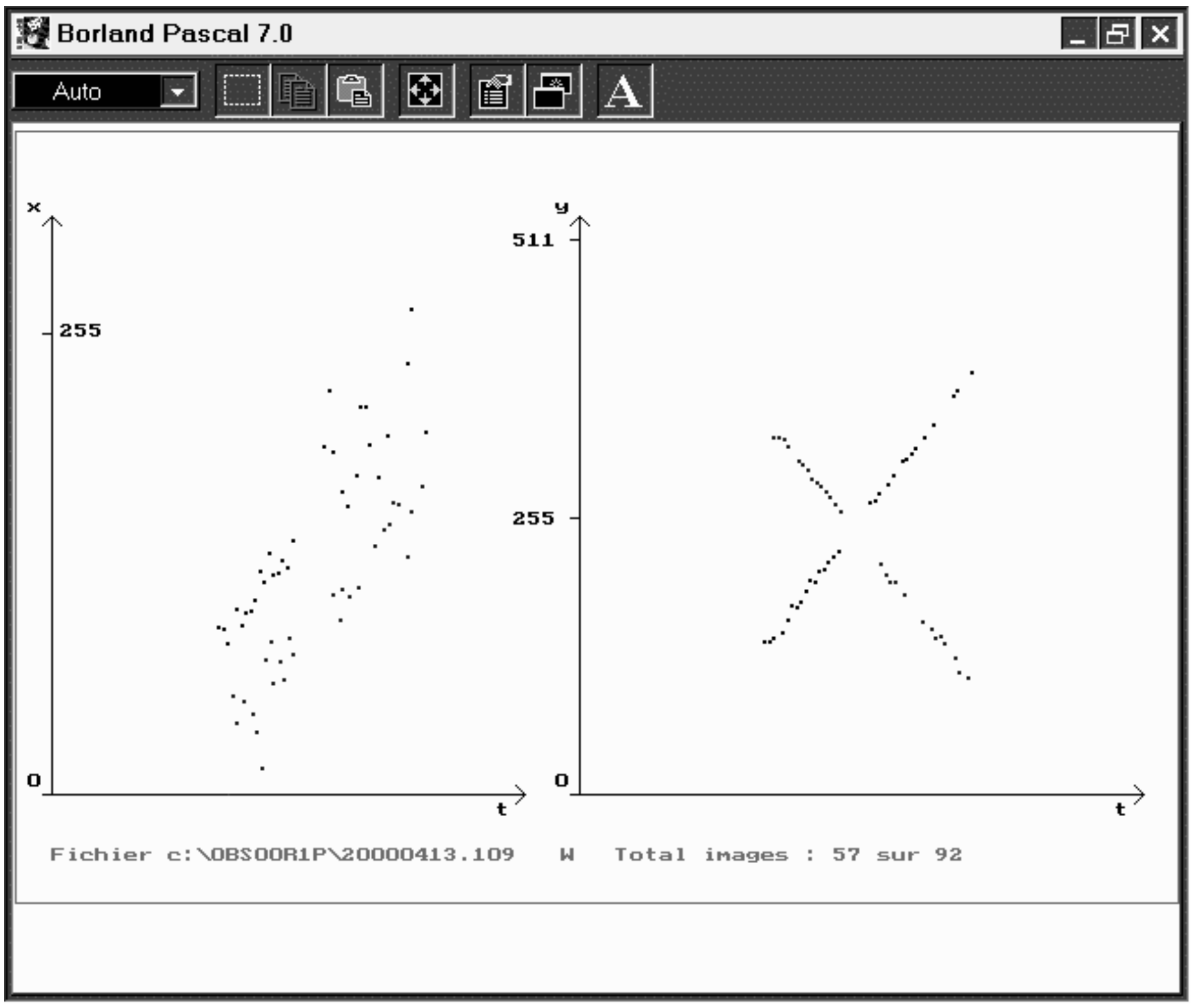

Figura 2.7. Evolução temporal das coordenadas $X_j$ (a esquerda) e $Y_j$ (a direita) dos vértices das parábolas ajustadas das imagens direta e refletida de cada *frame*.
Notar que não há pontos em torno do instante do trânsito.

O mesmo procedimento é executado para os dois bordos (superior e inferior). O diâmetro vertical observado do Sol é então determinado a partir da diferença entre estes dois instantes.

Para tanto são aplicadas as correções astrométricas às curvaturas do paralelo do almucântar e o diâmetro solar é reduzido para 1 U.A. É também aplicado a correção de primeira ordem à refração

# 2.4 Flat Field

## 2.4.1 Discussão

A diferença espacial de sensibilidade do CCD e da óptica instrumental, assim como o ruído típico do equipamento eletrônico, diminuem a precisão com que os pontos de inflexão da curva de luminosidade são determinados e consequentemente aumentam a dispersão dos resultados. Para se conhecer qualitativamente este efeito sobre o resultado final da redução, um teste de correção utilizando máscaras de *corrente de escuro* e *flat field* foi realizado.

## 2.4.2 Descrição

Um conjunto de programas e procedimentos foi desenvolvido para conversão das imagens do formato astrolábio para o formato *FITS* (*Flexible Image Transport System*) e vice-versa e para a aplicação de uma máscara de correção nas imagens antes da redução.

Para a máscara de *corrente de escuro* foram tomadas imagens com o instrumento e a câmara CCD fechados, seguindo os mesmos procedimentos de uma observação normal, com cada arquivo imagem contendo 46 *frames*. Utilizando as rotinas IRAF, estes *frames* foram combinados para a obtenção de uma única máscara composta pelas contagens médias. Para a máscara de *flat field* foram testadas imagens do fundo de céu, do disco solar (perto do centro e livre de manchas) e luz solar refletida num anteparo branco, todas obtidas e combinadas separadamente como descrito anteriormente.

## 2.4.3 Resultados

A máscara de *corrente de escuro* não foi de nenhuma utilidade, pois como a câmera possui ganho automático de sinal, a máscara resultante era composta de ruído eletrônico amplificado e de nada adiantaria ser subtraída das imagens originais. O mesmo aconteceu com a máscara de *flat field* do fundo de céu.

Entre a máscara obtida do anteparo e máscara de *flat field* tirada do disco solar, esta última demostrou melhor resultado. Esta máscara (F) era a combinação de 92 *frames* e apresentava a mais alta relação sinal/ruído, por isso, foi utilizada para os testes.

Para investigar algum efeito sistemático, foi testada também a utilização de mais duas máscaras derivadas deste *flat* F representadas por superfícies polinomiais de grau 5 (F5) e grau 20 (F20), obtidas através de rotinas IRAF. O intervalo mínimo e máximo dos pixels representam, em média, 10% da escala de contagens.

As figuras a seguir trazem as representações destas máscaras.

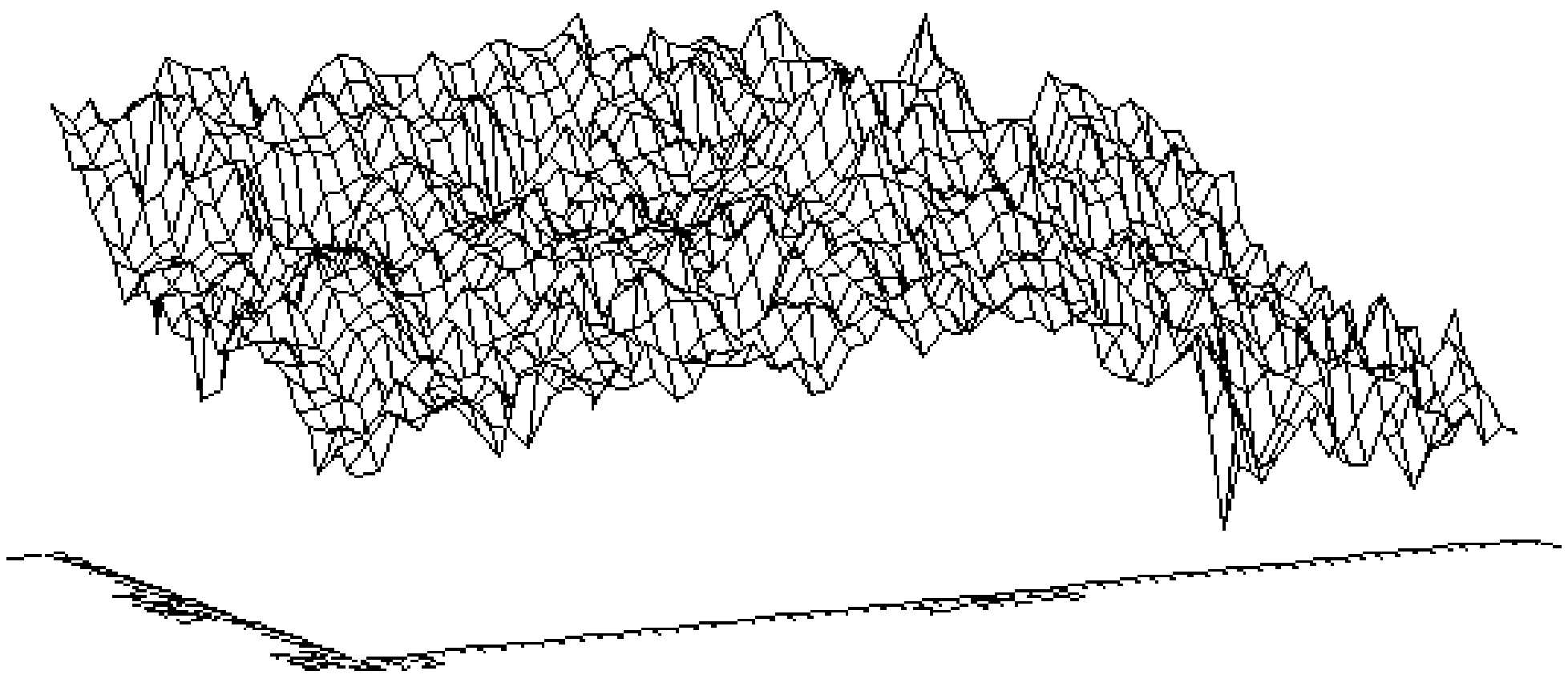

Figura 2.8. Representação da máscara de *flat field* do disco solar, gerada e plotada utilizando-se as rotinas IRAF.

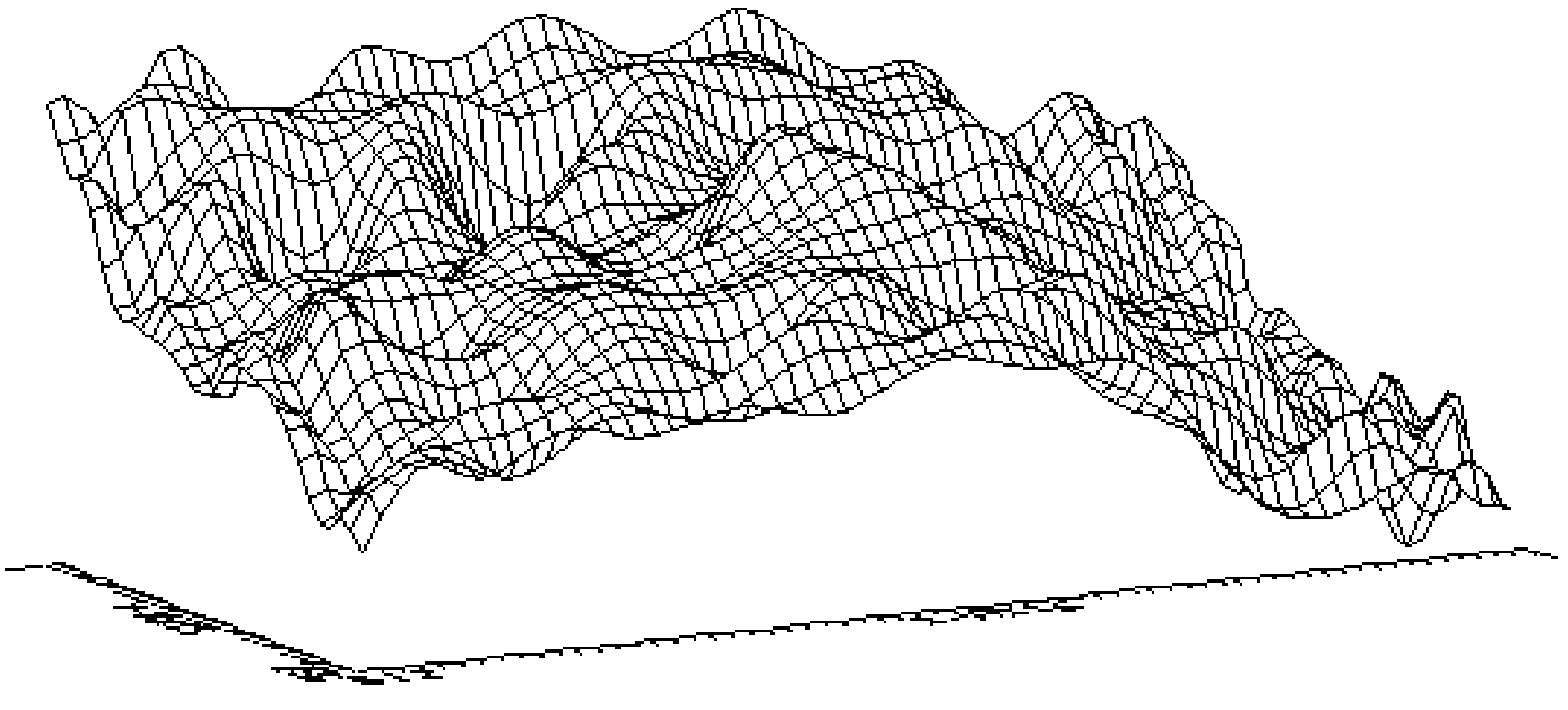

Figura 2.9. Representação polinomial (grau 20) da máscara de *flat field* do disco solar, gerada e plotada utilizando-se as rotinas IRAF.

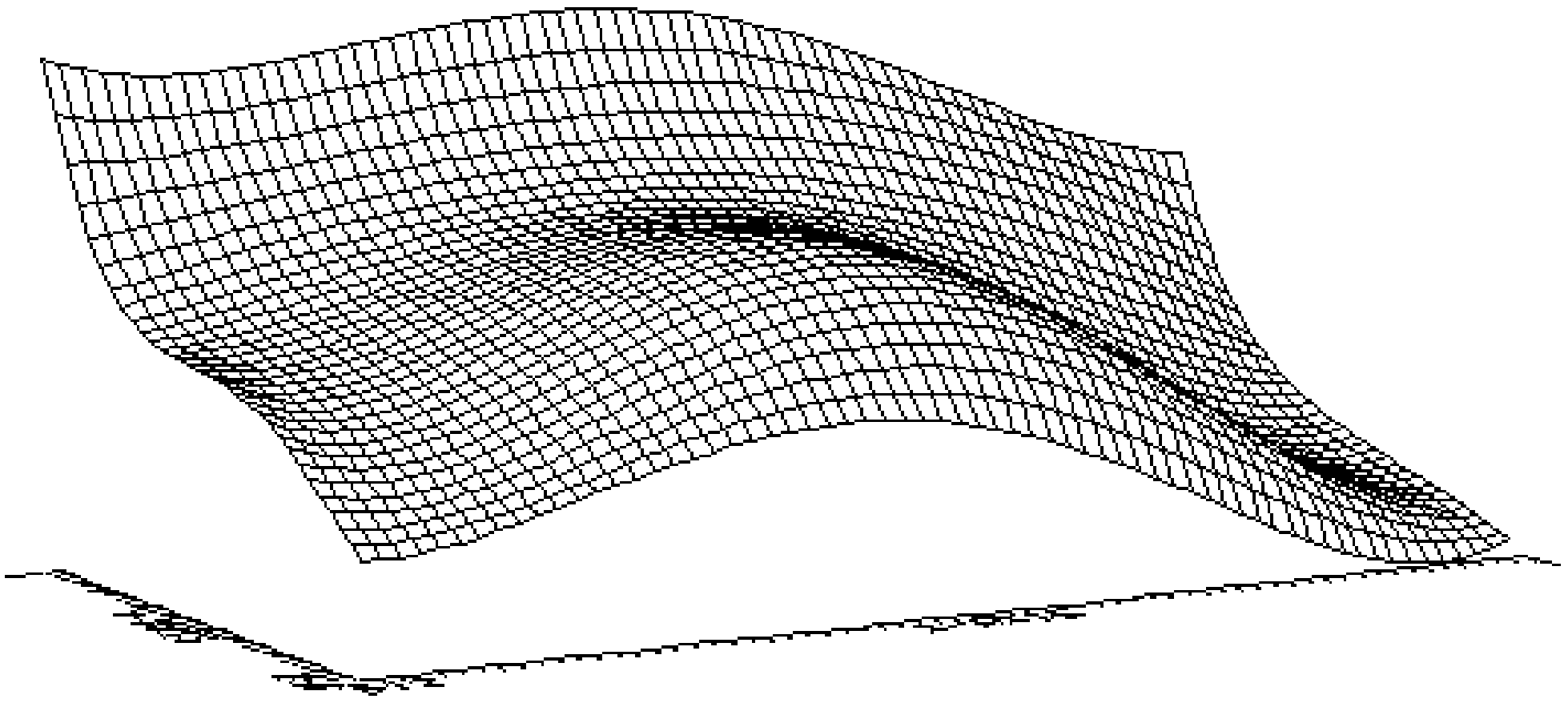

Figura 2.10. Representação polinomial (grau 5) da máscara de *flat field* do disco solar, gerada e plotada utilizando-se as rotinas IRAF. O alisamento da imagem evidencia o formato global do mapa de sensibilidade do CCD.

Como amostra de teste para as correções por *flat field*, foi utilizado um grupo de 6 sessões a Leste, contendo 76 observações e 8 sessões a Oeste, contendo 100 observações. Estas observações foram retiradas preferencialmente de dias em que houve dupla sessão, com boa qualidade observacional e próximos à época da confecção das máscaras. Em cada um dos 16192 *frames* a distribuição de contagens foi dividida pelas três máscaras normalizadas, produzindo 3 × 16192 *frames* corrigidos por *flat field*.

As observações foram então re-reduzidas.

Foram analisados vários aspectos resultantes da nova redução:

- O semidiâmetro médio da amostra (< SD >);
- O semidiâmetro médio de cada dia (< SDdia >);
- A dispersão do semidiâmetro médio diário ($\sigma_{< SDdia >}$)
- A média do erro formal do cálculo do semidiâmetro (< erro >);
- A largura média do bordo da imagem direta e refletida (LmBD e LmBR);
- A dispersão média do ajuste das parábolas ($\sigma_{pd}$ e $\sigma_{pr}$);
- A dispersão média do ajuste das retas da imagem direta e refletida ($\sigma_{rd}$ e $\sigma_{rr}$).

| Amostra | Máscara | < SD > | < SDdia > | $\sigma_{< SDdia >}$ | < erro > | < ΔSD > |
|---|---|---|---|---|---|---|
| Leste | c/ *flat* F | 959,053 | 959,074 | 0,421 | 0,149 | -0,288 |
| | c/ *flat* 20 | 959,311 | 959,351 | 0,441 | 0,142 | -0,030 |
| | c/ *flat* 5 | 959,279 | 959,316 | 0,460 | 0,143 | -0,061 |
| | s/ *flat* | **959,341** | **959,387** | **0,437** | **0,140** | |
| Oeste | c/ *flat* F | 958,439 | 958,456 | 0,441 | 0,162 | -0,275 |
| | c/ *flat* 20 | 958,675 | 958,678 | 0,433 | 0,159 | -0,040 |
| | c/ *flat* 5 | 958,669 | 958,671 | 0,418 | 0,160 | -0,046 |
| | s/ *flat* | **958,715** | **958,718** | **0,434** | **0,158** | |
| Completa | c/ *flat* F | 958,704 | 958,721 | 0,433 | 0,156 | -0,281 |
| | c/ *flat* 20 | 958,950 | 958,967 | 0,436 | 0,152 | -0,035 |
| | c/ *flat* 5 | 958,932 | 958,948 | 0,436 | 0,152 | -0,053 |
| | s/ *flat* | **958,985** | **959.004** | **0,435** | **0,150** | |

Tabela 2.1. Resultado dos teste de aplicação das máscara de *flat* F, *flat* 20 e *flat* 5 na amostra de observações, completa e dividida por lado da passagem. Todos os valores estão em segundos de arco e os valores em negrito referem-se à imagens originais. ΔSD é a diferença entre o semidiâmetro da nova redução e o originalmente encontrado.

| Amostra | Máscara | LmBD | LmBR | $< \sigma_{pd} >$ | $< \sigma_{pr} >$ | $< \sigma_{rd} >$ | $< \sigma_{rr} >$ |
|---|---|---|---|---|---|---|---|
| Leste | c/ *flat* F | 5,48 | 6,16 | 1,02 | 1,02 | 4,48 | 4,49 |
| | c/ *flat* 20 | 5,12 | 5,74 | 1,01 | 1,03 | 4,87 | 4,87 |
| | c/ *flat* 5 | 5,10 | 5,73 | 1,00 | 1,02 | 4,84 | 4,86 |
| | s/ *flat* | **5,08** | **5,71** | **1,00** | **1,04** | **4,88** | **4,91** |
| Oeste | c/ *flat* F | 4,51 | 5,06 | 1,04 | 1,04 | 5,10 | 5,02 |
| | c/ *flat* 20 | 4,74 | 5,27 | 1,02 | 1,03 | 5,14 | 5,10 |
| | c/ *flat* 5 | 4,72 | 5,26 | 1,03 | 1,03 | 5,10 | 5,10 |
| | s/ *flat* | **4,71** | **5,27** | **1,02** | **1,04** | **5,14** | **5,11** |
| Completa | c/ *flat* F | 4,93 | 5,53 | 1,03 | 1,04 | 4,83 | 4,79 |
| | c/ *flat* 20 | 4,91 | 5,47 | 1,02 | 1,02 | 5,02 | 5,00 |
| | c/ *flat* 5 | 4,89 | 5,46 | 1,02 | 1,03 | 4,98 | 5,00 |
| | s/ *flat* | **4,87** | **5,46** | **1,03** | **1,02** | **5,03** | **5,02** |

Tabela 2.2. Resultado dos teste de aplicação das máscara de *flat* F, *flat* 20 e *flat* 5 na amostra de observações, completa e dividida por lado da passagem. Todos os valores estão em pixels e os valores em negrito referem-se à imagens originais.

Da aplicação das três máscaras, a correção utilizando-se o *flat* F foi a que resultou em imagens menos ruidosas que as originais. A contagem de regiões da imagem contendo apenas o céu mostrou que a redução da dispersão foi de 0",05, em média. As máscaras de *flat* alisadas, F20 e F5, não reduziram a dispersão das contagens.

Dos resultados mostrados acima, o que mais chama a atenção é a redução no valor do semidiâmetro, quando o *flat* F é aplicado às imagens. A redução de ≈ 0",3 é observada nos dois conjuntos, Leste e Oeste. O ajuste das retas que materializa a trajetória dos bordos melhorou e os demais parâmetros testados mudam pouco.

Para verificar a influência destas correções sobre as coordenadas dos pontos de inflexão da função de luminosidade em cada uma das 256 linhas de uma imagem, foi escolhido o quinto *frame* de duas imagens astrolábio de um dia da amostra em que houve dupla sessão Verificou-se que a coordenada $y_i$ dos pontos de cada linha *i*, tendiam sistematicamente para o centro da imagem, como mostrado na tabela abaixo.

| Sessão | Imagem | Desvio médio da coordenada x do ponto de inflexão (pixel) | | |
|---|---|---|---|---|
| | | *flat* F | *flat* 20 | *flat* 5 |
| Leste | direta | +0,08 | +3,06 | +1,83 |
| | refletida | -2,72 | -0,73 | -0,46 |
| Oeste | refletida | +4,72 | +0,16 | +1,38 |
| | direta | -3,09 | -1,32 | -1,25 |

Tabela 2.3. Desvio médio do ponto de inflexão da coordenada x das imagens tratadas com as máscaras de *flat field*.
O sinal positivo significa que o ponto se deslocou para a direita, e no sentido oposto para o sinal negativo. As imagens direta e refletida invertem posição no CCD.

## 2.4.4 Conclusões

A aplicação dos *flats* alisados pouco mudou o resultado das novas reduções. O *flat* F, por sua vez, torna as imagens menos ruidosas e deve corrigir o valor do semidiâmetro observado, fazendo-o diminuir, consistentemente, cerca de 0",3 para a amostra estudada.

A re-redução de toda a série para a correção de *flat field* seria inexeqüível ou, no mínimo, muito trabalhosa.

Como será mostrado nos capítulos seguintes, outros tratamentos de dados, mais viáveis, também corrigem a série, de forma a extrair dela informações satisfatórias e de grande precisão, comparáveis a outros resultados publicados na literatura.

Outra máscara de *flat field* tirada do disco solar foi obtida, da mesma forma já explicada, seis meses depois deste teste e mostrou-se quase idêntica a máscara de *flat* F, comprovando a estabilidade do sistema de aquisição de imagem.

# Capítulo 3

## O Conjunto de Medidas

### 3.1 Quantidade de medidas e Pré-tratamento

No campus do Observatório Nacional, as medidas do diâmetro do Sol são realizadas em ambos os lados da passagem, sempre que as condições climáticas permitem. Assim, ao longo da campanha de 1998, 1999 e 2000, mais de 10800 medidas individuais de monitoramento do diâmetro do Sol foram realizadas. Por outro lado, o bordo observado é determinado a partir de 46 *frames* contendo as imagens direta e refletida, cujos pontos do perfil são pesquisados ao longo de 256 linhas da matriz do CCD. Portanto, para cada medida 47104 porções elementares do limbo solar contribuem para o resultado final.

Em ambos os casos, há uma grande quantidade de determinações unitárias, permitindo a utilização de métodos estatisticamente robustos de filtragem, isto é, sem utilizar modelos apriorísticos.

O critério de remoção se baseia num teste de quartis, com o fator multiplicativo variando entre 1.5 (mínima rejeição) e 3 (máxima rejeição). Se o número de pontos utilizados for inferior a 50, então nenhuma parábola é ajustada para o respectivo bordo. Quando das observações iniciais, em 1997, foram definidos três níveis de critérios: 1.7, 2.0 e 2.5. Se as condições observacionais forem boas, três soluções são obtidas, caso contrário, apenas as soluções 1.7 e 2.0 alcançam resultado ou, muito mais raramente, apenas a primeira.

As figuras a seguir apresentam estas três soluções obtidas para as séries brutas observadas a Leste e a Oeste da passagem meridiana, para todo o período. As curvas foram alisadas segundo uma média corrida de 600 pontos apenas para melhor visualização.

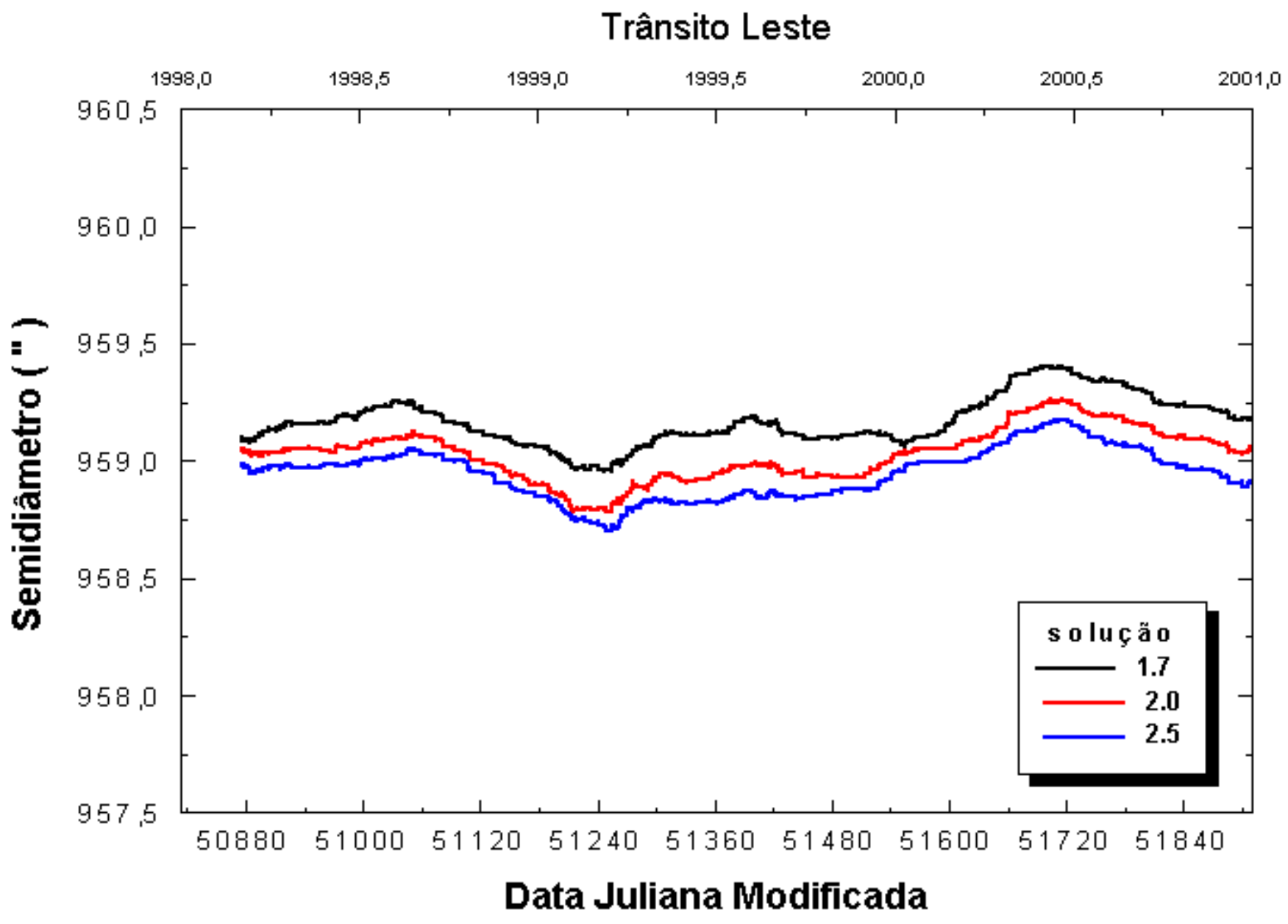

Figuras 3.1 As três soluções obtidas com a modificação do valore do critério de rejeição dos pontos que serão usados para definir a parábola para as observações a Leste.

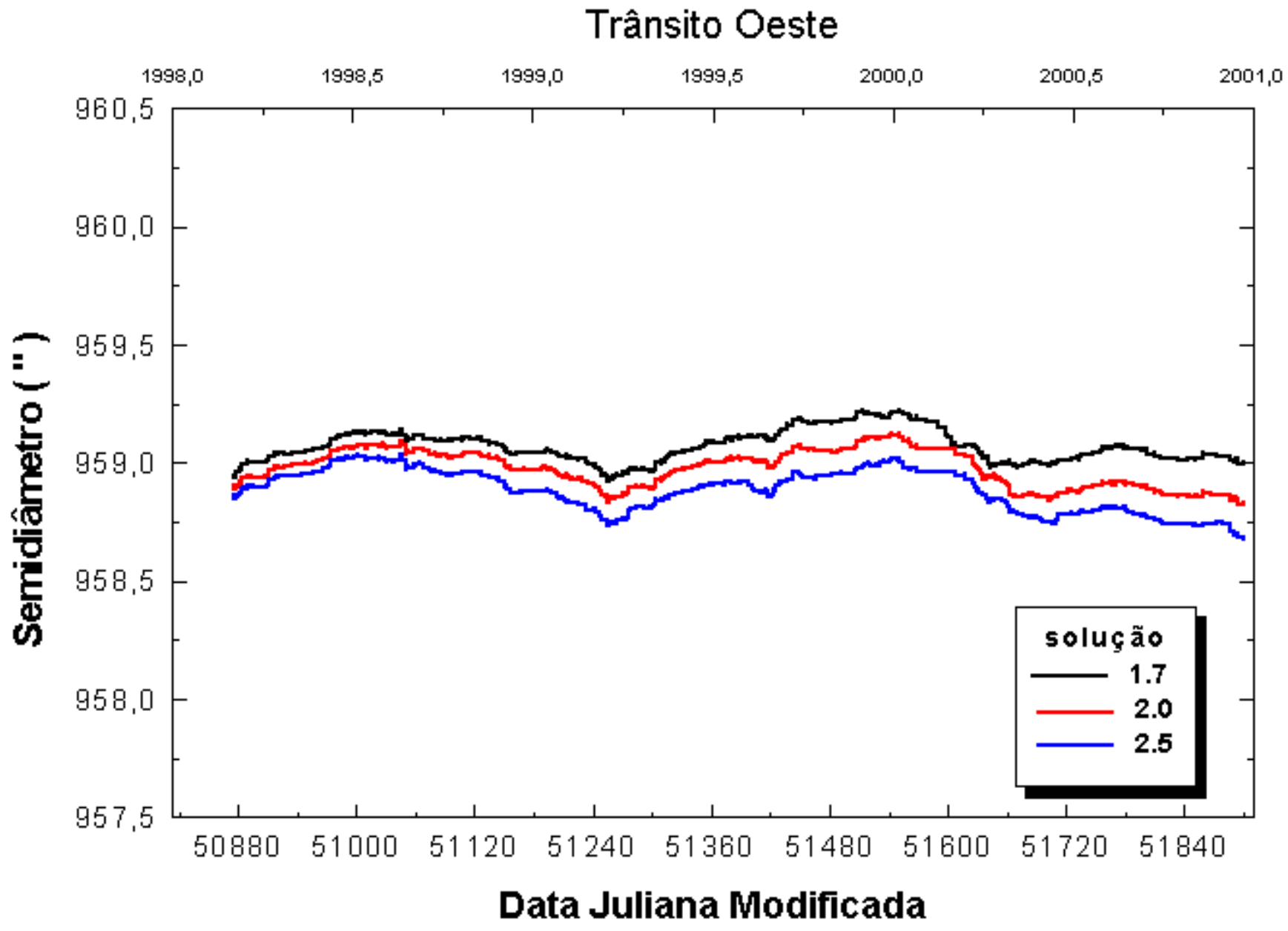

Figuras 3.2 As três soluções obtidas com a modificação do valore do critério de rejeição dos pontos que serão usados para definir a parábola para as observações a Oeste.

As soluções possuem entre si, em média, uma diferença de cerca de 0",10. O valor absoluto do semidiâmetro do Sol, neste trabalho, passa a ser uma questão de escolha da solução ou das soluções que comporão o resultado final, uma vez que as diferentes soluções poderiam ser normalizadas, adotando-se um modelo para a janela de coerência atmosférica. Como se verá adiante, a janela de coerência (medida pelo parâmetro de Fried) e a largura do bordo serão utilizadas para a consideração de efeitos sistemáticos. Vale notar, porém, que as mesmas variações do semidiâmetro solar (evolução temporal) estão presentes nas três curvas, sendo que a correlação entre os pontos originais das três soluções é acima de 0,9. No entanto, evidentemente, a dependência climática faz com que o número de dias e de observações considerados seja diferente:

| **Solução** | **1.7** | **2.0** | **2.5** |
|---|---|---|---|
| **Dias** | 450 | 434 | 434 |
| **Observações** | 10807 | 10315 | 9716 |

Tabela 3.1. Numero de dias e observações considerados, segundo as três soluções obtidas, ao longo de todo o período de 1998, 1999 e 2000.

A solução menos criteriosa (1.7) foi escolhida para a análise das observações a Leste e a Oeste e da série como um todo, pois além de apresentar comportamento semelhante às demais, retém um número maior de dias e pontos.

A tabela 3.2 a seguir traz os números de dias observados para cada ano da campanha, assim como o número total de observações correspondentes:

| | ANO | | | TOTAL |
|---|---|---|---|---|
| | 1998 | 1999 | 2000 | |
| Leste (dias) | 111 | 129 | 116 | 356 |
| Leste (observações) | 1647 | 2039 | 1492 | 5178 |
| Oeste (dias) | 123 | 131 | 125 | 379 |
| Oeste (observações) | 1847 | 2108 | 1674 | 5628 |
| **Dias** | **142** | **155** | **153** | **450** |
| **Observações** | **3494** | **4147** | **3166** | **10807** |

Tabela 3.2 Número de dias observados e o número de medidas individuais.

Em média, a campanha de monitoramento do diâmetro solar do Observatório Nacional realizou nesses três anos:

150 dias de observação/ano

**3602 observações individuais/ano**

24 observações/dia ou ≈15 observações/sessão

A comparação com outras campanhas permite estimar a dimensão destes números:

- Toulmonde (1997) compilou os dados de medições do diâmetro solar dos últimos três séculos, obtidos com variados métodos, num total de 71000 observações de 1660 até 1991. Média de **208 observações/ano**.

- Laclare *et al.* (1996) publicou os resultados de medições visuais do diâmetro solar, usando um astrolábio. De 1975 até 1994 foram 5153 observações. Média de **271 observações/ano.**

- Golbasi *et al.* (2001), publicou os resultados obtidos com o astrolábio solar do Observatório Nacional de Tubiak. De 1996 até 2000, foram 1304 observações utilizando-se uma câmera CCD adaptada ao instrumento. Média de **261 observações/ano.**

Tendo em vista a grande quantidade de medidas independentes e a escolha pela solução mais "relaxada" de ajuste da parábola que descreve o limbo solar, alguns critérios foram adotados para se eliminar, do conjunto total, possíveis observações discrepantes, ou de qualidade duvidosa. Em primeiro lugar foram descartados os pontos que se afastavam de mais de $2{,}5\sigma$ do valor médio de toda a série. A seguir, porque neste caso os termos de 2ª ordem tornam-se excessivamente grandes, foi adotado o critério de desconsiderar os resultados provenientes das reduções onde o ângulo de orientação da matriz CCD ultrapassava 2º, com respeito ao horizonte. Por fim, foram rejeitadas as medidas referentes às sessões (Leste ou Oeste) com menos de 4 observações.

Após esta primeira filtragem dos pontos, o número de observações descartadas foi de 1694, restando um total de 9112 medidas individuais e independentes utilizadas neste estudo. A tabela 3.3 mostra o resultado do emprego progressivo dos critérios citados anteriormente:

| Observações | Totais | $\sigma < 2{,}5$ | $\theta < 2^{o}$ | $N \geq$ de 4 |
|---|---|---|---|---|
| Leste | 5178 | 4960 | 4301 | 4246 |
| Oeste | 5628 | 5507 | 4923 | 4866 |
| SÉRIE COMPLETA | 10806 | 10467 | 9224 | **9112** |

Tabela 3.3. O número de pontos totais das séries por lado da passagem e da série completa com os critérios aplicados cumulativamente. $\theta$ é o ângulo de correção da inclinação das linhas do CCD e N o número de observações por sessão.

## 3.2 Correções Instrumentais

Durante o período entre abril e outubro de 2000 um procedimento falho de manutenção fez com que o ângulo do prisma frontal (i.e., a distância zenital instrumental) apresentasse uma evolução durante as sessões. Durante as observações, após o ajuste da distância zenital, o sistema de molas do prisma de reflexão deslizava um pouco diminuindo a distância zenital observada. Para as observações a Leste este fenômeno fazia com que a imagem do Sol demorasse um pouco mais para cruzar o almucantar, aumentando o diâmetro calculado. O oposto acontecia para as observações a Oeste, que apresentava uma diminuição significativa nos semidiâmetros calculados. A comparação entre as observações a Leste e a Oeste feitas durante este intervalo permitiu encontrar uma correlação entre o valor do semidiâmetro observado e a duração da respectiva medida. O coeficiente encontrado através de mínimos quadrados foi de $(8{,}3\pm0{,}6)\times10^{-4}$ "/s.

Aplicada a correção, os dados deste período específico foram recuperados.

Para o período da correção o valor médio do semidiâmetro era de 959",180 e $\sigma$ igual a 0",596. Após a correção o valor médio do semidiâmetro subiu para 959",184 enquanto o $\sigma$ diminuiu para 0",572. A média das diferenças absolutas entre os valores brutos e corrigidos das séries foi de 0",189.

O gráfico a seguir mostra toda a série (mais uma vez mostrada com um alisamento de 600 pontos), com destaque para o período onde foi necessário aplicar a correção.

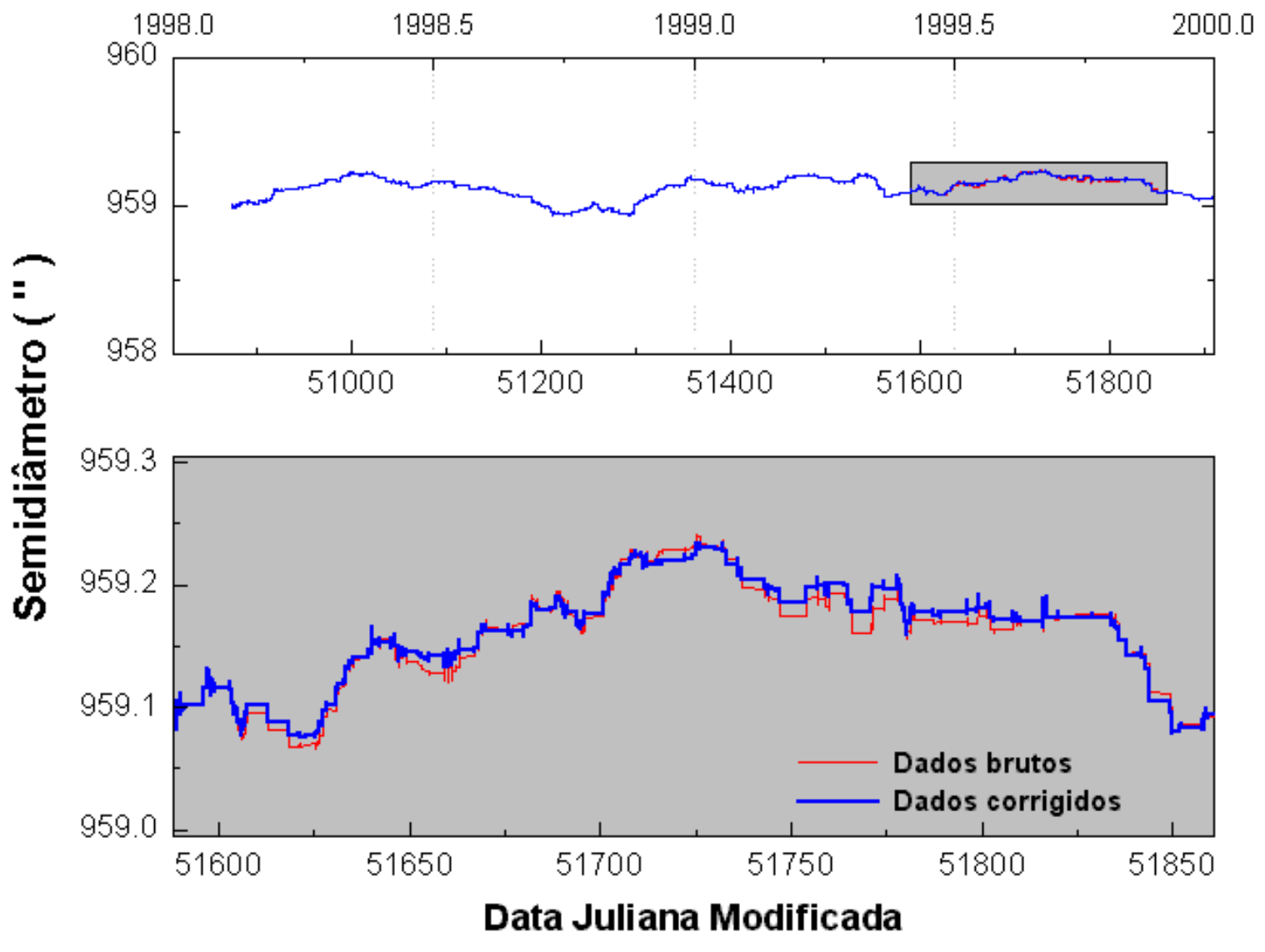

Figura 3.3. No gráfico superior, destaque para o período onde foi necessária a correção. No gráfico inferior, o detalhe deste período em particular, podendo se notar que a correção pouco alterou os valores originais da série.

## 3.3 Correções Observacionais

Paradoxalmente, a estrela mais próxima a Terra é um dos objetos para o qual a observação apresenta maior número de dificuldades. Se o sinal é superabundante, igualmente o é o nível de ruído, com a agravante de apresentar características marcantemente sistemáticas. As observações solares com astrolábios se iniciaram há quase 25 anos, voltadas para a definição do sistema de referência astronômico (eclítica, equador e obliquidade da eclítica). Embora a acurácia requerida pelos sistemas FK (*FK4 Catalogue* - Fricke e Kopff, 1963 e *FK5 Catalogue* - Fricke *et al.*, 1988) fosse da ordem do centésimo de segundo de arco, evidenciou-se a capacidade do astrolábio de manter sua planejada excelência metrológica mesmo para as difíceis condições da observação diurna. No entanto, Penna *et al.* (1997), mostram que as variáveis meteorológicas e suas variações devem ser levadas em conta, inclusive quanto à direção de visada. Da mesma forma, Irbah *et al.* (1994) mostram que o tamanho da janela de coerência atmosférica influi na determinação do diâmetro solar, tanto pelo espalhamento da imagem, como pelos termos em baixa freqüência (minutos), comensuráveis com a duração das observações. Finalmente, no curso deste projeto, as experiências para a determinação da correção de *flat field* mostraram que a deterioração da qualidade da imagem obtida se traduz sobre a largura do bordo e na dispersão desta, levando ultimamente a variações no diâmetro solar medido.

Deste modo, a preocupação com possíveis modulações que as condições observacionais, ou suas variações durante as sessões pudessem estar introduzindo nos valores do semidiâmetro levou à consideração de diversos parâmetros:

- temperatura;
- variação de temperatura (durante a sessão);
- pressão atmosférica;
- variação de pressão atmosférica (durante a sessão);
- distância zenital;
- evolução da distância zenital (durante a sessão);
- tangente da distância zenital;
- azimute;
- evolução do azimute (durante a sessão);
- fator de Fried.

Além destes parâmetros, conforme discutido acima, mais um foi considerado: o desvio padrão do ajuste das parábolas que descrevem o limbo solar, tanto na imagem direta, quanto na imagem refletida pela superfície de mercúrio.

Desta forma, se obteve uma verificação detalhada dos fatores que poderiam estar influenciando a dispersão das imagens e/ou sistematicamente desviando o ponto de inflexão da curva de obscurecimento do limbo solar, e conseqüentemente o valor do semidiâmetro calculado.

Cada um dos parâmetros foi testado separadamente, na forma de três equações com as seguintes condições:

a) Um único semidiâmetro médio e um valor único para o coeficiente paramétrico, tanto para as observações a Leste quanto para a Oeste:

$$SD = SD_{obs} - \left(k_{par} \cdot \text{parâmetro}\right)$$

b) Um único semidiâmetro médio, mas dois coeficientes paramétricos independentes para as observações Leste e Oeste:

$$SD = SD_{obs} - \left(k_{par}^{Leste} \cdot \text{parâmetro}^{Leste}\right) - \left(k_{par}^{Oeste} \cdot \text{parâmetro}^{Oeste}\right)$$

c) Semidiâmetros e coeficientes paramétricos independentes para as observações a Leste e a Oeste:

$$\begin{cases} SD^{Leste} = SD_{obs}^{Leste} - \left(k_{par}^{Leste} \cdot \text{parâmetro}^{Leste}\right) \\ SD^{Oeste} = SD_{obs}^{Oeste} - \left(k_{par}^{Oeste} \cdot \text{parâmetro}^{Oeste}\right) \end{cases}$$

Para selecionar os parâmetros mais relevantes vários critérios foram impostos a serem satisfeitos simultaneamente:

- Que o desvio padrão diminuísse em relação à série não tratada;
- Que o coeficiente paramétrico fosse estatisticamente significante (acima de $3\sigma$);
- Que o semidiâmetro ajustado (corrigido) se mantivesse compatível com o valor original da série não tratada (dentro de $3\sigma$);
- Que as soluções para as três equações teste fossem compatíveis entre si para os parâmetros testados.

Uma análise através de mínimos quadrados revelou que os parâmetros mais relevantes, ou seja, os que se mostraram influentes no resultado final da redução, foram a temperatura, a evolução da temperatura ao longo da sessão, o fator de Fried e o desvio padrão do ajuste da parábola da imagem refletida.

Laclare *et al.* (1996), na analise dos dados do Observatório de Calern, encontrou uma dependência sistemática do semidiâmetro com a distância zenital, no sentido de crescimento do valor do semidiâmetro com a diminuição da distância zenital. A série aqui estudada apresenta, ao contrário, uma tendência de aumento do semidiâmetro com o aumento da distância zenital, no valor de 0",003±0",001/º. A correlação entre estes dois valores não se mostrou relevante para o resultado final da redução.

Os parâmetros foram normalizados por seus valores médios e desvios padrão, de forma que os coeficientes de correção associados a cada parâmetro traduzissem a influência sobre a série original em termos de segundos de arco.

A tabela dos resultados dos testes com todos os parâmetros está no Anexo I.

Os resultados das três equações testes são mostrados na tabela 3.4.

| | Parâmetro | Temperatura | Δ temperatura | fator de Fried | σ parábola |
|---|---|---|---|---|---|
| **Teste (a)** | σ | 0,550 | 0,551 | 0,552 | 0,550 |
| | **⟨SD⟩** | 959,108±0,006 | 959,108±0,006 | 959,108±0,006 | 959,108±0,006 |
| | **$K_{par}$** | -0,043±0,006 | +0,033±0,006 | +0,017±0,006 | -0,053±0,006 |
| **Teste (b)** | σ | 0,550 | 0,551 | 0,552 | 0,550 |
| | **⟨SD⟩** | 958,794±0,046 | 959,143±0,009 | 959,223±0,045 | 959,108±0,013 |
| | **$K_{par}$ (L)** | -0,039±0,006 | +0,039±0,007 | +0,021±0,007 | -0,050±0,007 |
| | **$K_{par}$ (O)** | -0,042±0,006 | +0,024±0,007 | +0,010±0,005 | -0,056±0,007 |
| **Teste (c)** | σ | 0,550 | 0,551 | 0,552 | 0,549 |
| | **⟨SD⟩ (L)** | 958,875±0,064 | 959,158±0,011 | 959,380±0,078 | 959,068±0,017 |
| | **⟨SD⟩ (O)** | 958,716±0,065 | 959,120±0,013 | 959,159±0,054 | 958,947±0,018 |
| | **$K_{par}$ (L)** | -0,049±0,009 | +0,025±0,009 | +0,011±0,009 | -0,078±0,009 |
| | **$K_{par}$ (O)** | -0,032±0,008 | +0,031±0,008 | +0,026±0,008 | -0,034±0,008 |

Tabela 3.4. Resultados dos testes para a correção dos valores observacionais da série. Todos os valores estão em segundos de arco. Para comparação, o valor médio do semidiâmetro da série original é de 959",107 ± 0",006 e σ igual a 0",570.

Uma vez que os resultados fornecidos pelas três equações de condição são concorrentes, a solução (a) é a que melhor descreve os efeitos sistemáticos, além de fornecer a mais direta descrição física do sistema. As correções foram obtidas de uma solução, por mínimos quadrados, que incluía todas as observações e todos os quatro parâmetros simultaneamente, além de um termo linear dependente do tempo (data Juliana modificada) e um termo constante, isto é, o semidiâmetro para dj = $dj_0$.

$$SD_{obs} = SD_0 + \left[k_{dj} \cdot (dj - dj_0)\right] + (k_T \cdot T) + (k_{\Delta T} \cdot \Delta T) + (k_{ff} \cdot ff) + (k_{\sigma_{par}} \cdot \sigma_{par})$$

Onde: $SD_{obs}$ - semidiâmetro calculado/observado;

$SD_0$ - semidiâmetro para $dj = dj_0$;

$dj$ e $dj_0$ - data Juliana e a data Juliana do começo da série;

$T$ - temperatura no momento da observação;

$\Delta T$ - variação da temperatura durante a sessão;

$ff$ - valor do fator de Fried

$\sigma_{par}$ - valor do sigma do ajuste da parábola refletida;

$k's$ - são os coeficientes de correção dos respectivos parâmetros.

A tabela 3.5 a seguir mostra os valores obtidos para os coeficientes da solução simultânea e suas incertezas:

| **Parâmetro** | **coeficiente** |
|---|---|
| Temperatura | - 0",032 ± 0",006 |
| ΔTemperatura | + 0",022 ± 0",006 |
| fator de Fried | + 0",023 ± 0",006 |
| σ do ajuste da parábola | - 0",050 ± 0",006 |

Tabela 3.5. Valores dos coeficientes de correção dos respectivos parâmetros

As alterações introduzidas pelas correções simultâneas somam poucas centenas de segundo de arco e não modificam a série drasticamente. Isto demonstra que os efeitos sistemáticos sobre os resultados das observações são relativamente pequenos, apesar de relevantes.

A temperatura no momento da medida é o parâmetro que mais claramente influencia o resultado final. Os outros três variam de modo complexo e parecem perturbar a medida/cálculo do diâmetro somente quando assumem valores extremos.

Os gráficos a seguir mostram a evolução do valor do semidiâmetro de toda a série superposto com a evolução do valor dos parâmetros supracitados ao longo do mesmo período. Para melhor visualização todas as curvas foram alisadas por uma média móvel de 600 pontos.

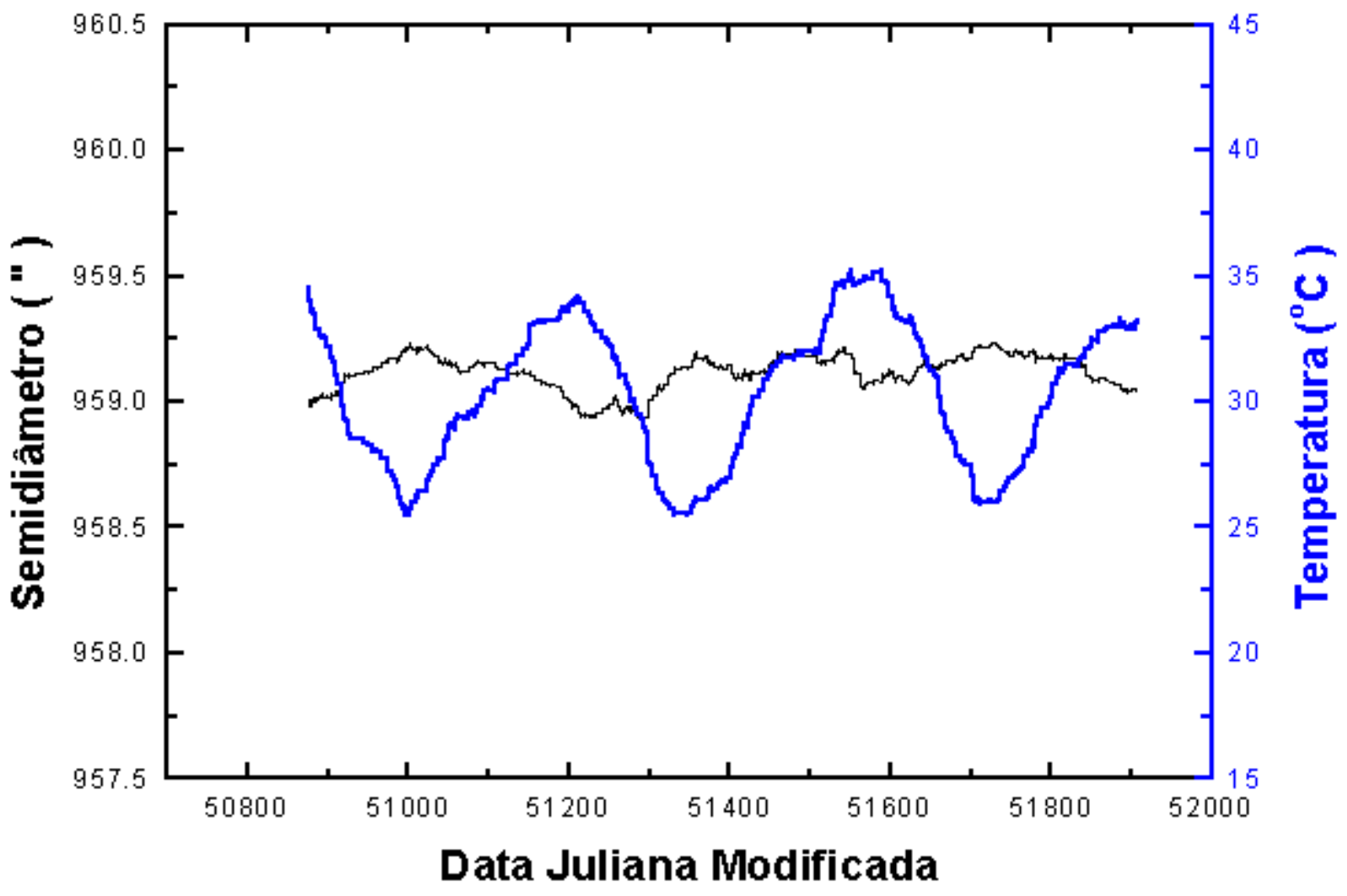

Figura3.4. Superposição da evolução semidiâmetro e temperatura ao longo de todo o período, alisados segundo uma média móvel de 600 pontos. A escala a esquerda está em segundos de arco e a direita em $^{o}$C.

Existe uma clara anticorrelação entre a temperatura no momento da medida e o valor do semidiâmetro observado.

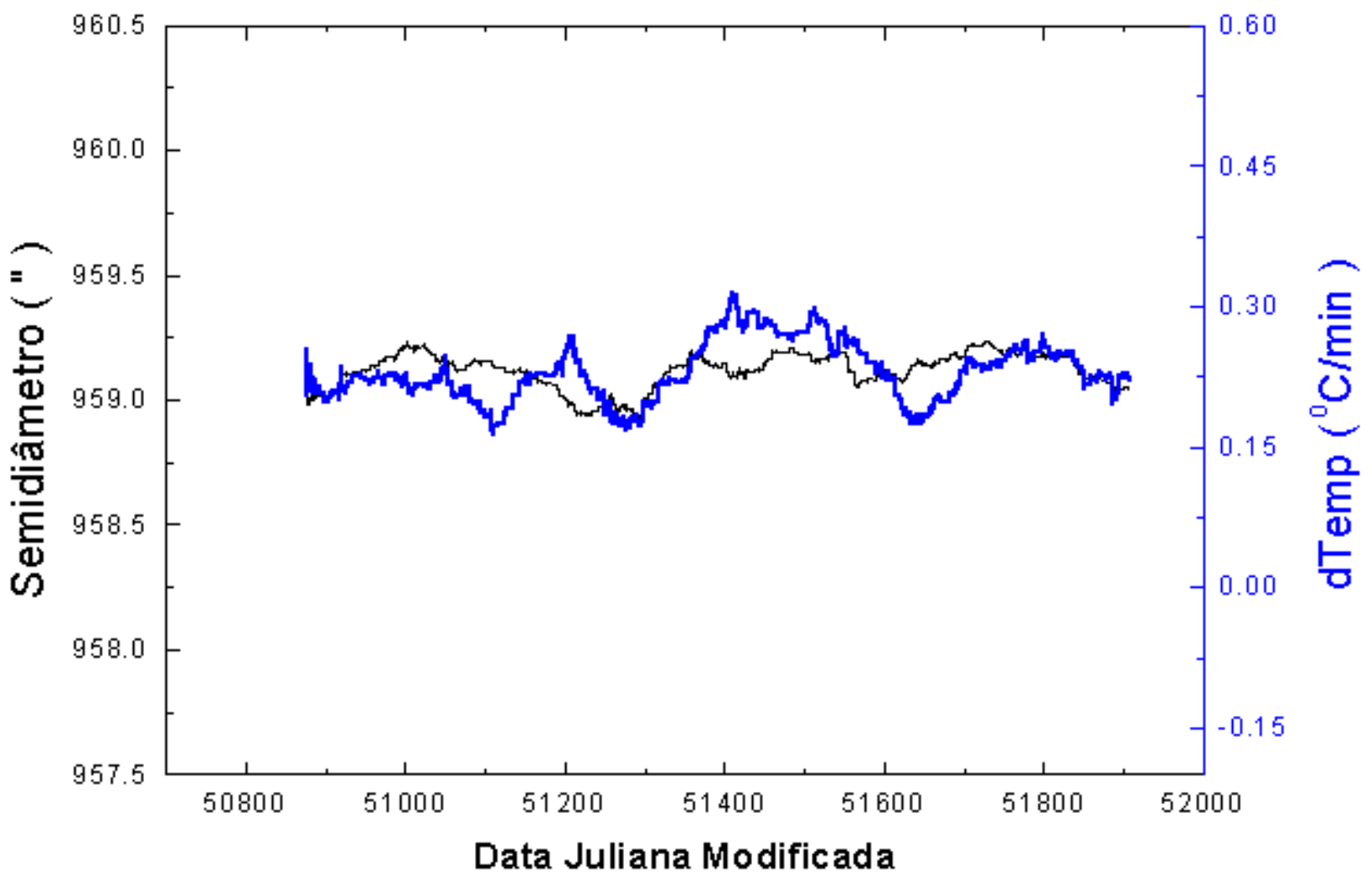

Figura 3.5. Superposição da evolução semidiâmetro e da variação da temperatura durante a sessão, ao longo de todo o período, alisados segundo uma média móvel de 600 pontos. A escala a esquerda está em segundos de arco e a direita em $^{o}$C/min.

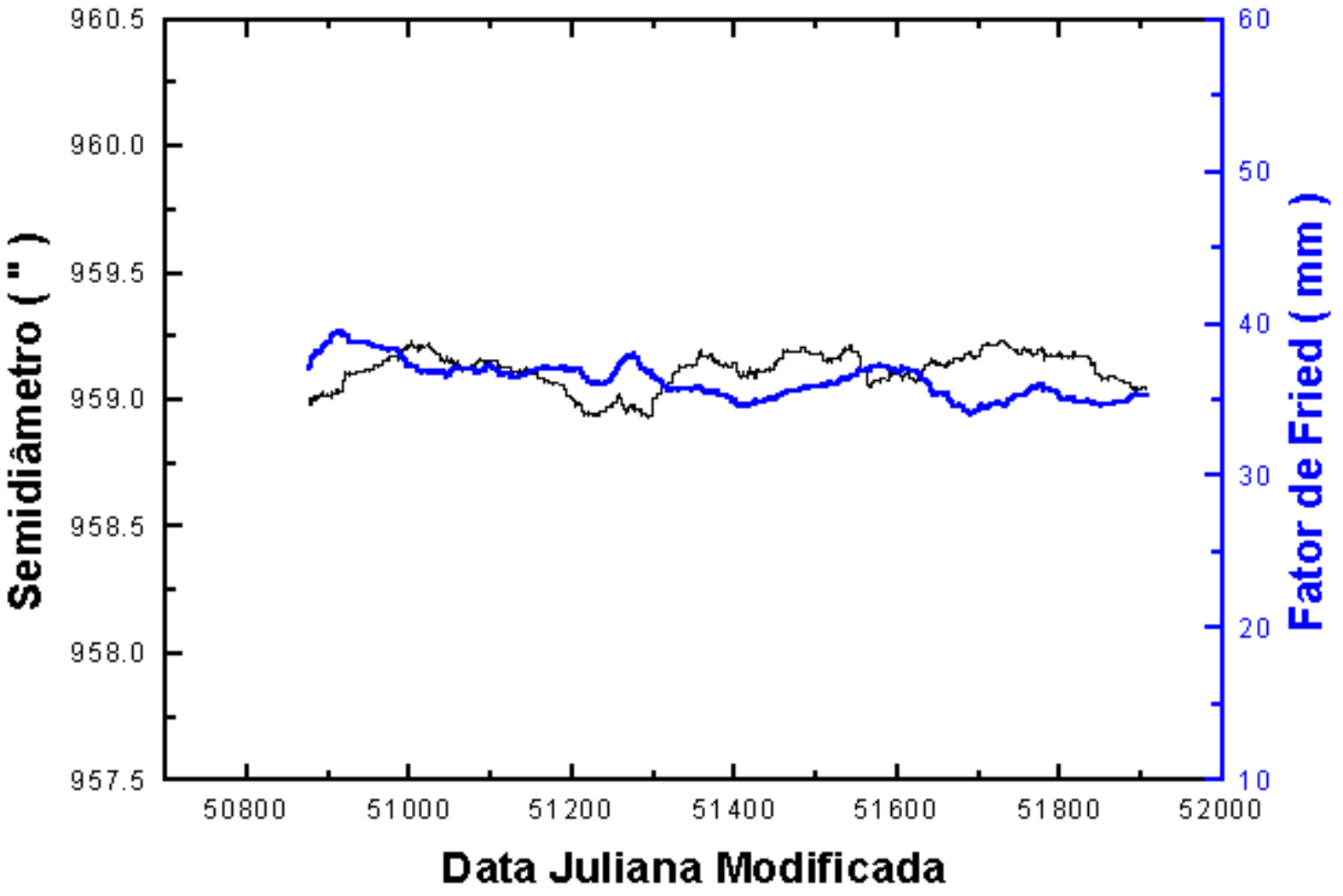

Figura 3.6. Superposição da evolução semidiâmetro e do fator de Fried, ao longo de todo o período, alisados segundo uma média móvel de 600 pontos. A escala a esquerda está em segundos de arco e a direita em mm.

O fator de Fried descreve a qualidade do *seeing* da atmosfera e está relacionado ao tamanho da área (janela) de coerência com que a frente de onda chega até o instrumento. Este fator é deduzido dos próprios dados observacionais (Irbah *et al.*, 1994) através da relação: $r_0 = 8{,}25 \times 10^5 \cdot D^{-1/5} \cdot \lambda^{6/5} \cdot \left(\sigma_s^2\right)^{-3/5}$, onde $D$ é o diâmetro da pupila de entrada, $\lambda$ é o comprimento de onda central observado e $\sigma_s$ pode ser entendido como a média, sobre toda a pupila de entrada, das flutuações do ângulo de chegada da frente de onda. $\sigma_s$ pode ser estimado pela dispersão dos vértices das parábolas de sua trajetória (Sinceac, 1998).

A evolução deste parâmetro mostra que seu valor diminuiu ao longo do período estudado.

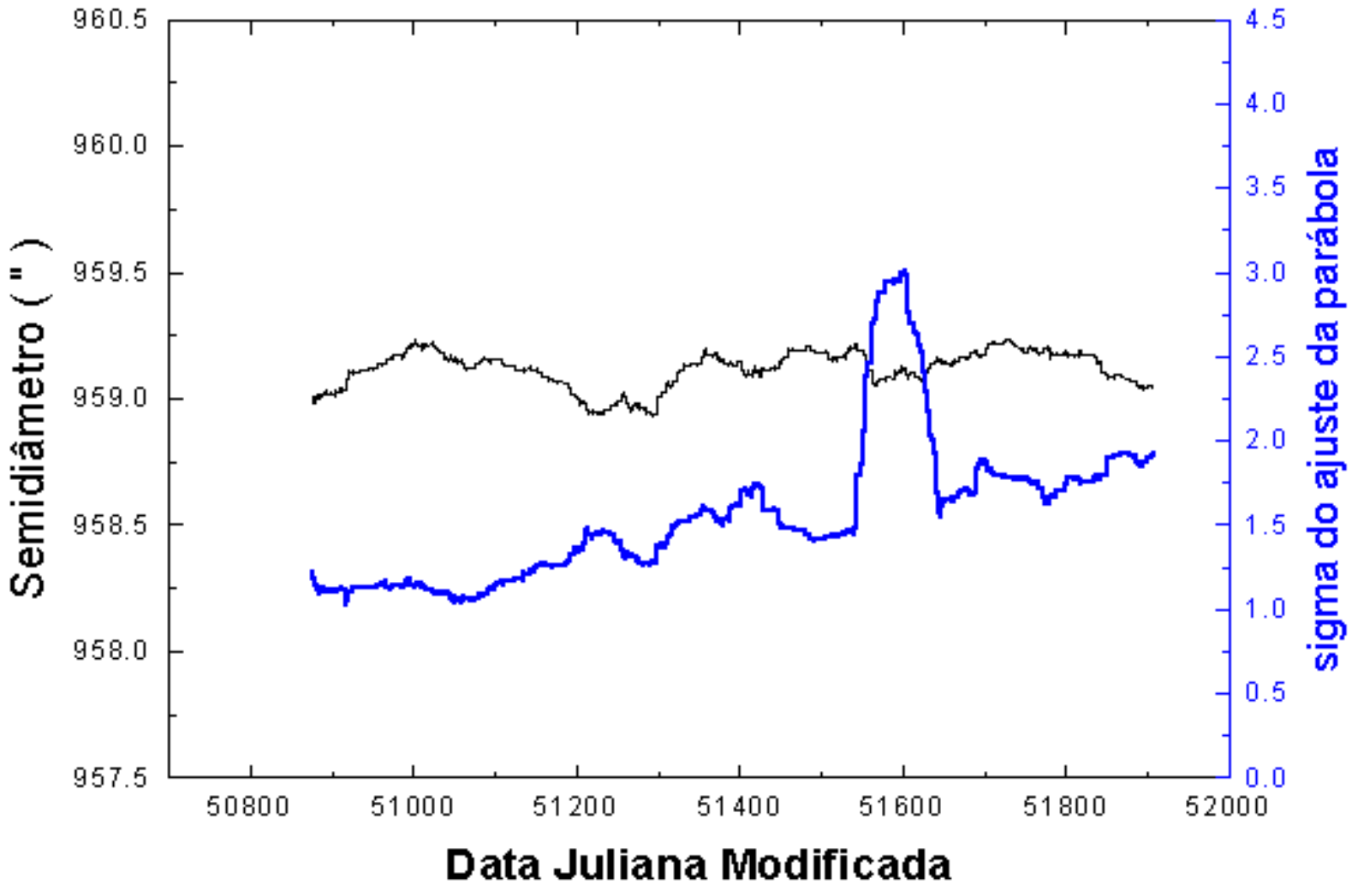

Figura 3.7. Superposição da evolução semidiâmetro e do sigma do ajuste da parábola refletida, ao longo de todo o período, alisados segundo uma média móvel de 600 pontos.

## 3.4 Série Corrigida

A figura 3.8 mostra a série antes e depois da aplicação da correção paramétrica, mas sem a correção da tendência linear. As curvas são praticamente idênticas e as variações dos valores são, em média, menores que 0,01% do valor original.

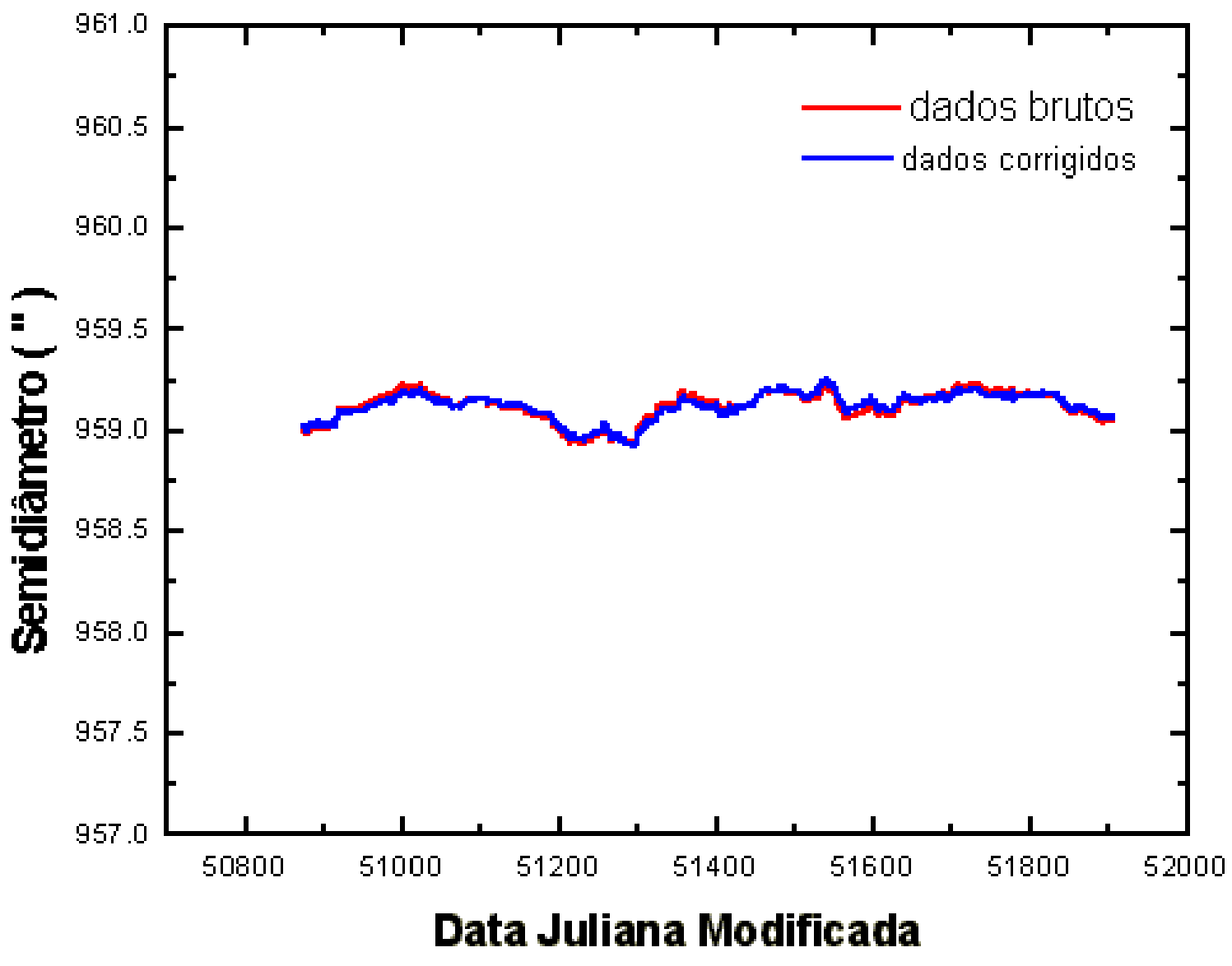

Figura 3.8. A série antes (vermelho) e depois (azul) da correção paramétrica, sem a correção do termo linear. Novamente aqui, uma média móvel de 600 pontos foi utilizada para melhorar a visualização dos resultados.

Uma comparação entre a diferença do valor corrigido para o valor original do semidiâmetro e a temperatura (figura 3.9) mostra claramente que este último foi o fator que mais influenciou o resultado final da medida do diâmetro solar:

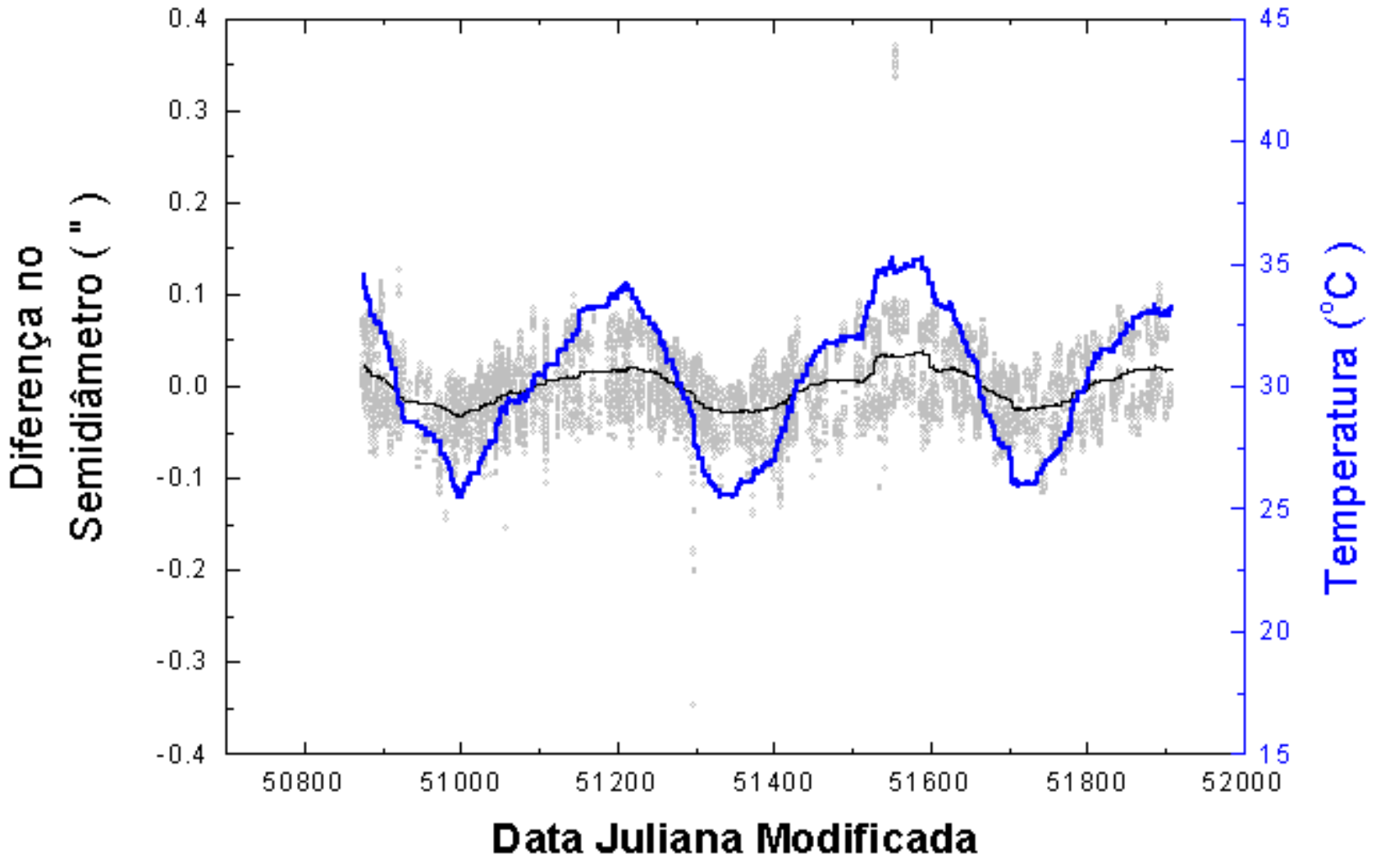

Figura 3.9. Variação da temperatura na hora da medida e a diferença entre os valores dos semidiâmetros corrigidos pelos parâmetros e os semidiâmetros da série original, ao longo de todo o período. As curvas contínuas são alisamentos por média móvel de 600 pontos.

O programa de correção paramétrica detectou uma tendência de aumento do semidiâmetro solar de 60,81±7,21 *mas*/ano, quando toda a série e período foram considerados. A Análise das observações visuais feitas entre 1997 e 1999, com o astrolábio em Santiago, Chile (Noël, 2001) mostrou uma tendência semelhante para o crescimento do valor do semidiâmetro, porém cerca de 3,5 vezes maior.

Para estudar melhor este termo linear, períodos menores da série também foram analisados.

A tabela abaixo compara os parâmetros encontrados e os erros associados ao ajuste linear da série completa para todo o período e para cada ano, assim como para cada lado da passagem.

| Ano | Série Completa (*mas*/ano) | Pass. Leste (*mas*/ano) | Pass. Oeste (*mas*/ano) |
|---|---|---|---|
| 1998/1999/2000 | **60,81±7,21** | 40,84±10,98 | 77,76±9,56 |
| 1998 | 190,97±44,58 | 154,60±67,15 | 254,65±59,92 |
| 1999 | 269,93±33,90 | 214,72±54,73 | 307,21±42,75 |
| 2000 | -129,23±35,64 | -54,43±50,92 | -196.92±49,64 |

Tabela 3.6. Parâmetros do ajuste linear anual e de toda a série, em negrito. Os parâmetros para cada ano e lado da passagem também são apresentados.

Na figura 3.10 as tendências lineares foram calculadas para intervalos de meio ano.

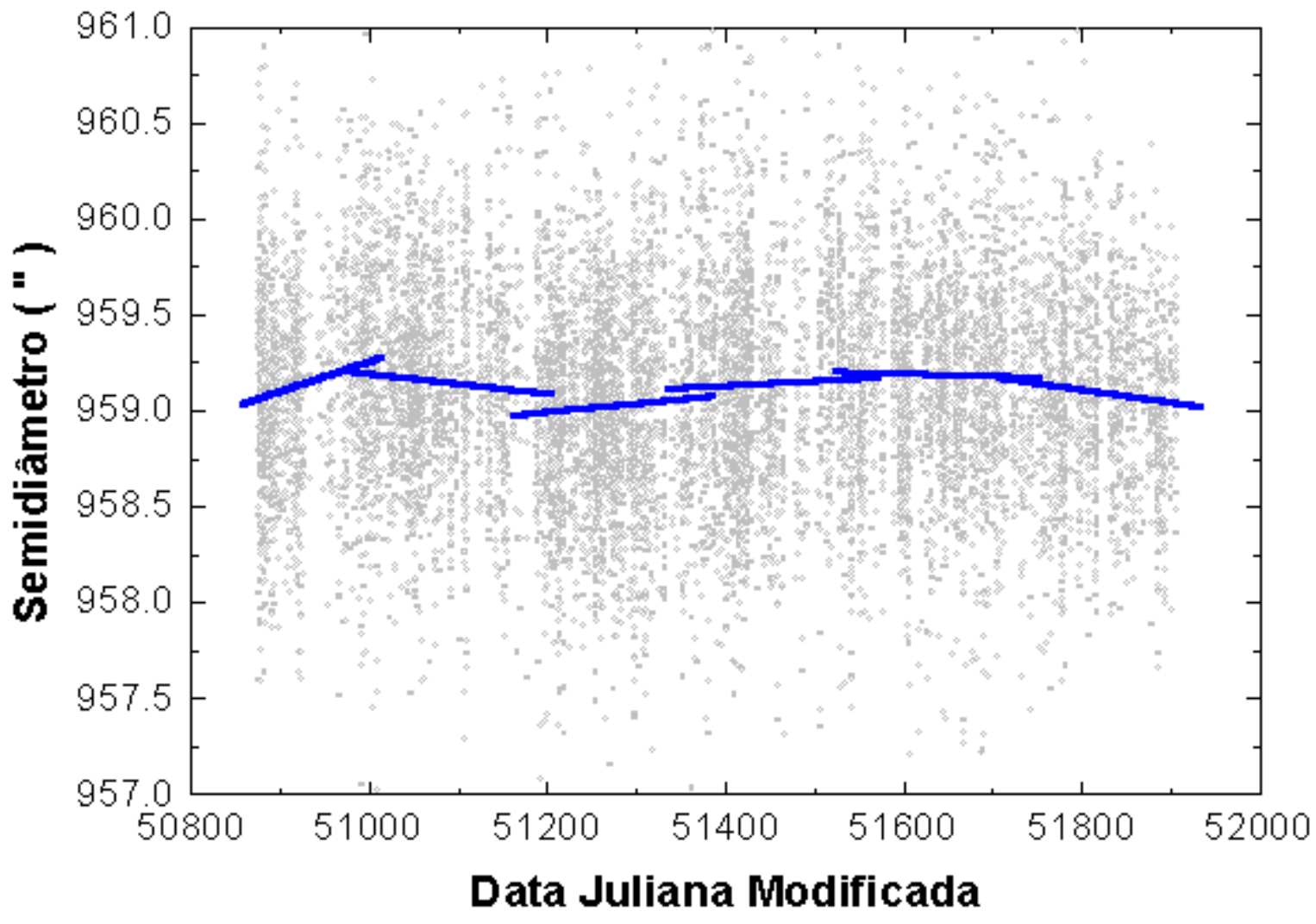

Figura 3.10. Tendências lineares da série em intervalos de meio ano (linhas azuis). Os pontos em cinza representam toda a série de valores obtidos para o semidiâmetro antes das correções paramétricas.

A tabela a seguir compara os parâmetros e os erros associados do ajuste linear para cada semestre da série.

| ano | Período: janeiro/junho (*mas*/ano) | Período: julho/dezembro (*mas*/ano) |
|---|---|---|
| 1998 | 563,19±153,51 | -188,97±101,76 |
| 1999 | 162,67±96,83 | 87,37±101,76 |
| 2000 | -46,54±98,01 | -239,98±105,63 |

Tabela 3.7. Parâmetros do ajuste linear a cada intervalo de seis meses da série. Para o ano de 1998, a análise da tendência linear começa no mês de março, por isso o parâmetro é pior definido neste semestre.

Os resultados do ajuste linear mostram que as observações feitas a Leste e a Oeste apresentam tendências semelhantes, seja considerando todo o período, seja considerando cada ano. Quando intervalos de tempo menores são investigados, o que se apresenta é uma grande variação deste valor com os erros associados crescendo muito rapidamente, tornando os coeficientes lineares encontrados estatisticamente questionáveis. O intervalo de três anos de medidas estudado neste trabalho parece ser ainda curto para se tirar conclusões sobre tendências de crescimento do semidiâmetro observado. Apesar da série estar homogeneamente distribuída dentro deste intervalo, os termos periódicos, reais ou introduzidos por modulações das condições observacionais, influenciam fortemente a tendência linear encontrada em cada semestre da campanha.

O número de manchas solares é considerado como o melhor estimador da atividade solar. A comparação entre a tendência linear de crescimento e o número de manchas solares durante o mesmo período (figura 3.11), mostra uma possível correlação entre o semidiâmetro observado e o crescimento atividade solar.

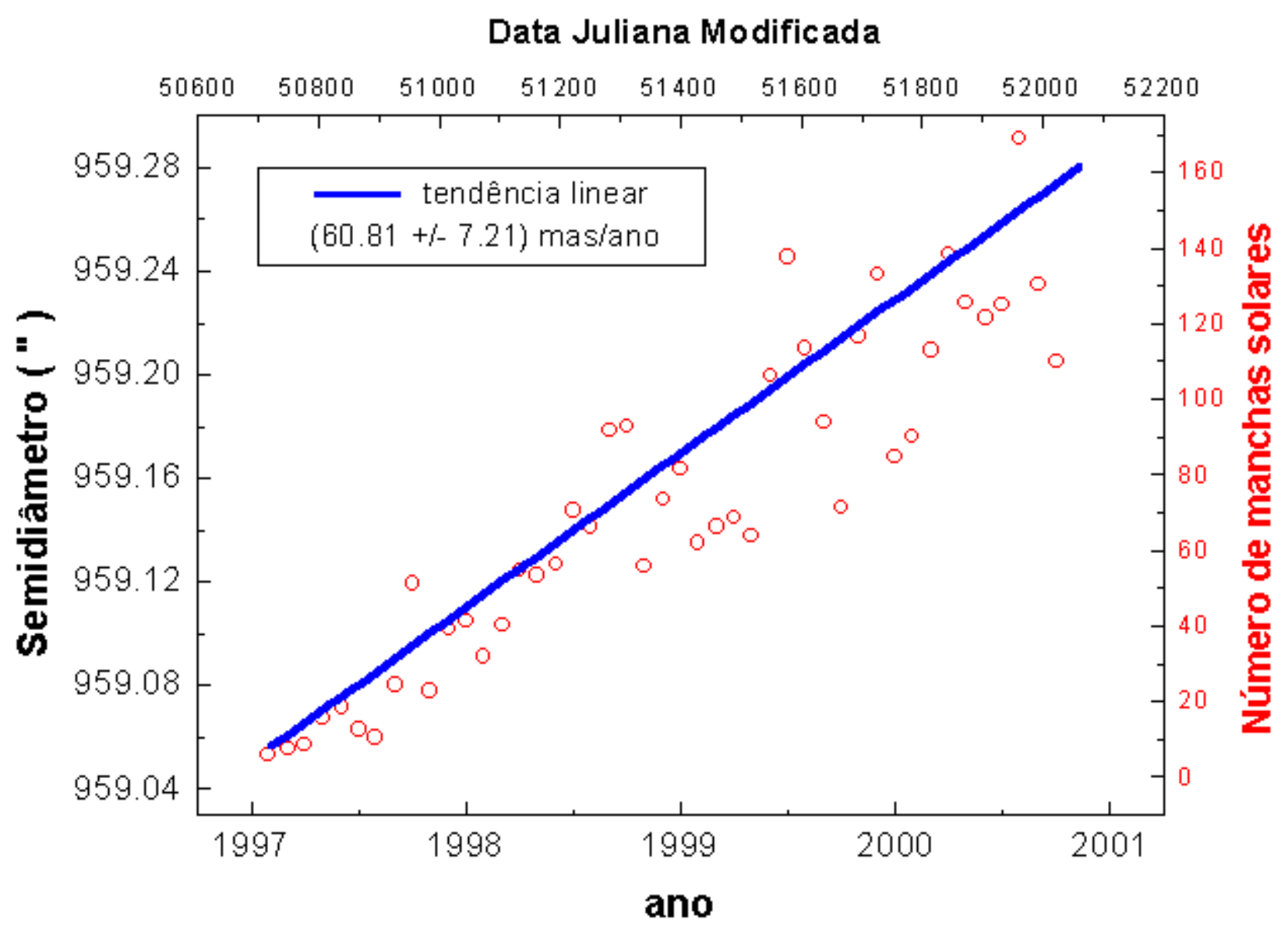

Figura 3.11. Os círculos vermelhos representam o número de manchas solares entre 1997 e 2001. Superposto ao gráfico a tendência linear de crescimento do semidiâmetro encontrado para o período entre 1998 e 2000.

Diversos autores referem-se à semelhante correlação na análise de medidas do diâmetro solar (Noël, 2001; Basu, 1998), porém anticorrelações também são encontradas na literatura (Laclare *et al.*, 1996).

A figura ilustra a complexidade envolvida em correlacionar, duas a duas, características solares num intervalo de tempo relativamente curto. Embora a correlação linear entre a tendência ajustada e a média mensal do número de manchas seja significativa, a correlação entre as médias diárias ou mensais do semidiâmetro medido e as médias correspondentes para o numero de manchas é estatisticamente nula. Dada a coerência da serie, resultados iguais (isto é, também estatisticamente nulos) são encontrados quando a serie observada é dividida por lado da passagem e/ou por ano.

Assim, conforme discutido, os dados analisados aqui não permitem ligar a tendência linear média encontrada à atividade solar. Para isto, um período mais longo é necessário, pelo menos compatível com o ciclo undecenial. De todo modo, a tendência linear média foi incorporada à correção simultânea dos parâmetros, como explicado no item 3.3. A figura 3.12 traz a comparação entre a série bruta e a série corrigida, agora incluíndo a tendência linear.

A diferença média entre os valores individuais das séries bruta e corrigida é de cerca de 0,05% para uma dispersão 0",004 menor.

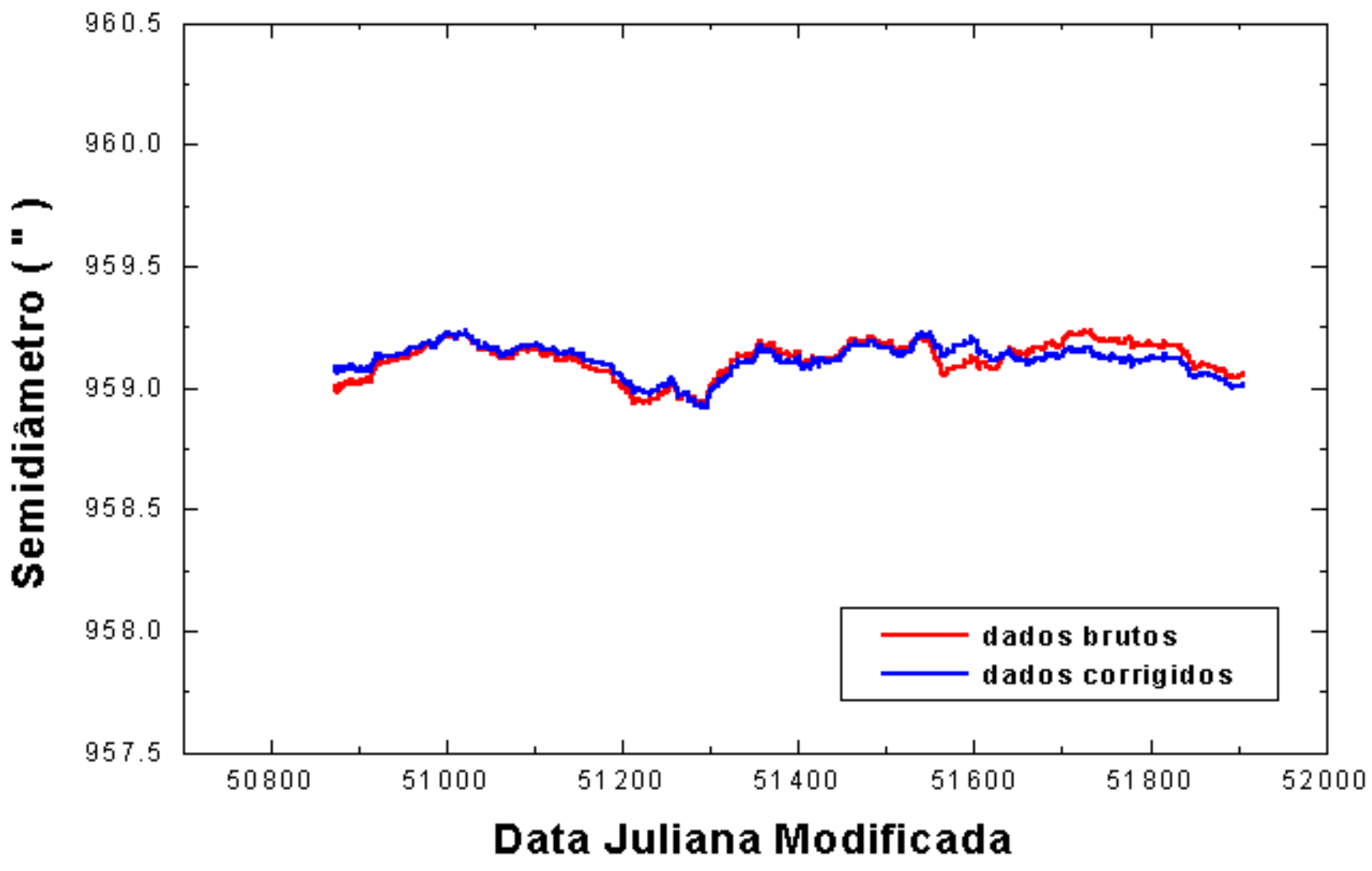

Figura 3.12. A série antes (vermelho) e depois (azul) da correção paramétrica, com a correção do termo linear. Uma média móvel de 600 pontos foi utilizada para melhor visualização da correção.

# Capítulo 4

## Análise da série

No capítulo 3 foram determinadas correções que minimizavam erros sistemáticos, de origem quer instrumental, quer ambiental. A série de medidas, assim tratada, permite a busca de termos intrinsecamente solares. Ao longo deste capítulo, tais termos serão divididos em dois tipos: aqueles que independem do modo de observação e aqueles para os quais uma explicação ligada às especificidades da observação com o astrolábio solar não pode ser completamente excluída. Os primeiros permitem obter informações diretas quanto à morfologia e a atividade solar, no período estudado, enquanto os segundos permitem inferências sobretudo sobre a função de luminosidade do limbo solar.

### 4.1 Histogramas

A distribuição das medidas da série e das sessões a Leste e Oeste podem ser vistas nas figuras a seguir. Verifica-se que os pontos possuem uma distribuição normal, para uma resolução da ordem de 500 *mas*. Uma curva gaussiana foi ajustada às distribuições. O coeficiente de correlação do ajuste é superior a 0,99 e $\chi^2$ da ordem de $10^{-6}$. Estes valores indicam que os erros dominantes são de natureza não-sistemática, tendendo a se cancelar frente ao grande número de medidas consideradas.

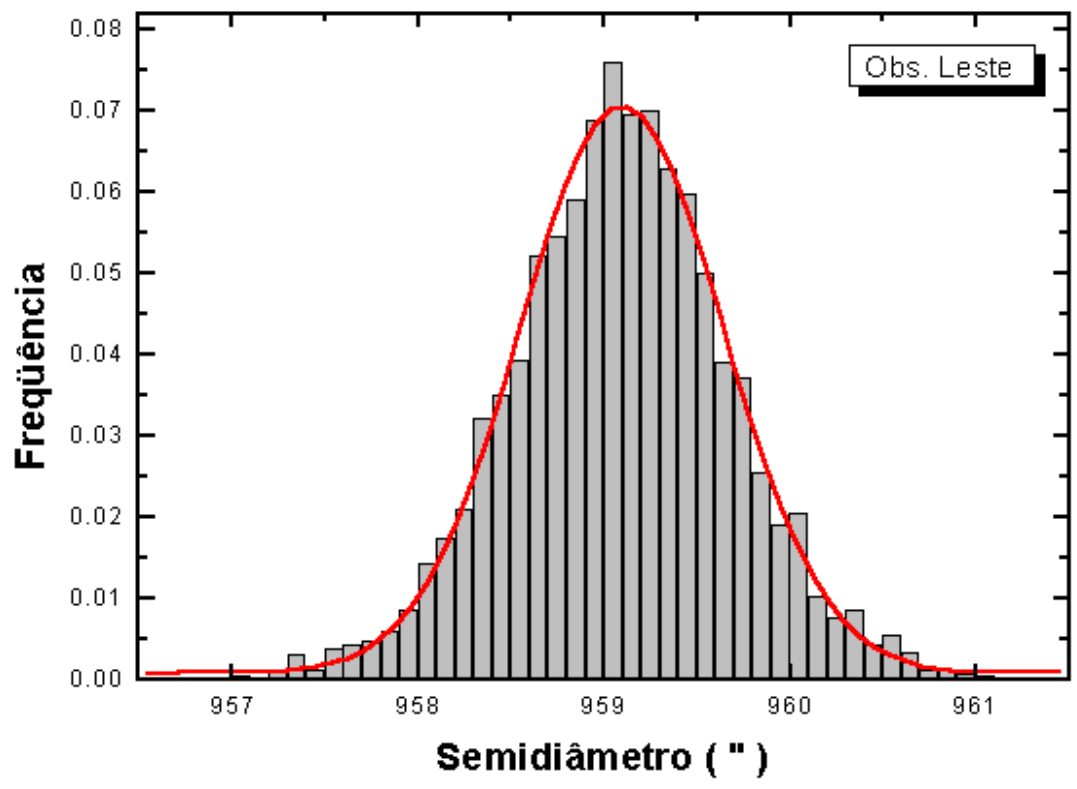

Figura 4.1. Distribuição do semidiâmetros medidos a Leste. A gaussiana ajustada tem valor médio de 959",099 ± 0",008 e σ igual a 0,560 ± 0”,010.

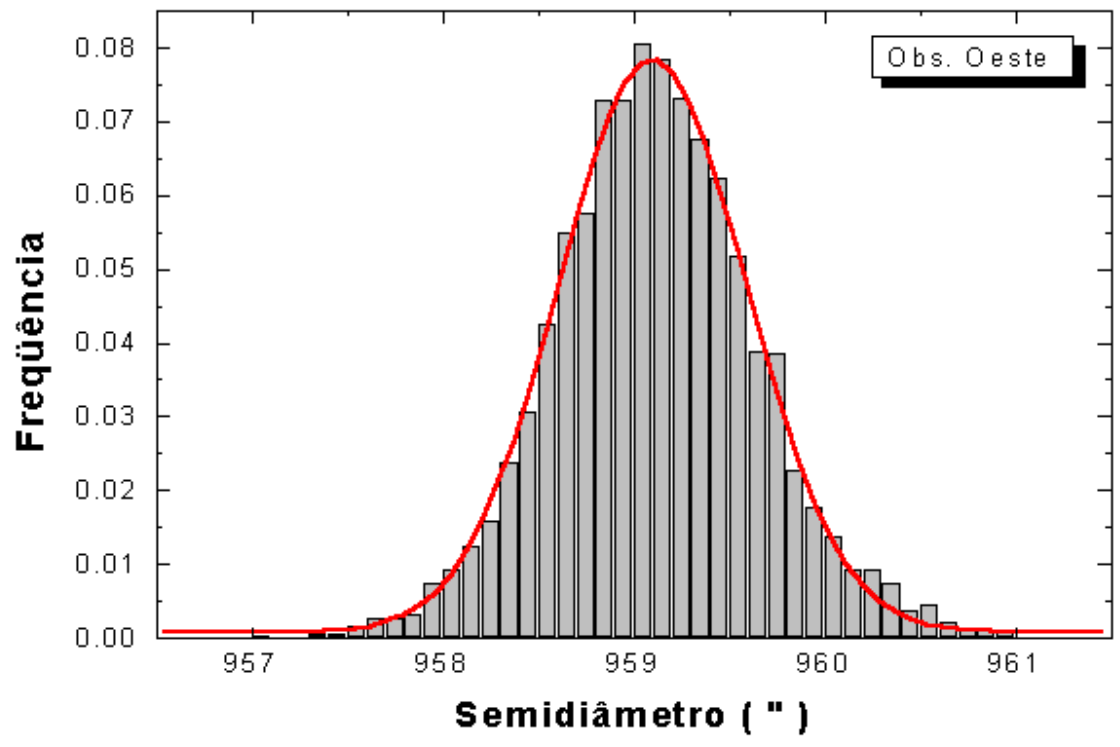

Figura 4.2. Distribuição do semidiâmetros medidos a Oeste. A gaussiana ajustada tem valor médio de 959",102 ± 0",006 e σ igual a 0",500 ± 0",008.

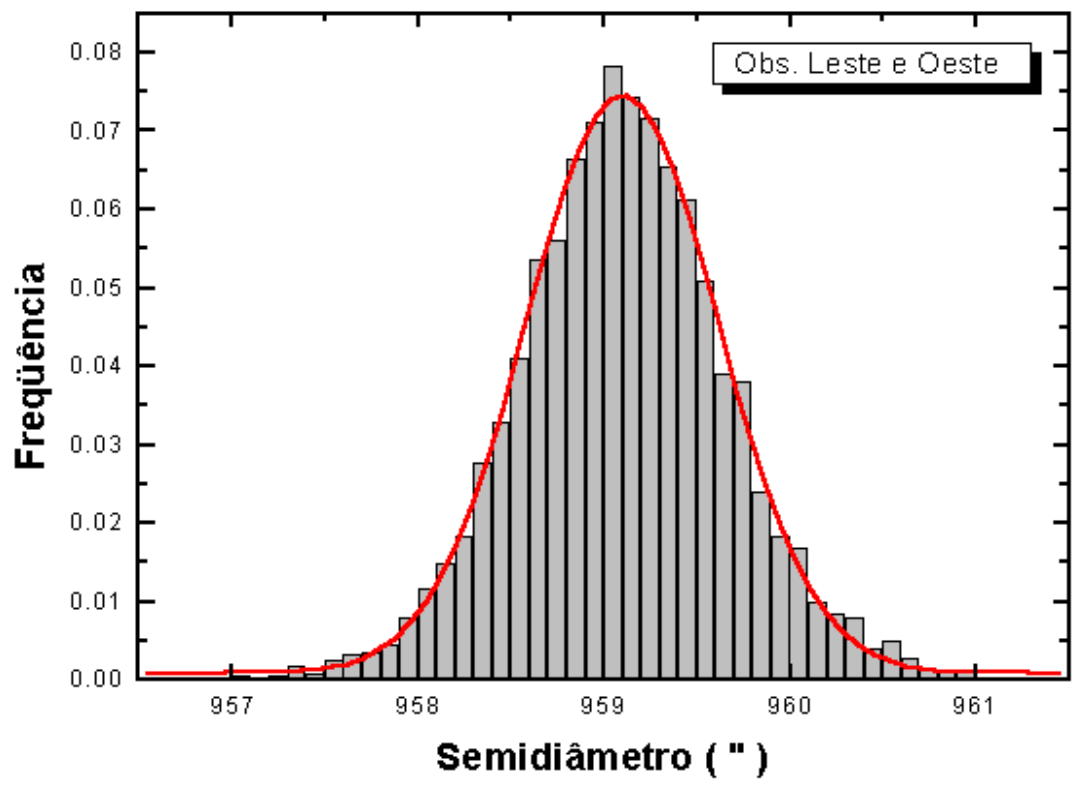

Figura 4.3. Distribuição do semidiâmetros medidos para toda a série. A gaussiana ajustada tem valor médio de 959",102 ± 0",005 e σ igual a 0",526 ± 0”,007.

## 4.1 Correlações

Variações globais do raio solar devem afetar de modo igual as observações temporalmente próximas, portanto um estudo das correlações entre as observações a Leste e a Oeste foi realizado. Esta premissa permite também verificar a qualidade das correções paramétricas. Para o teste, apenas os dias em que houve observações nos dois lados da passagem foram considerados (256 dias comuns). Além disso, tendo em vista o desvio padrão da série, determinado no item anterior, as medidas são alisadas em diferentes passos, agrupando-as em dias, por uma média móvel de 0 até 121 dias de 10 em 10 dias e de 0 até 1001 observações de 100 em 100 pontos.

A correlação entre os valores brutos das duas séries é baixa ($\approx$ 0.1) devido aos ruídos tanto instrumentais quanto observacionais. Esta correlação, entretanto, se torna superior a 0,5 quando as séries são agrupadas entre 80 e 120 dias e entre 50 e 250 pontos.

Quando as séries corrigidas pelos parâmetros e o termo linear são consideradas, a correlação é superior a 0,5, mesmo quando os pontos são considerados sem alisamento, chegando até próximo de 0,8. Isto é, as correções adotadas permitem um ganho efetivo para a relação sinal/ruído. Ressalte-se, ademais, que as características gerais do mapa de correlações são preservadas, evidenciando que as correções não impõem valores para as medidas obtidas.

Em resumo:

1. Apesar dos erros sistemáticos afetarem individualmente as medidas (vide a baixa correlação para as séries brutas), a coerência dos dados não é dominada por efeitos observacionais, uma vez que quando estes são removidos, aquela aumenta;

2. As correções aplicadas às séries foram eficazes, uma vez que a correlação entre as séries observadas a Leste e a Oeste aumentou;

3. As correções não modificaram o caráter geral da série, já que os mapas de correlação são semelhantes entre si.

As figuras a seguir mostram os mapas de correlação para as observações brutas, as corrigidas apenas pelos parâmetros e a correção final que inclui o termo linear.

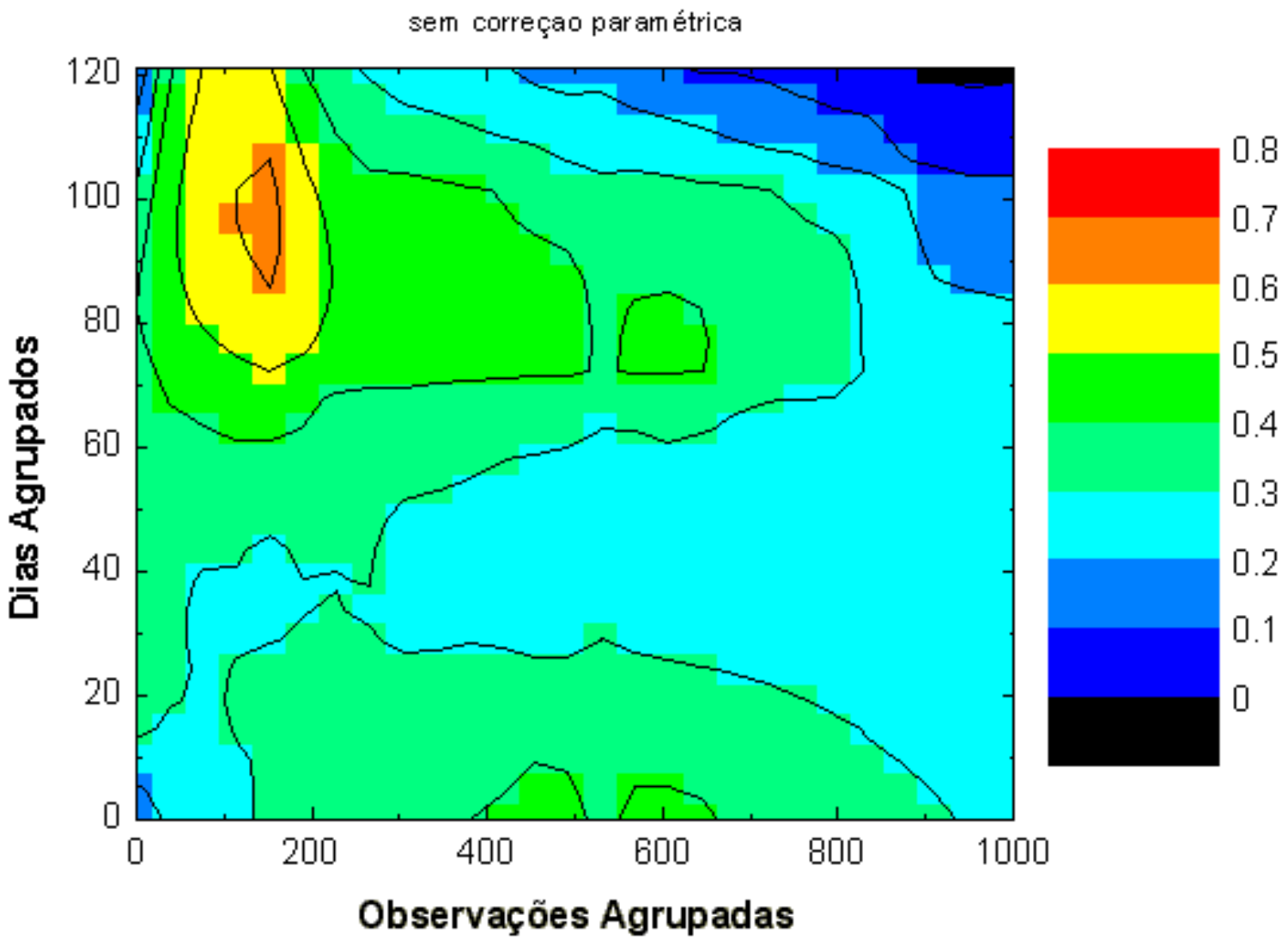

Figura 4.4. Mapa de correlação entre os dados brutos das observações feitas a Leste e a Oeste. Os dados estão agrupados em dias e em pontos. A média das correlações é de 0,31 chegando ao máximo de 0,62 na área laranja do mapa.

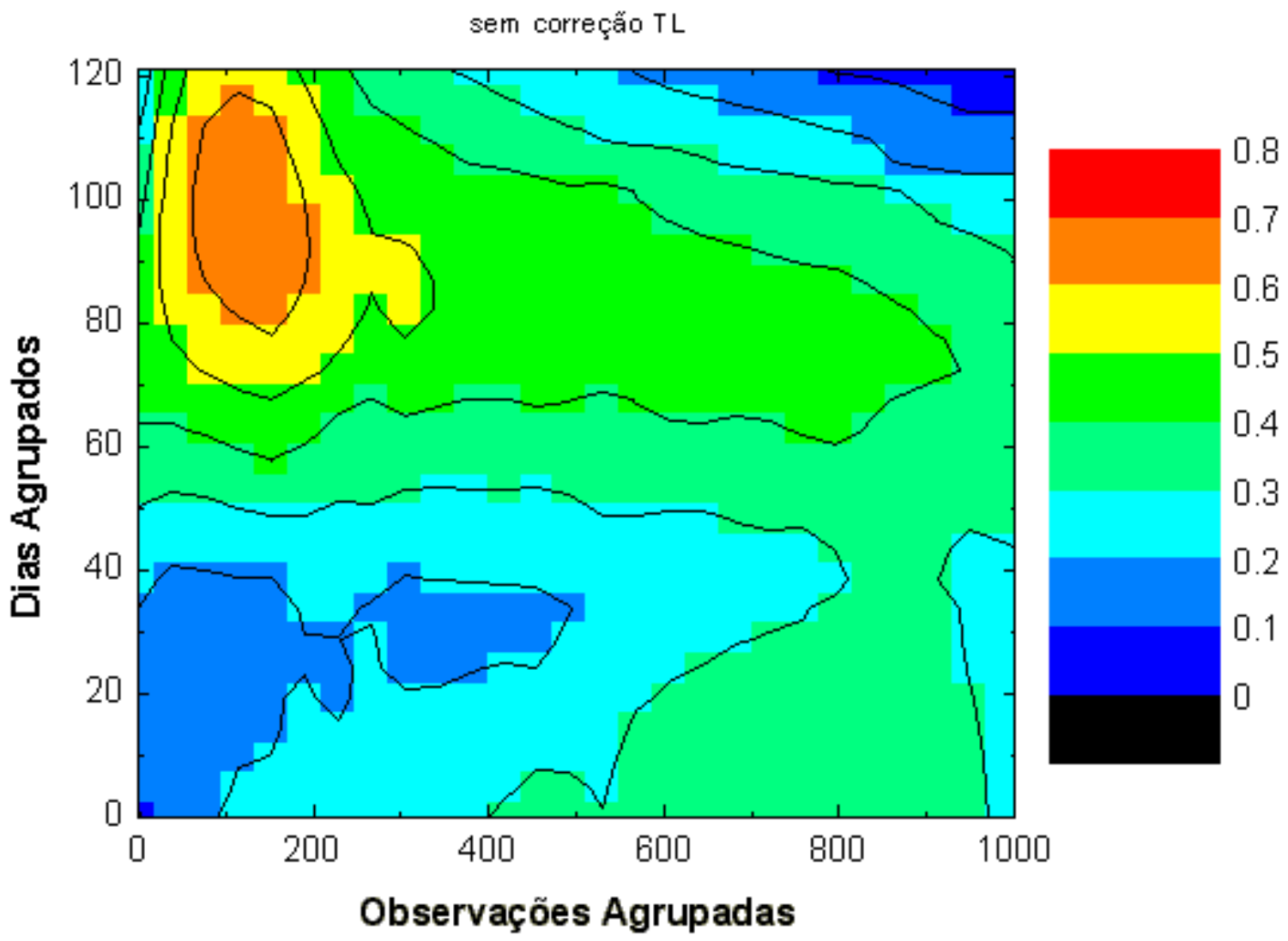

Figura 4.5. Mapa de correlação entre os dados corrigidos (sem a correção do termo linear) das observações feitas a Leste e a Oeste. Os dados estão agrupados em dias e em pontos. A média das correlações é de 0,33 chegando ao máximo de 0,67 na área laranja do mapa.

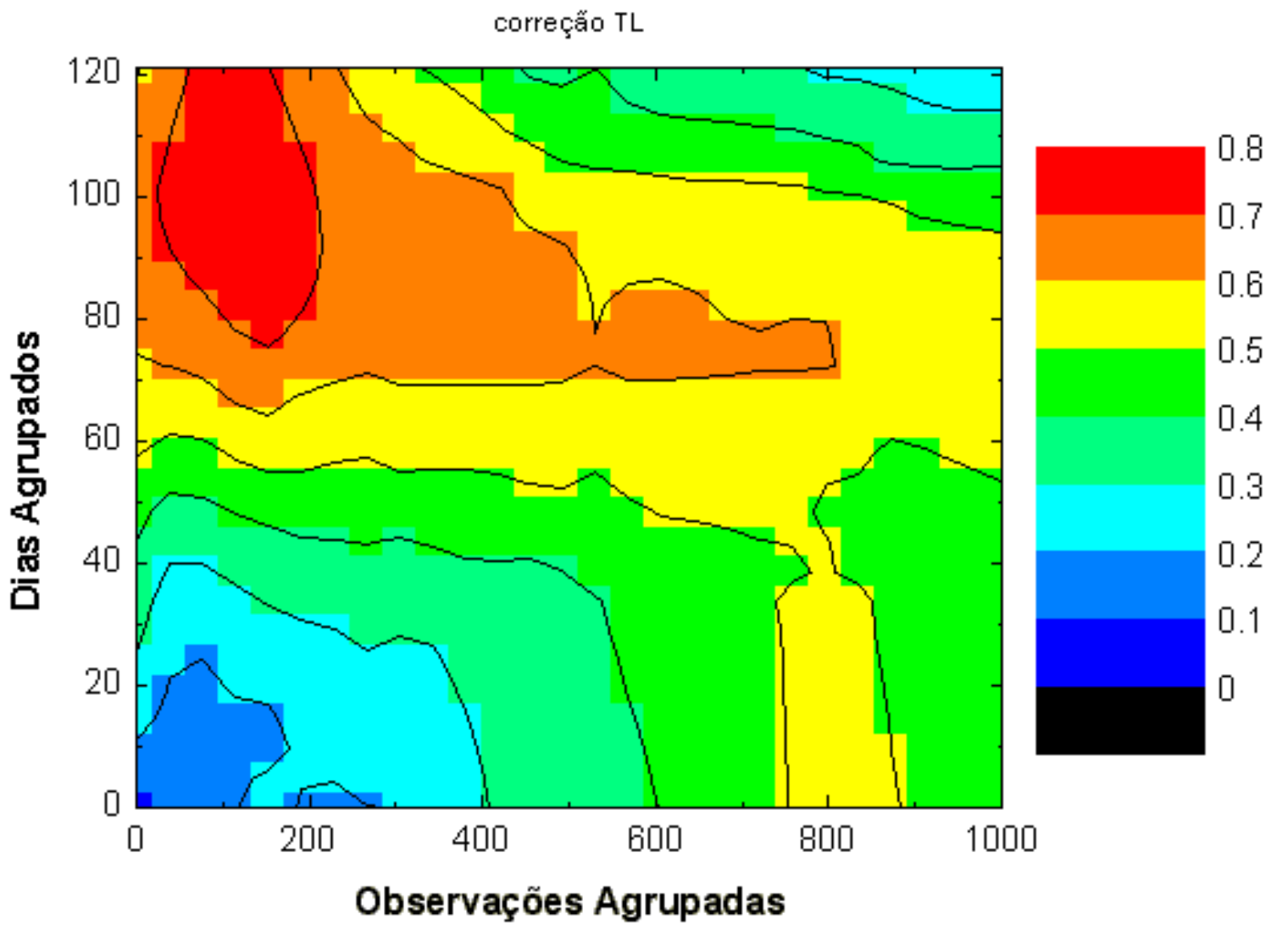

Figura 4.6. Mapa de correlação entre os dados corrigidos das observações feitas a Leste e a Oeste. Os dados estão agrupados em dias e em pontos. A média das correlações é de 0,47 chegando ao máximo de 0,77 na área vermelha do mapa.

## 4.3 Estatística da série corrigida

A seguir são apresentados o semidiâmetro médio por sessão e da série completa. A dispersão dos resultados foi computada para as séries como um todo e por sessão observacional. A tabela também traz a média destas dispersões por sessão e o erro de uma medida isolada. O maior valor para a dispersão do valor do semidiâmetro para toda a série, em comparação com a dispersão por sessão parece indicar a existência de termos periódicos. A série observada a Leste é mais ruidosa devido principalmente a uma maior variação das condições atmosféricas durante as sessões (o fator de Fried, que qualifica esta turbulência, apresenta uma dispersão 1,5 vezes maior para as sessões a Leste).

Os trabalhos de Neckel (1994) e Sinceac (1998) mostram que medidas do semidiâmetro numa atmosfera turbulenta tendem a ter um valor menor devido a um deslocamento sistemático do ponto de inflexão da curva de escurecimento do bordo. Isto se traduz por um progressivo afastamento entre o baricentro e o centro geométrico na figura que representa a primeira derivada da função de luminosidade, cujo pico representa o ponto de inflexão daquela função, o qual é escolhido para delimitar o limbo solar.

No caso de série presentemente analisada não existe diferença sistemática do valor médio do fator de Fried (≈36 mm) para ambos os lados da passagem. Coerentemente, a diferença entre os valores médios entre Leste e Oeste é estatisticamente inconclusiva, podendo de modo mais simples, ser explicada pelas variações periódicas e de heliolatitude, estas últimas analisadas no capítulo seguinte.

| | Séries | | |
|---|---|---|---|
| Estimador | Obs. Leste ( " ) | Obs. Oeste ( " ) | **Toda a série ( " )** |
| SD | 959,092±0,009 | 959,120±0,008 | **959,107±0,006** |
| $\sigma_{SD}$ | 0,60 | 0,54 | **0,57** |
| $\sigma_{SD/sessão}$ | 0,29 | 0,22 | **0,26** |
| $\langle\sigma_{sessão}\rangle$ | 0,52 | 0,48 | **0,50** |
| $\varepsilon$ | 0,21 | 0,18 | **0,19** |

Tabela 4.1. Precisão e médias das sessões e da série completa. SD é o valor médio do semidiâmetro observado, $\sigma_{SD}$ é o desvio padrão correspondente, $\sigma_{SD/sessão}$ é o desvio padrão do valor médio do semidiâmetro por sessão, $\langle\sigma_{sessão}\rangle$ é a média do desvio padrão dos resultados e $\varepsilon$ é precisão nominal de cada medida individual.

## 4.4 Periodicidades

Conforme apresentado, as observações a Leste e a Oeste são realizadas sob condições observacionais diferentes que influenciam estatisticamente as medidas, inserindo ruídos e erros sistemáticos. Estas influências, porém, podem ser removidas com sucesso. Os dados possuem uma distribuição gaussiana e o grande número de medidas homogeneamente distribuídas pode ser usado para verificar variações no semidiâmetro observado da ordem de 6 *mas*. Este limite é dado pela razão entre o desvio padrão da série e a raiz do número total de medidas.

Para investigar periodicidades nas séries, dois algoritmos foram utilizados por serem capazes de lidar com dados irregularmente espaçados: o CLEAN (Roberts *et al.*, 1987) e o *Date-Compensate Discrete Fourier Transform* (DCDFT) (Ferraz-Mello, 1981). Os ruídos observacionais somados ao sinal real da série podem afetar as respostas dos algoritmos, por isso, para avaliar o grau de confiabilidade dos picos do espectro de potência, primeiramente estes algoritmos foram testados com simulações. Estas foram feitas com uma centena de séries compostas de falsos dados (ruído) retirados de uma distribuição gaussiana de média em torno do zero, colocados ao longo dos mesmos dias e de mesmo desvio padrão da série verdadeira. A relação entre o sinal de entrada e o correspondente pico de saída foi determinada, utilizando-se sinais de amplitudes e freqüências bem definidas somadas a esta distribuição simulada. Períodos abaixo de 10 dias não foram considerados, tendo em vista o ruído e as desigualdades de espaçamento nas séries (reais). Também não foram considerados períodos acima de 2000 dias, uma vez que a série se estende por 1032 dias e assim este valor foi considerado como limite superior. Desta forma pode-se avaliar a influência que a distribuição temporal dos dados poderiam ter sobre o espectro de potências.

Resultados das simulações:

CLEAN:

- Em relação aos períodos encontrados em todas as simulações, nenhum apareceu em destaque, significando que, na série, as sessões em que não houve monitoramento não induzem o aparecimento de falsos períodos;
- A amplitude máxima destes períodos, em todas as simulações, não foi superior a 0''.012. Este valor dá uma estimativa, *grosso modo*, do erro associado aos componentes do espectro e foi considerado como o nível mínimo de significância para o estudo da série;
- O algoritmo mostrou-se eficiente na detecção dos sinais reais e na eliminação de seus harmônicos;
- Uma relação linear entre a amplitude de um sinal real de entrada e a amplitude do correspondente pico do espectro de potências foi encontrada.

DCDFT:

- Foram analisados todos os períodos entre 10 e 2000 dias, com intervalos de 1 dia e todos apresentaram coeficiente de correlação e grau de confiabilidade nulos;
- O algoritmo mostrou-se eficiente na detecção dos sinais reais;
- Não foi possível encontrar uma relação linear entre um sinal de entrada real e bem definido e a amplitude do pico do espectro de potências.

Uma vez definidas as restrições dos algoritmos as séries antes e depois da correção paramétrica foram analisadas, assim como separadas por lado da passagem. A primeira tabela traz os períodos mais relevantes obtidos através do CLEAN, e suas respectivas amplitudes. Todas as amplitudes apresentadas aqui, de forma bem conservadora, possuem uma incerteza de ±7 *mas*. Este valor é encontrado comparando a razões entre dispersão da série e amplitudes. Por um lado tem-se que a dispersão da série, contra a amplitude máxima dos períodos fictícios. Pelo outro lado a dispersão efetiva da série (isto é, descontada a soma quadrática das amplitudes dos períodos reais), fornecendo a estima do valor do erro sobre os períodos.

| | **Série bruta** | | | **Série tratada** | | |
|---|---|---|---|---|---|---|
| Período (dias) | Amplitude (*mas*) | | | Amplitude (*mas*) | | |
| | Pass. Leste | Pass. Oeste | **Série Completa** | Pass. Leste | Pass. Oeste | **Série Completa** |
| **10,10** | - | - | **-** | - | 42 | **-** |
| **11,44** | 42 | 42 | **28** | 50 | - | **32** |
| **16.09** | 28 | 30 | **32** | - | - | **-** |
| **20,60** | 30 | - | **-** | - | - | **-** |
| **25,75** | - | - | **-** | 46 | - | **-** |
| **26,41** | 40 | - | **34** | - | - | **26** |
| **29,43** | 24 | 40 | **30** | 30 | 38 | **32** |
| **44,78** | 24 | - | **42** | - | - | **40** |
| **46,82** | - | 56 | **-** | - | 58 | **-** |
| **49,05** | 30 | - | **28** | 32 | - | **30** |
| **54,21** | - | 38 | **-** | - | 42 | **-** |
| **60,59** | - | 40 | **-** | - | 30 | **-** |
| **64,38** | 48 | - | **24** | 52 | - | **24** |
| **68,67** | 36 | - | **-** | 42 | - | **-** |
| **79,23** | 28 | - | **-** | - | - | **-** |
| **171,67** | 36 | - | **-** | - | - | **-** |
| **206,00** | - | 52 | **28** | 20 | 60 | **32** |
| **343,33** | 60 | 50 | **56** | 30 | 20 | **24** |
| **515,00** | 38 | 40 | **32** | 70 | 60 | **60** |
| **1030,00** | 28 | 62 | **40** | - | 36 | **-** |

Tabela 4.2. Períodos e suas respectivas amplitudes encontradas utilizando-se o algoritmo CLEAN. As séries foram analisadas completas e por lado da passagem, antes e depois da correção.

- O termo de longo período (1030 dias) já havia sido encontrado por outros autores (Laclare *et al.*, 1983; Leister e Benevides Soares, 1990; Leister *et al.*, 1996) mas com amplitudes entre 4 e 5 vezes maiores e com amplitude semelhante (Moussaoui, *et al.*, 2001). Na série estudada aqui, este período parece estar ligado ao termo linear encontrado na correção, já que ele é minimizado quando a sessão a Oeste é analisada, e desaparece quando toda a série é levada em consideração;
- O período que aparece com maior amplitude é o de 515 dias. Períodos similares, em torno de 500 dias podem ser encontrados na literatura (Gavryusev *et al.*, 1994; Emílio, 2001), sem referências de amplitudes;
- O período quase anual de 343,3 dias está ligado a um efeito residual da correção observacional (sua amplitude cai menos da metade após o tratamento dos dados), ou ainda à variação da heliolatitude da observação, que será tratada no capítulo 5.

Também é citado por vários autores como Gavryusev *et al.* (1994) e Emílio, (2001). Moussaoui, *et al.* (2001), encontraram períodos de 348 e 357 dias com 0",05 e 0",04 respectivamente;

- O período de 206 dias. Não há referência deste período por outros autores;
- Período próximo ao períodos de 64 dias é mencionado por Moussaoui, *et al.* (2001), com amplitude semelhante. Este período pode estar ligado ao tempo de vida médio de grupos de manchas solares;
- O período de 44,7 dias, como o anterior, pode ter correlação com o tempo de vida das manchas solares individualmente;
- Os períodos de 29 e 26 dias têm correlação com a rotação diferencial do Sol. Uma vez que as observações se dão no limbo solar, qualquer objeto mais ou menos brilhante nesta região, causa modulação no diâmetro observado;
- E finalmente o curto período de 11 dias pode estar ligado a rápidas flutuações da irradiância, mas como também está muito próximo da freqüência de Nyquist, pode ser simplesmente reflexo de ruído observacional.

Os periodogramas trazem a comparação entre os períodos encontrados nas séries antes e depois da correção simultânea dos parâmetros. Os gráficos estão superpostos para facilitar o confronto entre os resultados obtidos.

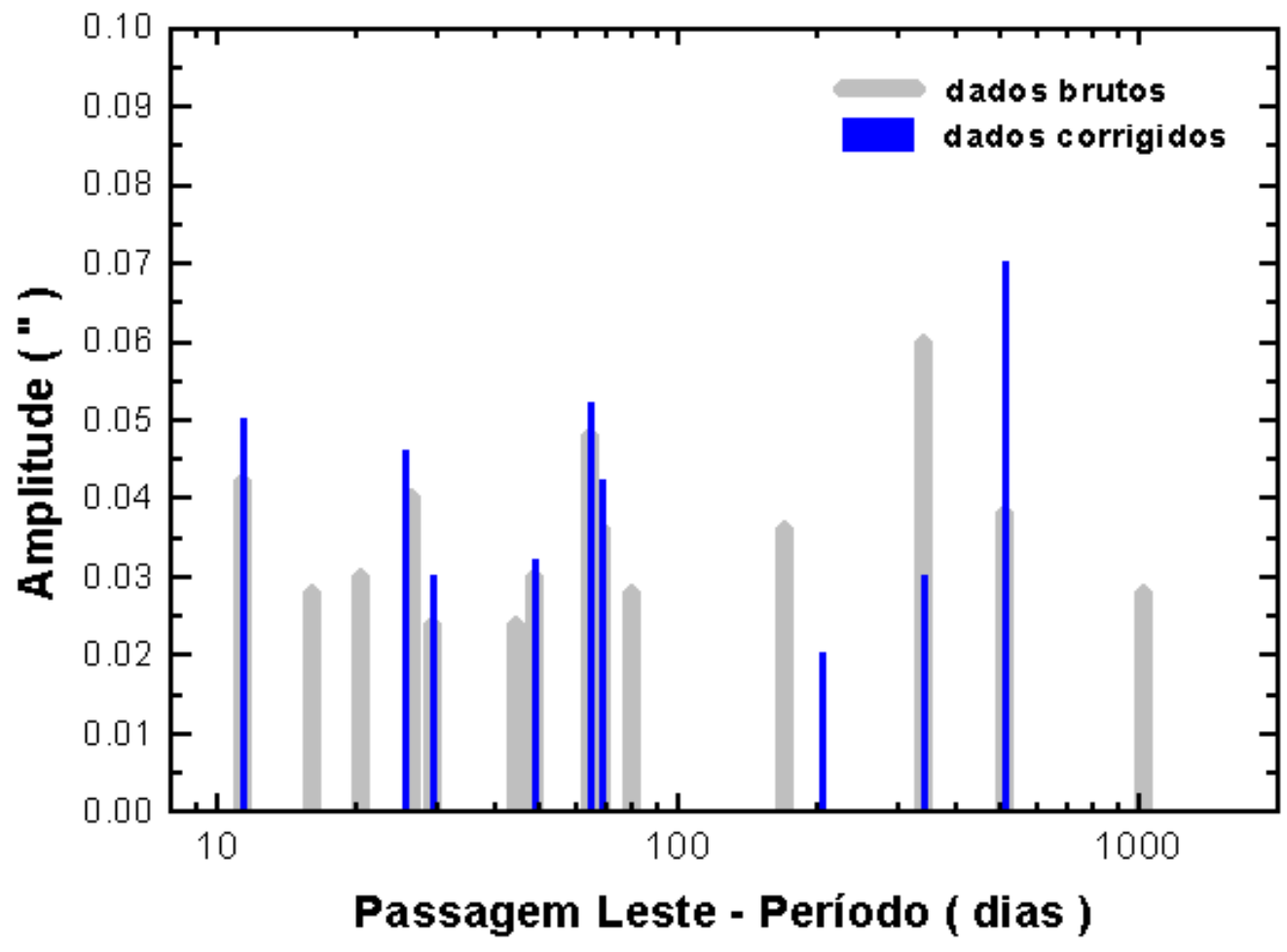

Figura 4.7. Periodograma comparativo entre os períodos encontrados nas séries antes (cinza) e depois das correções (azul) somente para a passagem Leste.

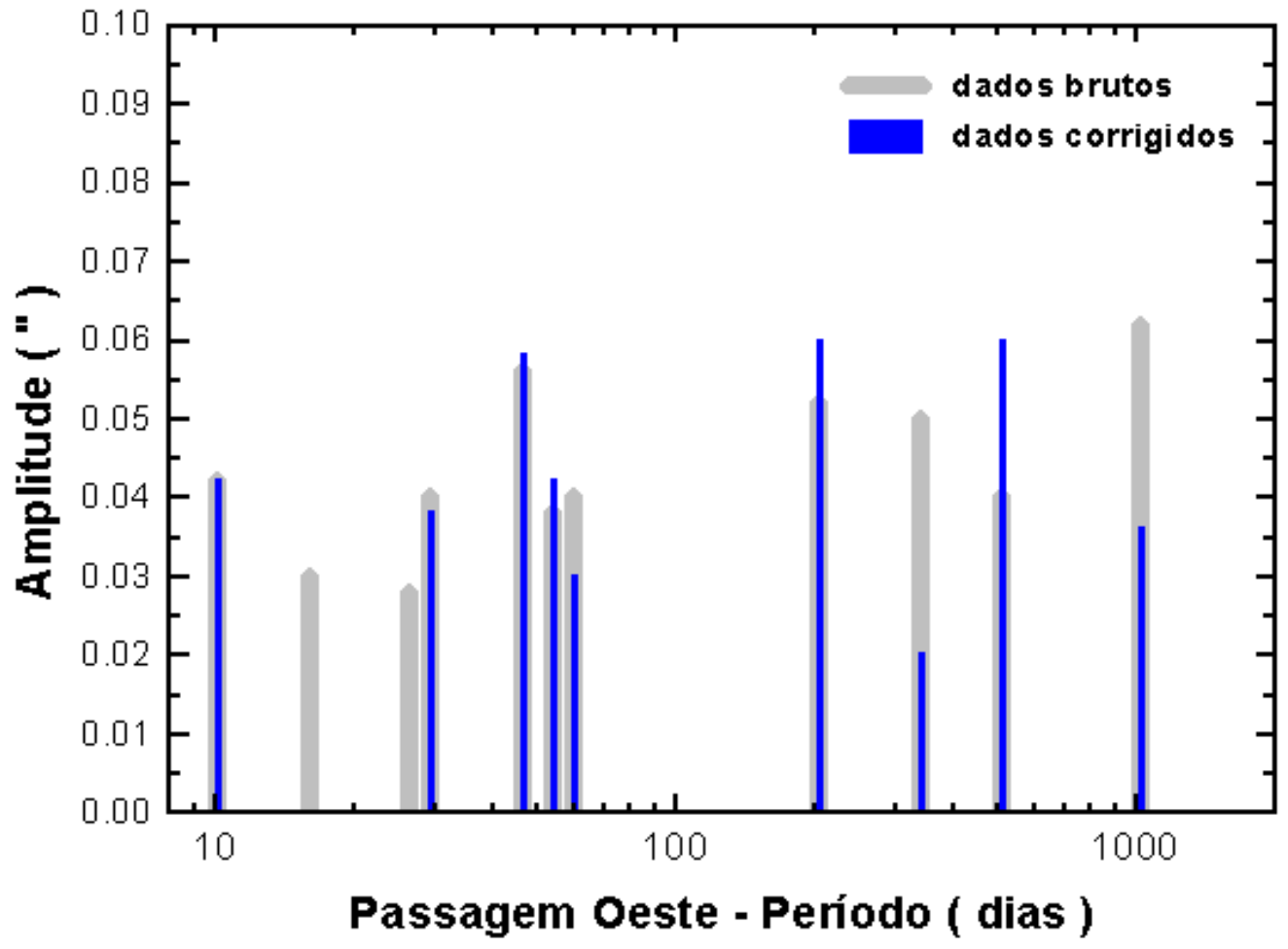

Figura 4.8. Periodograma comparativo entre os períodos encontrados nas séries antes (cinza) e depois das correções (azul) somente para a passagem Oeste.

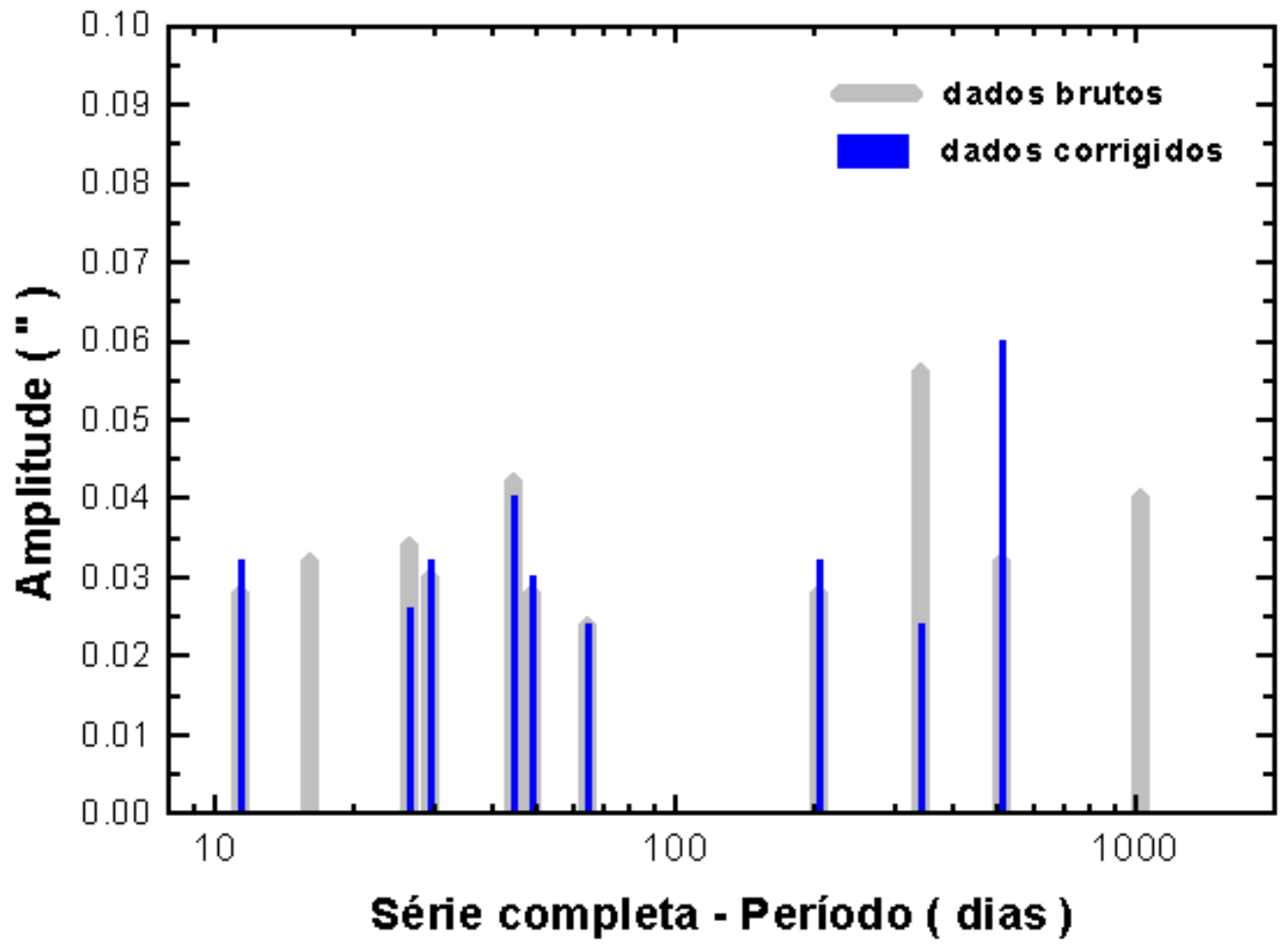

Figura 4.9. Periodograma comparativo entre os períodos encontrados na série completa, antes (cinza) e depois das correções (azul).

Os periodogramas mostram claramente que existe concordância entre os períodos encontrados para as sessões a Leste e a Oeste e consequentemente com a série completa. O fato de possuírem características semelhantes, apesar das condições observacionais distintas, sugere que as variações no semidiâmetro observadas têm origem verdadeiramente solar e/ou advêm de características do perfil da imagem do Sol.

Utilizando-se o DCDFT, não foi possível obter informação sobre amplitudes, mas a simples comparação entre os períodos encontrados por dois algoritmos conceitualmente diferentes dá mais segurança à afirmação de que estes períodos sejam reais.

| Sem correção paramétrica | | | | | |
|---|---|---|---|---|---|
| Passagem Leste | | Passagem Oeste | | **Toda Série** | |
| DCDFT | CLEAN | DCDFT | CLEAN | DCDFT | CLEAN |
| 10,0 | - | - | - | - | - |
| - | 11,44 | - | 11,44 | - | **11,44** |
| - | 16,09 | - | 16,09 | - | **16,09** |
| - | 20,60 | - | - | - | - |
| - | - | - | - | - | - |
| - | 26,41 | - | - | **27,0** | **26,41** |
| - | 29,43 | - | 29,43 | - | **29,43** |
| - | 44,78 | 44,0 | - | **44,0** | **44,78** |
| - | 49,05 | 47,0 | 46,82 | **47,0** | **49,05** |
| - | - | 56,0 | 54,21 | - | - |
| - | - | - | 60,59 | - | - |
| 65,0 | 64,38 | 66,0 | - | **66,0** | **64,38** |
| - | 68,67 | - | - | - | - |
| - | 79,23 | - | - | - | - |
| 171,0 | 171,67 | - | - | - | - |
| - | - | 198,0 | 206,00 | **193,0** | **206,00** |
| 349,0 | 343,33 | 348,0 | 343,33 | - | **343,33** |
| - | 515,00 | - | 515,00 | - | **515,00** |
| - | 1030,00 | - | 1030,00 | - | **1030,00** |

Tabela 4.3. Comparação entre os períodos, para a série não tratada, encontrados através dos dois algoritmos utilizados, vistos lado a lado. Em cinza estão agrupados os períodos semelhantes encontrados.

| Com correção paramétrica | | | | | |
|---|---|---|---|---|---|
| Passagem Leste | | Passagem Oeste | | **Toda Série** | |
| DCDFT | CLEAN | DCDFT | CLEAN | DCDFT | CLEAN |
| 10,0 | - | - | 10,10 | - | - |
| - | 11,44 | - | - | - | **11,44** |
| - | - | - | - | - | - |
| 18,0 | - | - | - | - | - |
| - | 25,75 | - | - | - | - |
| - | - | - | - | - | **26,41** |
| - | 29,43 | - | 29,43 | - | **29,43** |
| - | - | 44,0 | - | **44,0** | **44,78** |
| - | 49,05 | 47,0 | 46,82 | **47,0** | **49,05** |
| - | - | 56,0 | 54,21 | - | - |
| - | - | - | 60,59 | - | - |
| 65,0 | 64,38 | 66,0 | - | **66,0** | **64,38** |
| 68,0 | - | - | - | - | - |
| - | - | - | - | - | - |
| - | - | - | - | - | - |
| - | 206,00 | 198,0 | 206,00 | **197,0** | **206,00** |
| - | 343,33 | - | 343,33 | **348,0** | **343,33** |
| 509,0 | 515,00 | - | 515,00 | **521,0** | **515,00** |
| - | - | - | 1030,00 | - | - |

Tabela 4.4. Comparação entre os períodos, para a série tratada, encontrados através dos dois algoritmos utilizados, vistos lado a lado. Em cinza estão agrupados os períodos semelhantes encontrados.

Os gráficos a seguir trazem a superposição da série temporal completa (novamente mostrada segundo uma média móvel de 600 pontos) e os períodos de 515 dias, 206 dias, a soma destes e suas respectivas amplitudes.

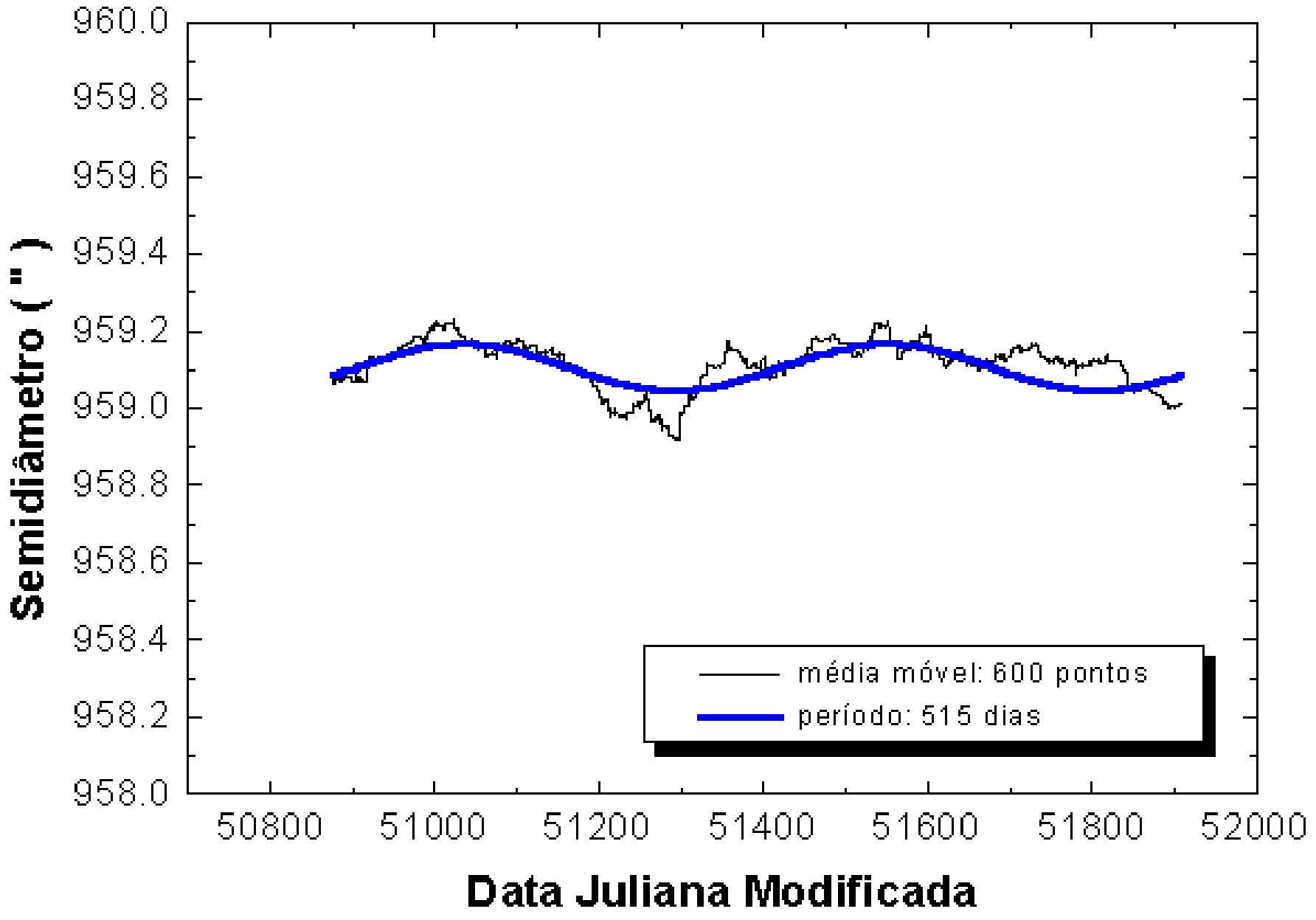

Figura 4.10. A curva em preto corresponde à média corrida de 600 pontos de toda a série corrigida. Superposto, em azul e na mesma escala, está o período de 515 dias (maior amplitude encontrada).

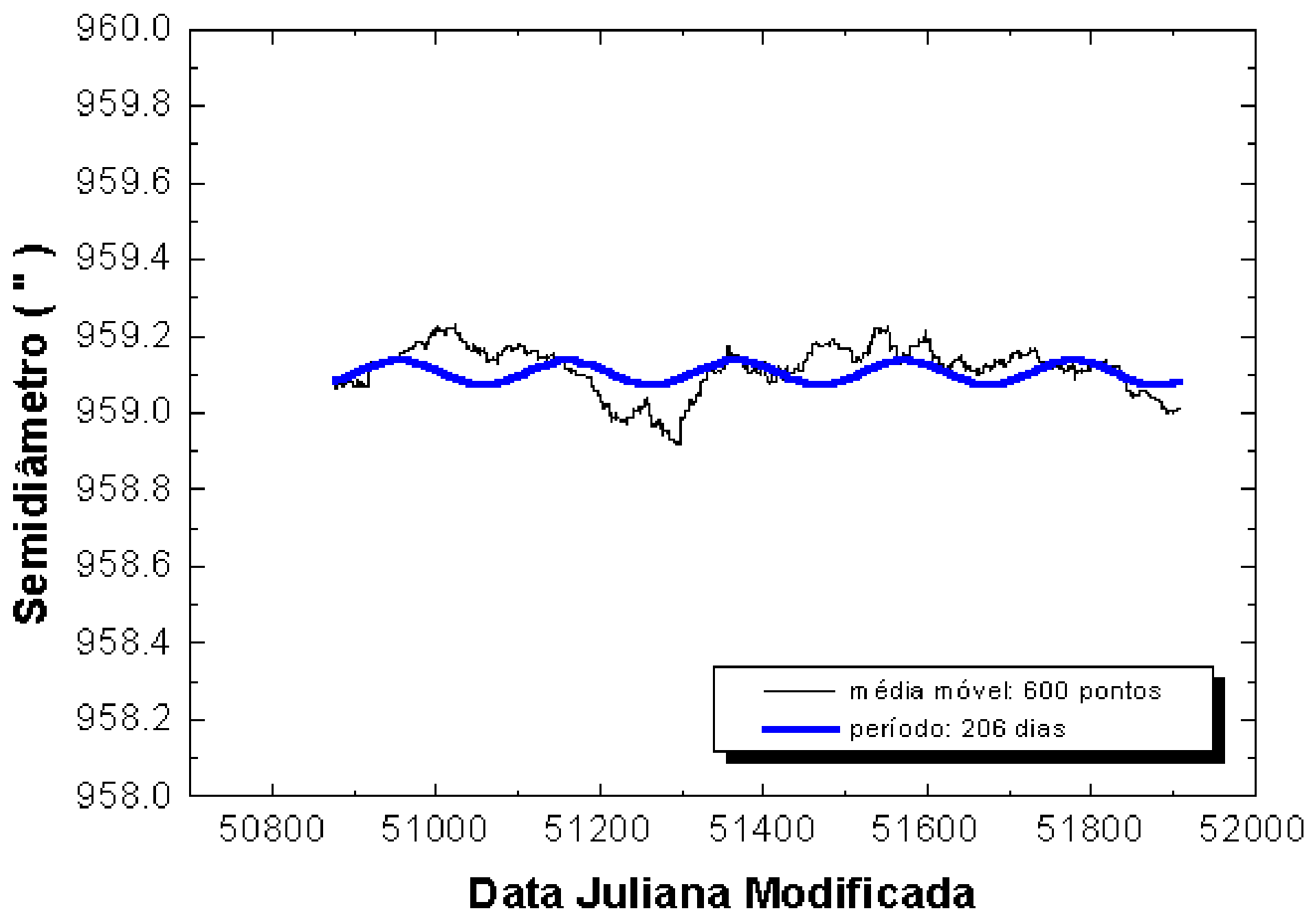

Figura 4.11. A curva em preto corresponde à média corrida de 600 pontos de toda a série corrigida. Superposto, em azul e na mesma escala, está o período de 206 dias.

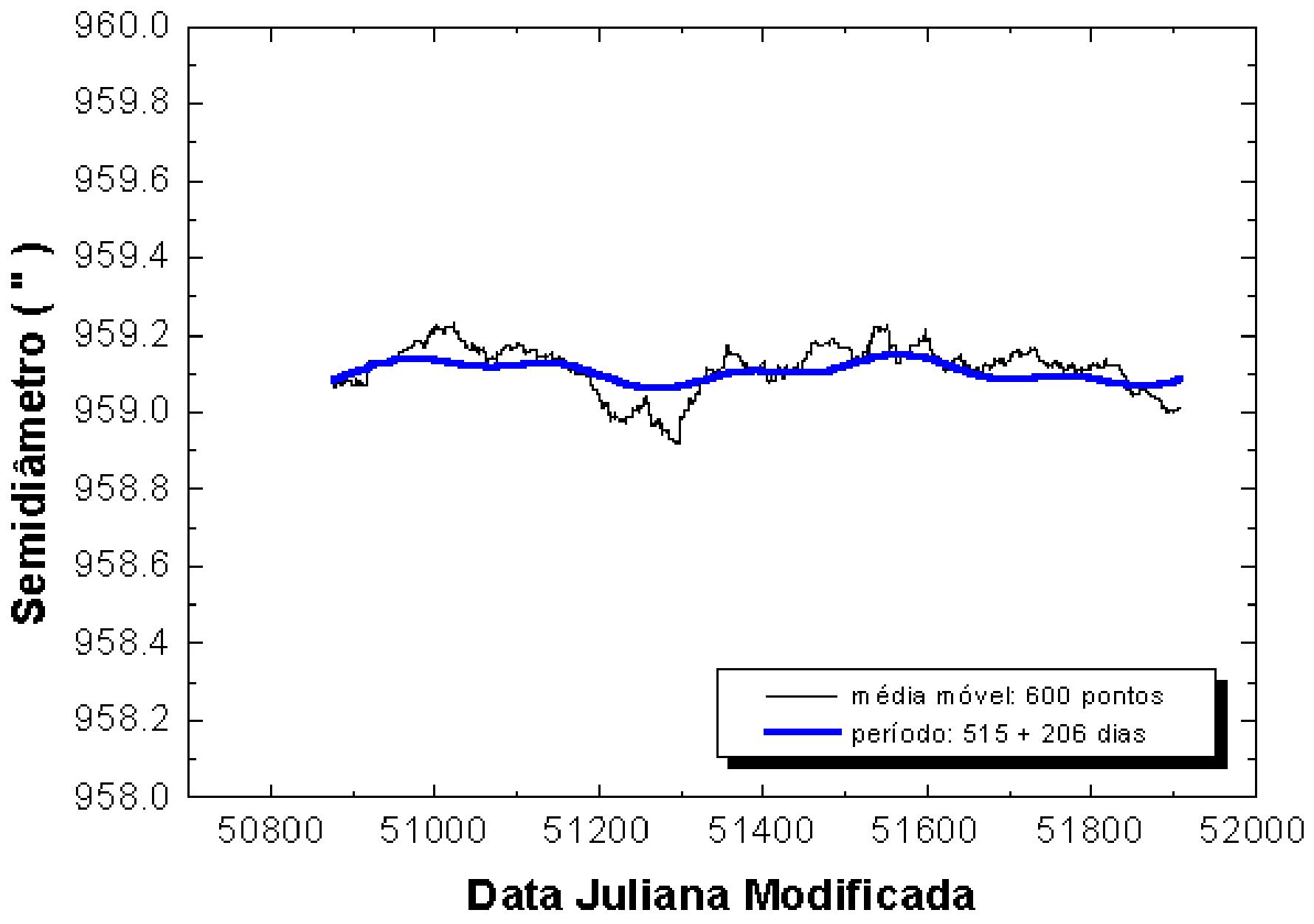

Figura 4.12. A curva em preto corresponde à média corrida de 600 pontos de toda a série corrigida. Superposto, em azul e na mesma escala, está a soma dos períodos de 515 dias e 206 dias.

# Capítulo 5

# Heliolatitudes

## 5.1 Cálculo e Evolução temporal

Devido ao método observacional, a heliolatitude dos pontos interceptados pelo diâmetro aparente do Sol varia, não só ao longo do ano, como de acordo com a distância zenital e é, ainda, geralmente diferente para lados opostos do meridiano central.

A latitude heliográfica é determinada por:

$$L = 90° - (C + S) \text{ (figura 5.1)}$$

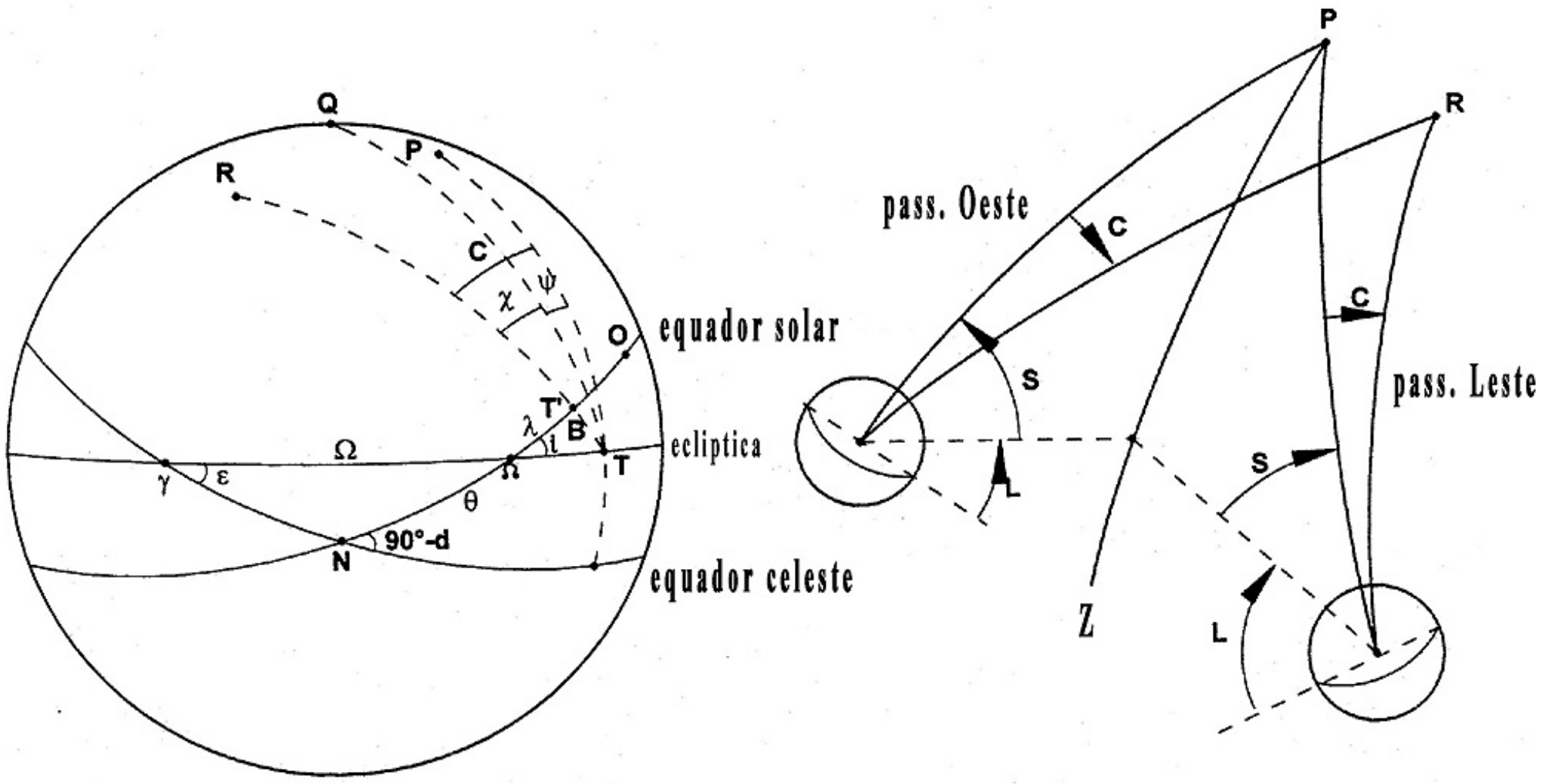

figura 5.1 C é o ângulo de posição do eixo do Sol, e S é o ângulo paralático. O ângulo S, contado positivamente para passagens ao Leste, é calculado ao mesmo tempo que o diâmetro, e o ângulo C é determinado pelas fórmulas da teoria da rotação do Sol (Danjon 1980).

Onde:

$$\cos\odot = \cos\alpha \cdot \cos\delta$$
$$\mathrm{tg}\psi = -\mathrm{tg}\varepsilon \cdot \cos\odot$$
$$\mathrm{tg}\chi = -\mathrm{tg}\iota \cdot \cos(\odot - \Omega)$$

$\odot$ - longitude do Sol
$\alpha$ - ascensão reta
$\delta$ - declinação

$\varepsilon$, $\iota$ e $\Omega$ são calculados para a época *t* da observação segundo as seguintes relações:

$$\varepsilon = 23^{o},446 - 0^{o},00013 \cdot (t - 1950,0)$$
$$\iota = 7^{o},252 + 0^{o},00002 \cdot (t - 1950,0)$$
$$\Omega = 74^{o},962 + 0^{o},01295 \cdot (t - 1950,0)$$

$\varepsilon$ - inclinação do equador celeste
$\iota$ - inclinação do equador solar
$\Omega$ - longitude do nodo ascendente

O programa original não leva em conta a diferença entre a posição do ponto central do disco solar observado para o ponto no equador solar interceptado pelo meridiano solar central, no instante. A diferença resultante é sempre pequena, da ordem do grau ou menor, mas deve ser levada em conta, devido à sua natureza sistemática e ao grande número de observações tratadas simultaneamente.

Um programa de correção foi desenvolvido e as heliolatitudes foram recalculadas para todo o período, segundo o formulário já apresentado acima. Sendo o diâmetro solar o resultado das medidas, a hipótese de simetria norte-sul aparece naturalmente. Por outro lado, embora muito grande, o número de medidas dentro de um período de rotação do disco solar não permite o estudo preciso de variações com a longitude heliográfica. Assim, foi também feita a hipótese de simetria leste-oeste. Ou seja, todas as medidas referem-se ao primeiro quadrante e, para propósito de visualização somente, serão rebatidas para os demais. Note-se que a distinção no grau de atividade observado nos dois hemisférios solares, em altas energias, (Pevtsov *et al.*, 2001) não encontra correspondência na atividade fotosférica. Deste modo, as hipóteses de simetria tem sido adotadas por outros autores (Leister e Emílio, 1998).

A figura 5.2, a seguir, mostra a variação temporal da heliolatitude do diâmetro observado ao longo do período. Ressalta, novamente, a assimetria entre as observações a Leste e a Oeste, que contribui para minimizar o efeito dos erros de natureza observacional.

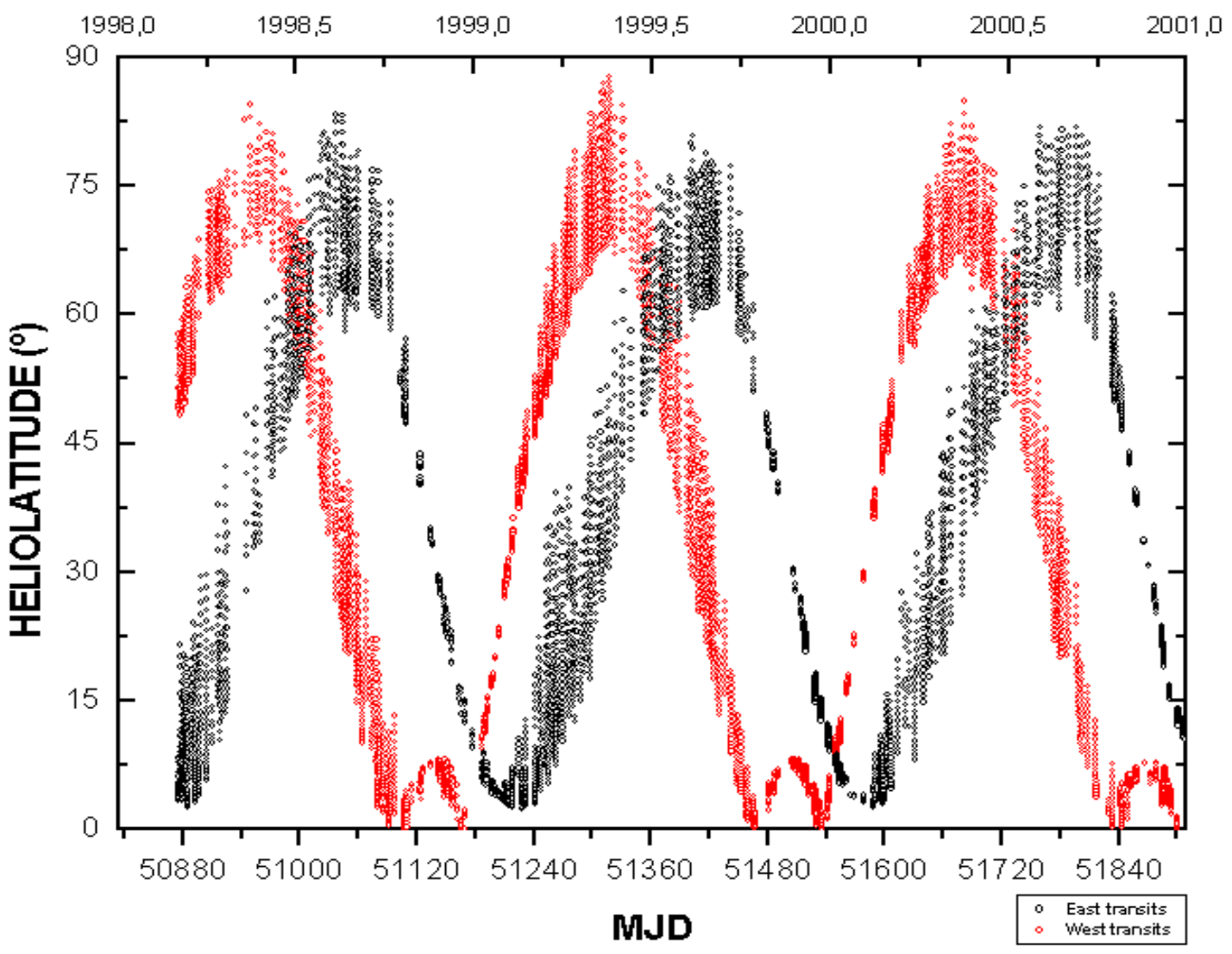

Figura 5.2. Evolução temporal das heliolatitudes dos diâmetros observados durante toda a série. As heliolatitudes tanto a Leste quanto a Oeste foram calculadas e rebatidas para o 1º

Durante o ano, são observados diâmetros em praticamente todas as latitudes do Sol. As medidas feitas a Oeste cobrem um intervalo aproximadamente 7º maior do que do que as medidas a Leste ($0°{,}06 \le hl_{oeste} \le 87°{,}73$, enquanto $2°{,}40 \le hl_{leste} \le 83°{,}31$). Se não levarmos em conta o rebatimento para o primeiro quadrante, esta diferença sobe para 15º.

Todas as heliolatitudes são adequadamente recobertas, isto é, qualquer que seja a região, o valor médio do diâmetro pode ser obtido com erro padrão na ordem do centésimo de segundo de arco, agrupando as medidas observadas em bandas de três graus de heliolatitude, no máximo. Por outro lado, conforme mencionado, uma das conveniências do sítio é que as diferenças entre o número de observações decorrem da geometria envolvendo as coordenadas locais do instrumento, o instante da medida e posição orbital da Terra, e nunca de dificuldades observacionais.

São apresentados, a seguir, os histogramas com os números de observações para cada heliolatitude observada durante o período de 1998 até 2000.

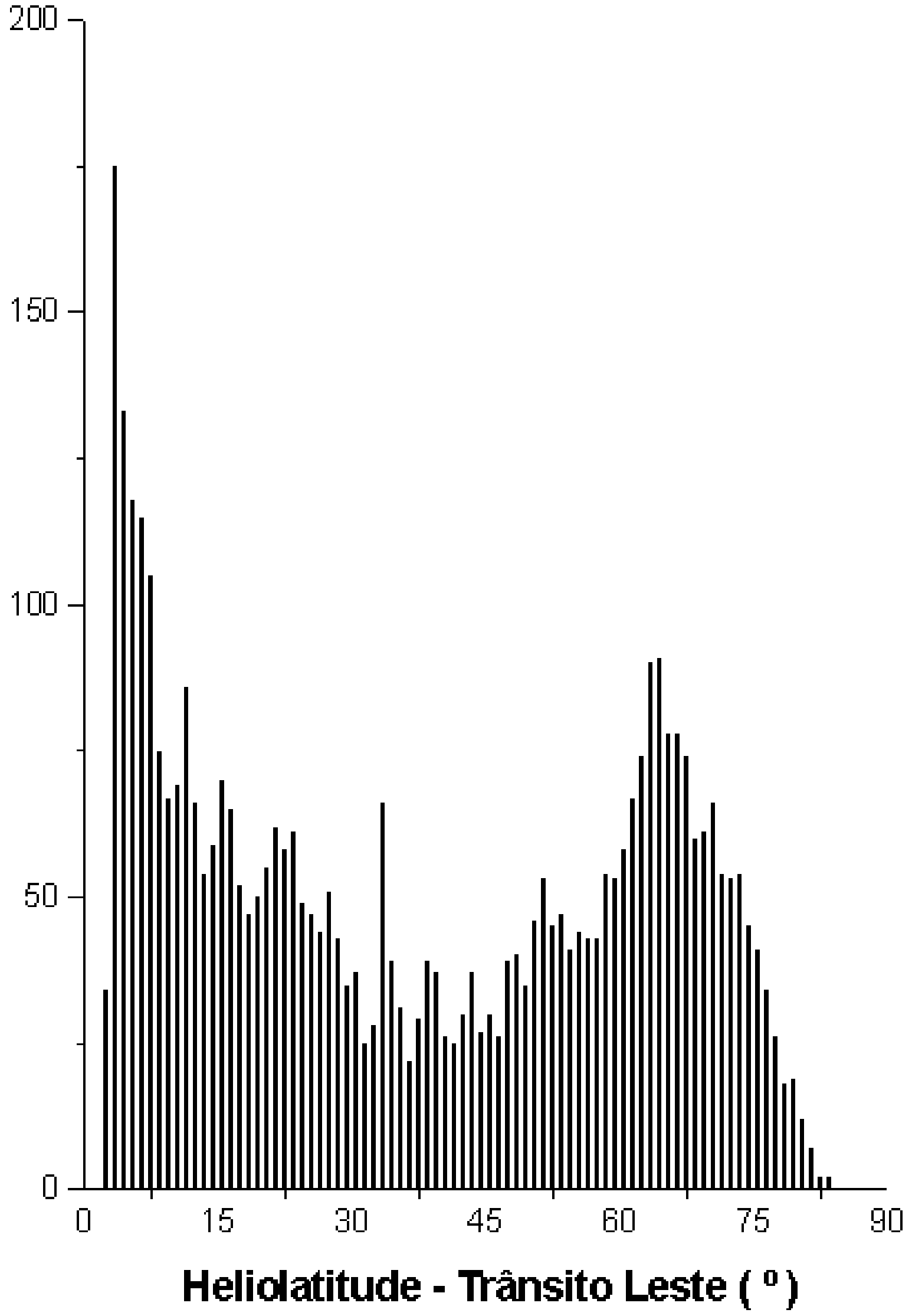

Figura 5.3. Heliolatitudes dos diâmetros solares observados para a passagem Leste durante toda a campanha de 1998, 1999 e 2000.

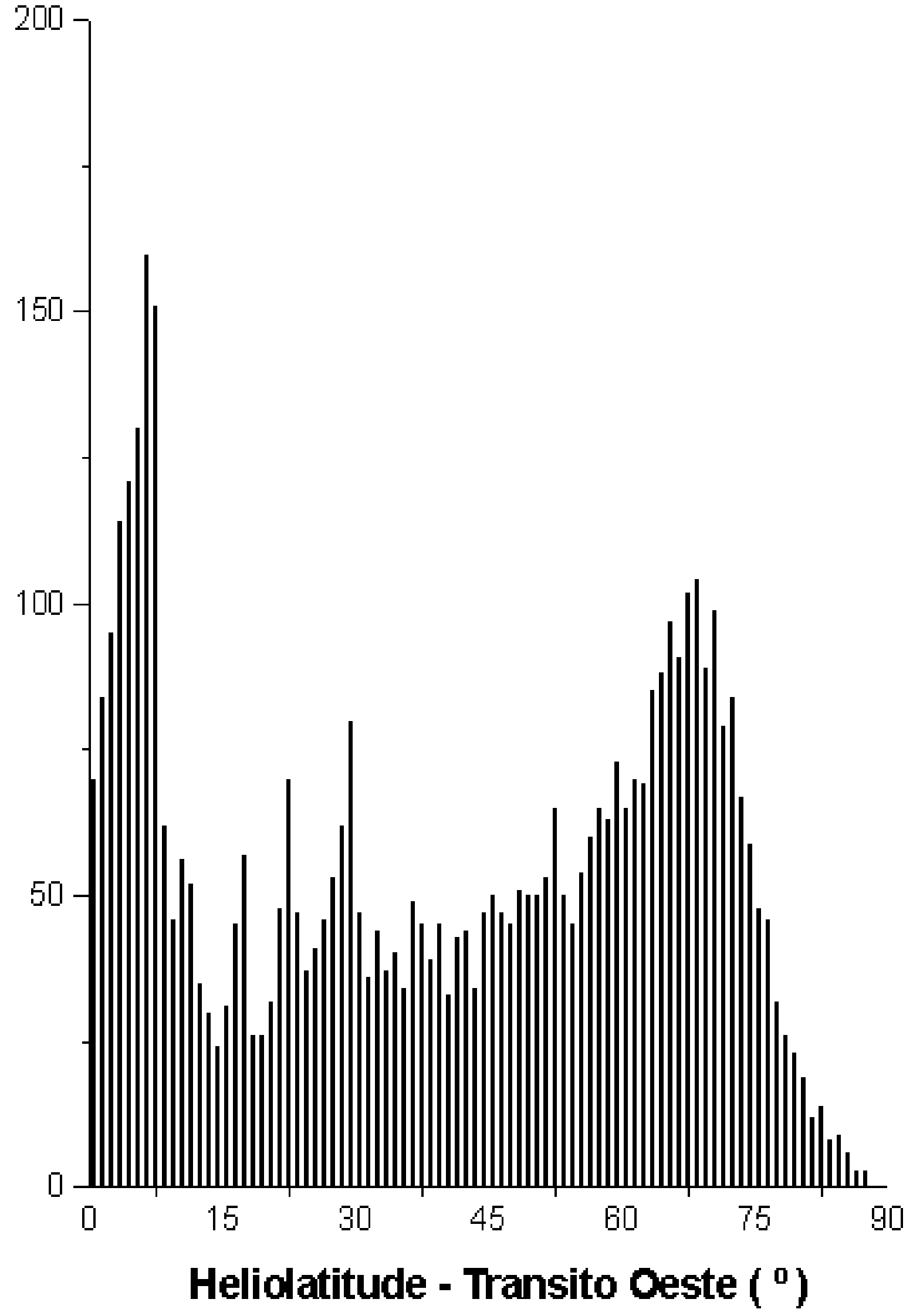

Figura5.4. Heliolatitudes dos diâmetros solares observados para a passagem Oeste durante toda a campanha de 1998, 1999 e 2000.

As contagens do números de observações, por heliolatitude, para as passagens Leste e Oeste de toda a campanha, estão no Anexo II.

Para todas as análises que se seguem, os resultados utilizados já foram tratados pelos parâmetros que descrevem os efeitos observacionais. As latitudes heliográficas usadas são aquelas recalculadas com maior precisão. A tendência linear também está removida. Desta maneira, variações temporais por heliolatitude não foram consideradas, mesmo porque, neste caso, o intervalo de tempo não é suficiente.

Nestas condições, o grande número de observações, cobrindo adequadamente o quadrante de heliolatitudes, permite investigar se existe ou não uma dependência do raio solar com sua latitude.

Espera-se que o Sol, por não ser um corpo rígido e possuir rotação, não seja uma esfera perfeita e sim algo muito próximo de um elipsóide de revolução (Hamy, 1889). Desvios da figura regular do Sol, tanto em termos de um achatamento ou pequenos "ressaltos" e "depressões", já foram teoricamente postulados. Godier e Rozelot (2000) discutiram a variação com a heliolatitude do ponto de vista de distorções do potencial gravitacional, levando em consideração a rotação diferencial. As amplitudes previstas, dependendo do mecanismo proposto, se situam entre frações de milésimos até algumas dezenas de milésimos de segundos de arco.

## 5.2 Forma do Sol

### 5.2.1 Ressaltos e Depressões

O modelo de Godier e Rozelot admite uma descrição em harmônicos esféricos da heliolatitude. As amplitudes entretanto são pequenas para serem detectadas neste estudo e, também, podem variar no tempo, neste caso de maneira pouco previsível. No modelo de Khun, que admite maiores amplitudes, as irregularidades da forma aparecem ligadas a variações de temperatura, levando a mudanças na curva de escurecimento do limbo solar. São, portanto, também temporalmente irregulares. A abordagem utilizada aqui constituiu em verificar a existência de irregularidades na forma, considerando-se todo o período analisado. Para investigar estas irregularidades na série tratada, foi calculada a média dos semidiâmetros em intervalos de 1º, segundo uma média móvel de 20 graus. A figura 5.5 traz o semidiâmetro para cada grau de heliolatitude.

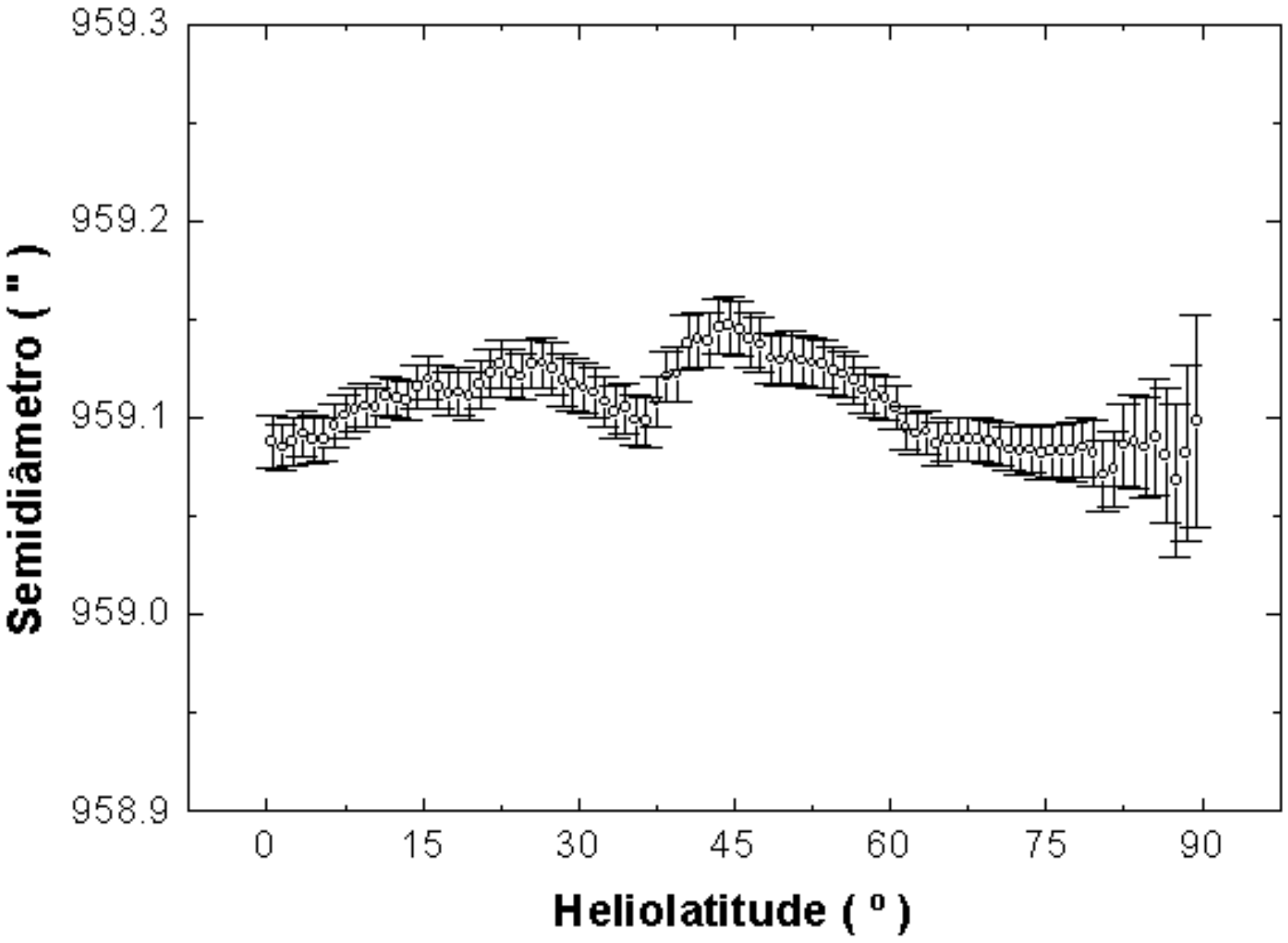

Figura 5.5. Dependência do semidiâmetro com a heliolatitude. Cada ponto representa o semidiâmetro médio a cada grau segundo uma média móvel de 20 graus.

Efetivamente, são encontradas irregularidades do semidiâmetro observado em certas heliolatitudes. Existe uma depressão relativa em $\text{hl} = 36°,5$ - $\langle sd \rangle = 959'',098 \pm 0'',013$ e dois máximos em torno $\text{hl} = 26°,5$ - $\langle sd \rangle = 959'',128 \pm 0'',013$ e $\text{hl} = 44°,5$ - $\langle sd \rangle = 959'',147 \pm 0'',014$.

Estas características concordam (em localização) com os perfis encontrados por Laclare *et al.* (1996) e Noël (1999). Naqueles trabalhos, porém, as amplitudes são maiores por um fator de 2 e 3 respectivamente.

Uma possível explicação para estas distorções poderia estar nas pequenas variações, em brilho, do limbo solar, encontradas por Kuhn *et al.* (1985) e que dependem também da heliolatitude observada. Kuhn pôde associar estas variações à mudanças de temperatura ao longo do limbo, da ordem de 0,5 K.

Uma comparação entre o perfil do excesso de temperatura versus heliolatitude obtido por Kuhn *et al.* (1984) mostra a partir de $\text{hl} \approx 36°$ uma anti-fase entre as curvas. Porém a máxima amplitude, relativa ao semidiâmetro médio, e cerca de 4 vezes maior do que as predições teóricas (Kuhn, 1988), no entanto, deve-se levar em conta que os presentes resultados foram obtidos na vizinhança do máximo do ciclo solar, enquanto que as predições referem-se a um período de mínimo. Também, a contribuição de fáculas contamina o cálculo do excesso de temperatura (Kuhn, 1985) dificultando a comparação dos resultados.

Como foi explicado antes, o diâmetro observado do Sol é obtido pelo intervalo com que o ponto de inflexão da função de escurecimento de bordos opostos cruza determinada distância zenital. Esta anticorrelação encontrada parece indicar que este ponto de inflexão depende do brilho e/ou temperatura do limbo, uma vez que o perfil da função escurecimento é modificado.

Entre 5º e 35º as curvas se apresentam em fase, sugerindo que entre estas latitudes outro fenômeno pode estar prevalecendo, por exemplo o efeito de achatamento devido à rotação.

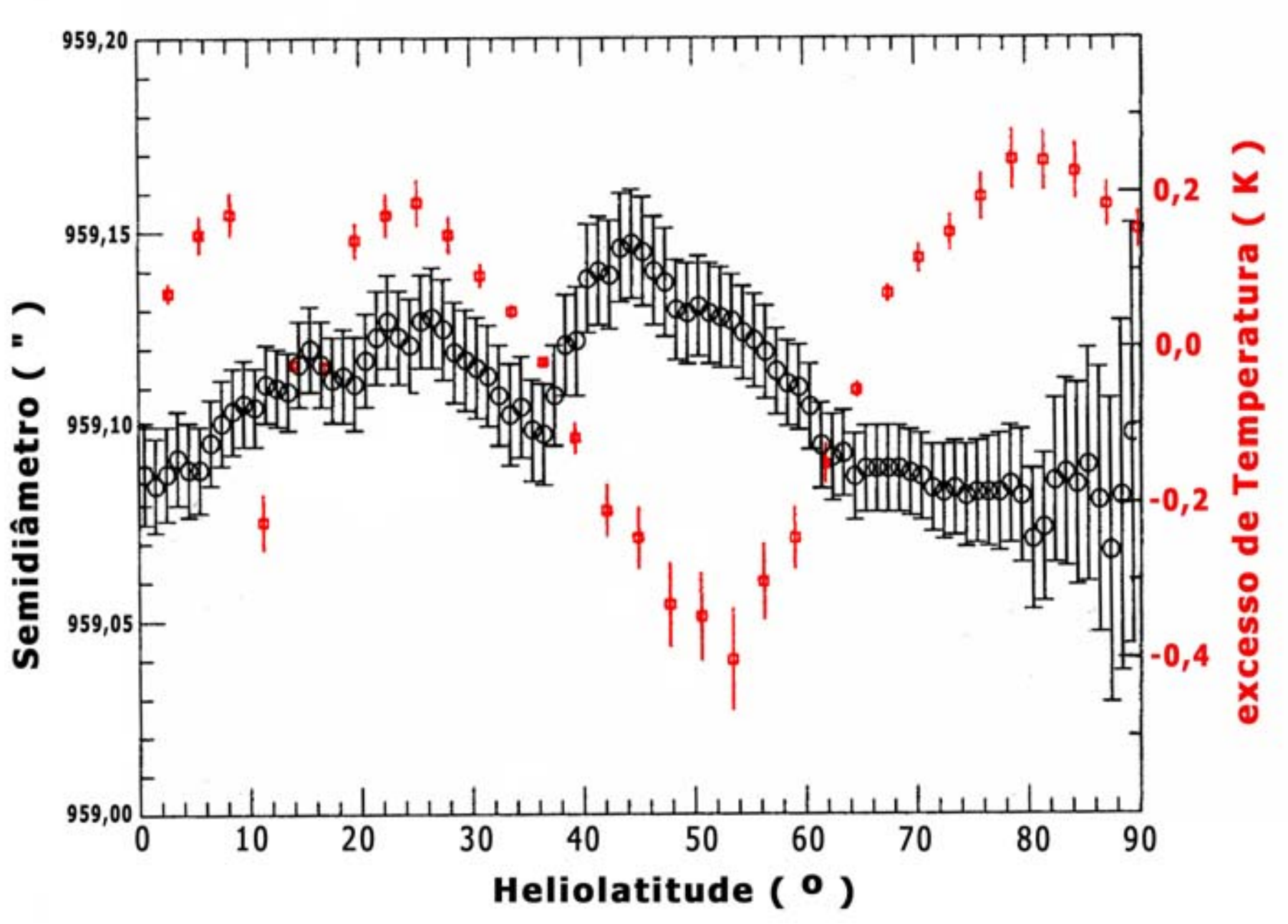

Figura 5.6. Superposição dos gráficos do semidiâmetro e o excesso de temperatura (pontos vermelhos, Kuhn *et al.*, 1984) em cada heliolatitude. A partir de 36º observa-se uma antifase entre as curvas.

### 5.2.2 Achatamento visível do Sol

A previsão teórica, para um corpo com as características solares em rotação, indica um limite máximo para a diferença entre o raio equatorial e polar da ordem da dezena de milisegundos de arco (Rozelot *et al.*, 2001). Para as medidas aqui apresentadas, portanto, algo como 50 vezes menor do que o desvio padrão da série de medidas. Resulta, então, que a coerência das medidas, sua distribuição por heliolatitude e a grande quantidade de pontos obtidos pelo projeto de monitoramento solar do ON, permite que esta grandeza possa ser investigada com suficiente grau de confiança.

O achatamento é definido segundo a expressão: $f = \frac{r_{eq} - r_{pol}}{r_{eq}}$, onde $r_{eq}$ é o raio equatorial e $r_{pol}$ é o raio polar

Outros autores utilizam uma expressão semelhante: $\varepsilon = \frac{r_{eq} - r_{pol}}{r_O}$, onde $r_O$ é o valor médio do raio do Sol (Dicke, 1970) e ε seria a 'oblacidade' do Sol.

De acordo com a teoria da Relatividade Geral, a precessão do periélio de Mercúrio deveria ser menos intensa do que se observa (-4 segundos de arco/ano). Um achatamento do Sol seria uma possível explicação para o fenômeno (Bauschinger, 1884). Uma medida precisa do achatamento visual do Sol seria significativo na determinação do seu campo gravitacional.

Vários autores determinaram teoricamente o valor desta grandeza:

- Kislik (1983) analisando os resíduos dos parâmetros orbitais dos planetas internos estimou os seguintes limites para o achatamento: $1{,}08 \times 10^{-5} < f < 2{,}69 \times 10^{-5}$

- Burša (1986), estimou o achatamento aplicando cálculos de equilíbrio hidrostático ao Sol. O valor do achatamento estaria entre $1{,}1x10^{-5}$ e $2.7x10^{-5}$, sendo que o limite inferior corresponde a um corpo quasi-homogêneo e o limite superior a um corpo com núcleo pesado.

- Rozelot, Godier e Lefebvre (2000), combinando medidas feitas pela heliossismologia e incluindo a rotação diferencial, obtiveram limites máximo e mínimo para a diferença entre os raios equatorial e polar e o achatamento:

$$(6{,}7 \pm 1{,}4) \times 10^{-6} < f < (11{,}0 \pm 0{,}3) \times 10^{-6}$$
$$(6{,}39 \pm 1{,}31) \times 10^{-3} < \Delta r < (10{,}54 \pm 0.25) \times 10^{-3}$$

- Lydon e Sofia (1996), baseados em dados do *Solar Disk Sextant* de 1992 e 1994, mediram o achatamento e encontraram os valores $(9{,}17 \pm 1{,}25) \times 10^{-6}$ e $(8{,}77 \pm 0{,}99) \times 10^{-6}$, respectivamente.

Além disso, Dicke, Kuhn e Libbrechet (1985) sugeriram que a magnitude do achatamento do Sol poderia ser função do ciclo solar.

Para obtermos os raios equatorial e polar do Sol, a fim de tirar partido da distribuição de nossos dados, uma idéia mais simples foi utilizada e a equação de uma elipse foi ajustada, por mínimos quadrados, à série tratada.

$$\frac{1}{r_{obs}^2} = \frac{\cos^2(hl)}{r_{eq}^2} + \frac{\text{sen}^2(hl)}{r_{pol}^2}$$, onde $r_{obs}$ é o semidiâmetro observado.

Os valores obtidos foram: $r_{eq} = 959'',113 \pm 0'',007$

$r_{pol} = 959'',100 \pm 0'',011$

O que leva a uma diferença entre os raios equatorial e polar de 13 mas. Este valor concorda com outros encontrados na literatura (Rozelot e Rösh, 1997 e Dicke, Kuhn e Libbrecht, 1985).

Novamente, a característica de independência das medidas, aliada à sua coerência e quantidade, permitiu estabelecer um número de sub-conjuntos. Através do artifício de impor um vínculo adicional de determinar o valor médio para todos os sub-conjuntos, obtem-se um ganho significativo na acurácia da medida do achatamento. Com tal fim, sub-conjuntos foram estabelecidos pela retirada progressiva dos pontos mais discrepantes. Onze sub-conjuntos foram estabelecidos, diferindo em passos de 0,1σ, retirando-se pontos desde 2σ até 3σ**,** relativamente ao ajuste de uma elipse. O número de medidas utilizadas varia desde 7556 até 9047. Os valores finais dos raios equatorial e polar são dados pela média dos valores obtidos para cada sub-conjunto. Os erros finais são calculados pela média quadrática dos quadrados dos erros obtidos para cada sub-conjunto mais os termos de covariância entre sub-conjuntos adjacentes. Do modo mais conservador, a covariância foi sempre tomada como unitária.

Os valores finais obtidos são: $r_{eq} = 959'',110 \pm 0'',002$

$r_{pol} = 959'',097 \pm 0'',003$ (1)

A mesma diferença é verificada ($\Delta r = 0'',013 \pm 0'',004$).

De posse destes valores, pode-se proceder ao cálculo do achatamento visual do Sol:

i) Substituindo os valores encontrados:

$$f = \frac{959{,}110 - 959{,}097}{959{,}110} \Rightarrow f = 1{,}355 \times 10^{-5}$$

ii) Calculando a incerteza do achatamento:

$$\sigma_f^2 = \left[\frac{\partial}{\partial r_{eq}}(f)\right]^2 \cdot \sigma_{r_{eq}}^2 + \left[\frac{\partial}{\partial r_{pol}}(f)\right]^2 \cdot \sigma_{r_{pol}}^2$$

$$\sigma_f^2 = \left[\frac{\partial}{\partial r_{eq}}\left(\frac{r_{eq} - r_{pol}}{r_{eq}}\right)\right]^2 \cdot \sigma_{r_{eq}}^2 + \left[\frac{\partial}{\partial r_{pol}}\left(\frac{r_{eq} - r_{pol}}{r_{eq}}\right)\right]^2 \cdot \sigma_{r_{pol}}^2$$

$$\sigma_f^2 = \left(\frac{1}{r_{eq}} - \frac{r_{eq} - r_{pol}}{r_{eq}^2}\right)^2 \cdot \sigma_{r_{eq}}^2 + \left(-\frac{1}{r_{eq}}\right)^2 \cdot \sigma_{r_{pol}}^2$$

iii) Substituindo os valores:

$$\sigma_f^2 = \left(\frac{1}{959{,}110} - \frac{959{,}110 - 959{,}097}{(959{,}110)^2}\right)^2 \cdot (0{,}002)^2 + \left(-\frac{1}{959{,}110}\right)^2 \cdot (0{,}003)^2 =$$

$$= 1{,}4132 \times 10^{-11}$$

$$\Rightarrow \sigma_f = \sqrt{1{,}4132 \times 10^{-11}} = 3{,}75926 \times 10^{-6}$$

iv) $f = (13{,}55 \pm 3.76) \times 10^{-6}$

Este valor está um pouco acima do valor máximo teórico (vide página 66) estabelecido por Rozelot *et al.*(2000).

A forma do disco solar obtido da série completa é mostrada na figura 5.7. Os pontos em cinza, no primeiro quadrante, correspondem a cada uma das medidas individuais analisadas neste trabalho, e foram rebatidos para os demais quadrantes. Uma média móvel de 600 pontos é mostrada na curva em vermelho. Para melhor visualização das irregularidades, estas estão ampliadas por um fator de $10^3$.

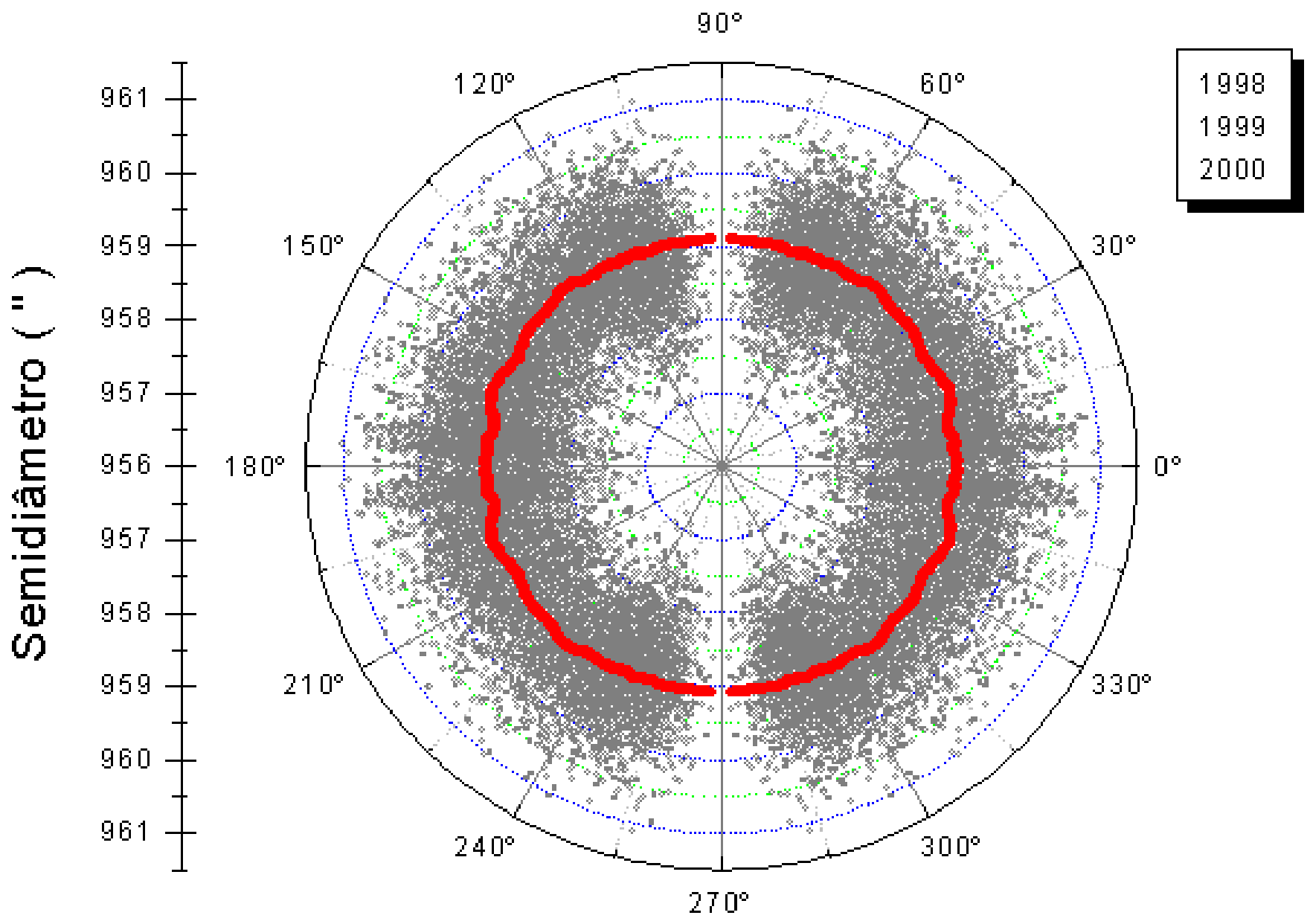

Figura 5.7. Perfil solar obtido dos dados analisados. As imperfeições estão exageradas por um fator de 1000. Todos os raios foram determinados para o primeiro quadrante e rebatidas para os outros para formar a figura.

## 5.2.3 Momento de quadrupólo ($J_2$)

O momento de quadrupólo pode ser determinado com grande aproximação a partir da forma observada da superfície do Sol, de uma maneira independente dos detalhes do seu interior (Dicke e Goldenberg, 1967).

Se um momento de quadrupólo gravitacional existir, mostrar-se-ia certamente na superfície visível; porém, um achatamento visual medido não indica com a mesma certeza a existência de momento de quadrupólo. Assim, uma medida do achatamento poria um limite superior pelo menos em $J_2$ ou, dependendo de considerações teóricas, alguma indicação de sua magnitude (Clayton, 1973).

Afanas'eva *et al.* (1990) analisando dados de observações por radar dos planetas internos durante 1964-1986 encontrou o valor de $(0{,}66 \pm 0{,}9) \times 10^{-6}$ para $J_2$.

Rozelot *et al.*, 2000 calcularam teoricamente o momento de quadrupólo, levando em conta todos os dados observacionais, o momento de quadrupólo do Sol como sendo $J_2 = -(6{,}84 \pm 3.75) \times 10^{-7}$ e $J_2 = -(3{,}49 \pm 1{,}86) \times 10^{-7}$ se só os dados heliossismológicos forem usados. O sinal negativo indica corretamente a forma oblata do Sol.

### 5.2.3.1 Cálculo de $J_2$

A oblacidade do Sol, no caso de uma rotação uniforme, é relacionada com o momento de quadrupólo do Sol através da expressão: $\varepsilon = \varepsilon_q + \varepsilon_s = \frac{3}{2} J_2 + \frac{1}{2} \frac{\Omega^2 r_O^3}{G M_O}$ (Dicke, 1970). Onde $G$ é a constante gravitacional, $M_O$ a massa do Sol e $\Omega$ a velocidade angular uniforme. A oblacidade seria portanto a soma de dois termos, o primeiro $\varepsilon_q$, seria a medida da distorção do potencial gravitacional produzida pela rotação interna e o segundo, $\varepsilon_s$ produzido pela rotação da superfície. Para o cálculo do momento de quadrupólo visual do Sol, usaremos uma aproximação mais simples onde $J_2$ está relacionado com a diferença entre os raios equatorial e polar segundo a expressão:

$$J_2 = \frac{2}{3} \frac{(\Delta r - \delta r)}{r_0} \qquad (2)$$

onde: $\Delta r = r_{eq} - r_{pol}$;

$\delta r = (7{,}8 \pm 2.1) \times 10^{-3}$, devido exclusivamente à rotação da superfície (Rozelot e Rösh, 1996);

$r_0 = \sqrt[3]{r_{eq}^2 r_{pol}}$, raio do Sol (ou seja, a melhor superfície esférica ajustada para $r_{eq}$ e $r_{pol}$, Rozelot, Godier e Lefebvre, 2001).

i) substituindo os valores obtidos em (1) em (2)

$$J_2 = \frac{2}{3}\frac{\left(959{,}110 - 959{,}097 - 7{,}8\times10^{-3}\right)}{\sqrt[3]{\left(959{,}110\right)^2\cdot 959{,}097}} =$$

$$= 3{,}61448\times10^{-6}$$

ii) calculando a incerteza de $J_2$:

$$J_2 = J_2\left(r_{eq}, r_{pol}, \delta r\right)$$

$$\Rightarrow \sigma_{J_2}^2 = \left[\frac{\partial}{\partial r_{eq}}\left(J_2\right)\right]^2\cdot\sigma_{r_{eq}}^2 + \left[\frac{\partial}{\partial r_{pol}}\left(J_2\right)\right]^2\cdot\sigma_{r_{pol}}^2 + \left[\frac{\partial}{\partial \delta r}\left(J_2\right)\right]^2\cdot\sigma_{\delta r}^2$$

$$\frac{\partial}{\partial r_{eq}}\left(\frac{2}{3}\frac{\left(r_{eq}-r_{pol}-\delta r\right)}{\sqrt[3]{r_{eq}^2 r_{pol}}}\right) = \frac{2}{3}\frac{1}{\sqrt[3]{r_{eq}^2 r_{pol}}} - \frac{4}{9}\frac{r_{eq}-r_{pol}-\delta r}{\left(\sqrt[3]{r_{eq}^2 r_{pol}}\right)^4} r_{eq} r_{pol} = 6{,}95089\times10^{-4}$$

$$\frac{\partial}{\partial r_{pol}}\left(\frac{2}{3}\frac{\left(r_{eq}-r_{pol}-\delta r\right)}{\sqrt[3]{r_{eq}^2 r_{pol}}}\right) = -\frac{2}{3}\frac{1}{\sqrt[3]{r_{eq}^2 r_{pol}}} - \frac{2}{9}\frac{r_{eq}-r_{pol}-\delta r}{\left(\sqrt[3]{r_{eq}^2 r_{pol}}\right)^4} r_{eq}^2 = -6{,}95093\times10^{-4}$$

$$\frac{\partial}{\partial \delta r}\left(\frac{2}{3}\frac{\left(r_{eq}-r_{pol}-\delta r\right)}{\sqrt[3]{r_{eq}^2 r_{pol}}}\right) = -\frac{2}{3}\frac{1}{\sqrt[3]{r_{eq}^2 r_{pol}}} = -6{,}9509\times10^{-4}$$

$$\Rightarrow \sigma_{J_2}^2 = \left(6{,}95089\times10^{-4}\right)^2\cdot\left(0{,}002\right)^2 +$$
$$+\left(-6{,}95093\times10^{-4}\right)^2\cdot\left(0{,}003\right)^2 +$$
$$+\left(-6{,}9509\times10^{-4}\right)^2\cdot\left(2{,}1\times10^{-3}\right)^2 = 8{,}41169\times10^{-12}$$

$$\Rightarrow \sigma_{J_2} = \sqrt{8{,}41169\times10^{-12}} = 2{,}90029\times10^{-6}$$

$$J_2 = \left(3{,}61 \pm 2{,}90\right)\times10^{-6}$$

Este valor é maior, de 5 a 10 vezes, do que o valor encontrado por Rozelot *et al.* (2001), mas concorda bem com outros encontrados na literatura.

A tabela a seguir traz alguns destes valores:

| | Data | $\Delta r$ (") ($10^{-3}$) | f ($10^{-6}$) | $\|J_2\|$ ($10^{-6}$) |
|---|---|---|---|---|
| **Este trabalho** | **1998/1999/2000** | **13 ± 4** | **13,55 ± 3,76** | **3,61 ± 2,90** |
| Rozelot & Rösh (1997) | 1996 | 13,1 ± 4,1 | - | 3,64 ± 2,84 |
| Rozelot (1996) | 1993/94 | 11,5 ± 3,4 | - | - |
| Rösh *et al.* (1996) | 1993/94 | - | 12,0 ± 3,5 | 2,57 ± 2,36 |
| Sofia *et al.* (1994) | 1994 | 8,21 ± 0,84 | 8,63 ± 0,88 | - |
| Dicke *et al.* (1987) | 1985 | 14,6 ± 2,2 | - | - |
| Beardsley (1987) | 1983 | - | 13,2 ± 1,5 | 3,4 ± 1,3 |
| Dicke *et al.* (1986) | 1984 | 5,6 ± 1,3 | 5,8 | - |
| Hill & Stebbins ( 1975) | 1975 | 18,4 ± 12,5 | 9,6 ± 6,5 | 1,0 ± 4,3 |
| Dicke & Goldenberg (1974) | 1966 | 43,3 ± 3,3 | 45,1 ± 3,4 | 24,7 ± 2,3 |

Tabela 5.1. Comparação entre os diversos valores para a diferença entre os raios equatorial e polar, o achatamento e o momento de quadrupólo do Sol, encontrados por diversos grupos.

# Capítulo 6

## Conclusões e Perspectivas

### 6.1 Sobre o trabalho

Indicamos, na introdução desta dissertação, o interesse astronômico, físico e geofísico que há na compreensão das variações do diâmetro e da forma do Sol. A esta multiplicidade de interesses corresponde uma multiplicidade de escolhas cientificamente interessantes para a abordagem do tema. Até recentemente havia uma lacuna quanto a observações prolongadas do diâmetro solar. Esta lacuna veio a ser parcialmente preenchida com o trabalho de Emilio *et al.* (2001), a partir de re-análises de dados do satélite SOHO. Mesmo aquele trabalho, no entanto, tem importantes limitações instrumentais que precisam ser confrontadas com outros experimentos de duração e freqüência comparável. Desta maneira, esta dissertação visa abordar as seguintes questões:

- As observações de solo podem ser tratadas dos ruídos observacionais, a fim de permitir resultados com precisão da ordem do milisegundo de arco?

- Em caso positivo, os princípios de funcionamento do Astrolábio Solar, do qual vem se formando uma pequena rede mundial, são adequados para esta tarefa?

- Com que precisão se pode determinar que, durante o tempo de monitoramento aqui estudado, o semi-diâmetro solar variou? E com que periodicidades e amplitudes?

- Com que precisão se pode determinar que, durante o tempo de monitoramento aqui estudado, o semi-diâmetro solar apresentou uma evolução de natureza secular?

- Em que medida a forma do Sol médio se afasta de uma perfeita esfericidade?

Este trabalho, portanto, teve como objetivo analisar os dados da campanha de monitoramento do diâmetro solar de 1998 até 2000. Durante este intervalo foram realizadas 10807 medidas individuais do diâmetro solar, divididas entre observações a Leste a Oeste do meridiano e homogeneamente distribuídas ao longo destes três anos. O instrumento utilizado foi um astrolábio Danjon adaptado para observações solares, equipado com um prisma refletor de ângulo variável e sistema de aquisição de imagens através de uma câmera CCD. A distância zenital de observação varia entre 25º e 55º, o que permite o monitoramento do diâmetro solar durante todo o ano. O instrumento está localizado no campus do Observatório Nacional ($\phi = -22^o 53'42''$, $\Lambda = +2^h 52^m 53^s$, $h = 33m$).

O sistema de aquisição e redução dos arquivos imagens é o mesmo desenvolvido no Observatório de Paris (Sinceac, 1998) e deriva o valor do diâmetro solar computando o intervalo da passagem dos bordos solares através de um almucântar. Este programa permite encontrar até três soluções (diferindo em média de 0",10 entre si), o que fez com que o valor absoluto de semidiâmetro do Sol tenha sido uma questão de critério escolhido, e não tenha sido uma preocupação primária nesta análise.

Um tratamento das imagens, levando-se em conta a *corrente de escuro* e o *flat field* da câmara foi testada. Para este tratamento desenvolveu-se um algoritmo para a conversão das imagens visando a utilização das rotinas IRAF possibilitando uma melhor relação sinal/ruído e ganho na definição do bordo solar, permitindo a discussão do modelo que o representa.

A correção das imagens astrolábio por tratamento de *flat-field* resultou numa diminuição do ruído das contagens e menor dispersão dos resultados. No entanto, tendo em vista a magnitude dos demais erros, sua contribuição foi pequena.

### 6.1.2 Sobre as Correções

Antes da análises dos dados, primeiramente uma varredura em toda a série foi feita para eliminar-se os pontos destoantes: pontos que estavam fora da faixa de $\pm 2{,}5\sigma$ da série completa, pontos resultantes de observações cuja orientação da matriz do CCD ultrapassava 2º e pontos provenientes de sessões com menos de 4 observações. Foi obtida então uma série contendo 9112 medidas independentes do semidiâmetro do Sol, dividida em 4246 observações feitas a Leste e 4866 feitas a Oeste.

Um estudo das influências sistemática que as condições observacionais têm sobre o resultado final da redução foi realizado. Foram testados diversos parâmetros refletindo as condições locais e atmosféricas: distância zenital, azimute, latitude heliográfica, temperatura, pressão, fator de Fried, a largura do bordo direto e refletido e o sigma do ajuste das parábolas do limbo direto e refletido. Também foram testados parâmetros derivados destes, como a variação dos valores durante a observação e a tangente trigonométrica da distância zenital. Além destes, um termo linearmente dependente do tempo foi incorporado à pesquisa.

A temperatura média na hora da observação, mostrou-se como o parâmetro mais influente sobre o resultado final. A variação da temperatura durante a sessão, o fator de Fried e o desvio padrão do ajuste da parábola refletida também apareceram mas com um complexo e menor grau de influência. O termo linear também foi incluído na correção, embora as características da série não permitam descrevê-lo como um termo secular. Por outro lado ele não é incompatível com a atividade solar. Os coeficientes de ajuste da série foram semelhantes para as sessões a Leste e a Oeste, por isso um único coeficiente pôde ser usado, por parâmetro, para todas as observações.

As correções introduzidas são da ordem de centésimos de segundo de arco, o que é dez vezes menor do que o erro associado a uma única observação.

O valor final encontrado para o semidiâmetro médio deste período foi de 959",107±0",006.

O método observacional demonstrou que possui qualidade astrométrica e que juntamente com o número de observações é suficiente para obter-se uma precisão nas medidas da ordem de poucos milisegundos de arco. A extensão da série permite a pesquisa de variações periódicas ou não do semidiâmetro solar observado da ordem de 10 até 1000 dias e amplitude mínima da ordem de dezena de segundos de arco.

As séries, completa e separada por sessões, seguem uma distribuição normal verificada a mais de 99%.

### 6.1.3 Sobre os Períodos

Variações globais do raio solar afetam de modo semelhante as observações temporalmente próximas, sejam tomadas a Leste quanto a Oeste. Seguindo essa linha foram efetuados dois testes de correlação (Spearman e Pearson) entre os valores observados para as duas sessões, servindo também para testar a qualidade das correções paramétricas. Os dias em que houve sessões duplas foram separados e testados ponto a ponto e com vários graus de alisamento. O teste sugeriu a presença de termos periódicos fornecendo uma indicação preliminar da amplitude dos termos globais, além de confirmar que o tratamento aumentou a correlação entre as sessões não alterando seu caráter geral.

Através do algoritmo CLEAN, os termos periódicos das variações do valor do semidiâmetro foram obtidos. Novamente foram comparados os resultados alcançados com a série antes e depois do tratamento paramétrico e também dividida entre as sessões Leste e Oeste. Períodos semelhantes estão presentes nas séries de ambas as sessões, reforçando a origem solar das variações encontradas, uma vez que estas séries são obtidas sob condições observacionais bem distintas. Após a correção paramétrica, o período de 1030 dias desapareceu, provavelmente por ser um reflexo do termo linear encontrado e subtraído da série. O período anual (343,3 dias), apesar de ter sua amplitude atenuada, continua presente. O período de 515 dias, após a correção, teve sua amplitude praticamente duplicada, fazendo deste o principal período contido na série. A utilização de um segundo algoritmo, o DCDFT (*Date-Compensate Discrete Fourier transform*), serviu de contraprova para os períodos encontrados.

### 6.1.4 Sobre a figura do Sol

Pela própria natureza do método observacional e qualidade das séries a procura da *não esfericidade* e o achatamento do Sol pôde ser executada. O diâmetro observado do Sol tem sua orientação variando com a distância zenital e com a época do ano. Devido a localização do sítio do Observatório Nacional ($\phi = -22^o 53'42'',50$) praticamente todas as heliolatitudes podem ser observadas ao longo do ano. Esta situação, aliada ao grande número de pontos, é ideal para se procurar por uma dependência do valor do diâmetro observado e sua respectiva latitude solar, já citada por diversos autores (Laclare *et al.,* 1996 e Noël, 1999).

Para minimizar o ruído observacional, a série foi alisada tomando-se a média dos semidiâmetros a cada intervalo de 1 grau, segundo uma média móvel de 20 graus. Observam-se irregularidades na forma do perfil do Sol, e notadamente um pequeno máximo para heliolatitude de 26º,5 seguido por uma depressão relativa em 36º,5 e outro máximo em 44º,5. Estas variações do semidiâmetro médio podem estar associadas à mudança de temperatura ao longo do limbo. A amplitude deste máximo, no entanto é cerca de 4 vezes maior do que variação prevista teoricamente.

As medidas do achatamento visual do Sol são difíceis de serem conseguidas, devido ao nível de ruído observacional que a turbulência atmosférica introduz nas medidas de solo. A partir das séries aqui tratadas, para se obter este valor, uma aproximação mais conservadora foi admitida e os dados foram ajustados, através de mínimos quadrados, à equação de uma elipse. Os ressaltos e depressões da figura solar foram considerados como parte do ruído. Os valores encontrados para o raio equatorial e polar foram respectivamente: 959",113±0",007 e 959",100±0",011. Estas medidas estão de acordo com outras determinações do achatamento da figura do Sol feita por outros autores. No entanto, os valores da diferença entre os dois raios e sua incerteza são da mesma ordem de grandeza. Para ganhar em precisão, tirando proveito da longa série de três anos e sua distribuição, retiraram-se progressivamente os pontos fora da faixa de 2σ até 3σ, em passos de 0,1σ (relativamente ao ajuste da elipse) retendo-se os resultados e as incertezas para cada subconjunto.

Os raios equatorial e polar encontrado são ligeiramente menores, porém mais precisos:

$R_{eq} = 959'',110 \pm 0'',002$ e $R_{pol} = 959'',097 \pm 0'',003$

Que corresponde a um $\Delta R = 0'',013 \pm 0'',004$.

Esta precisão também permite encontrar uma relação entre este achatamento e o momento de quadrupólo ($J_2$), descontada a rotação diferencial do Sol, (Rozelot *et al.*, 2001). O valor obtido foi $J_2 = (3{,}61 \pm 2{,}90) \times 10^{-6}$ que é superior ao valor encontrado por Rozelot *et al.* (2001) mas concorda bem com outros encontrados na literatura.

As causas e amplitudes das variações do diâmetro solar são ainda imperfeitamente compreendidas. Nossos estudos demostraram que as séries Este e Oeste de 1998 a 2000, aqui tratadas, depois de corrigidas dos erros observacionais e instrumentais, possuem resolução suficiente para obter valores para o raio , amplitude das variações, períodos, forma e achatamento solares. Os valores obtidos mostram-se equivalentes com outros encontrados na literatura e portanto podem concorrer, com adequado grau de precisão, para dirimir questões que ainda permanecem em aberto.

## 6.2 Perspectivas

A pesquisa do diâmetro solar ganhou um impulso com o início de uma série de outros projetos, como: PICARD, PICARD-SOL, DORAYSOL e MISOLFA.

Variações anuais e seculares do diâmetro solar, na escala de décimos de segundos de arco, têm sido detectadas em vários estudos. Variações rápidas, em menor escala, também estão presentes nas séries observadas, ainda que com uma relação sinal/ruído baixa.

Para melhorar as investigações de variações do diâmetro e suas relações com a constante solar, eventos do Sol e resposta climática é necessária precisão da ordem de milésimo de segundos de arco, em escala de tempo de poucos dias. Para realizar isto, um novo instrumento encontra-se em desenvolvimento no Observatório Nacional-MCT, num empreendimento apoiado pelo CERGA/França e a Universidade Estadual de Feira de Santana. As características mecânicas originais do astrolábio serão conservadas o mais possível. Diversas partes do astrolábio, anteriormente na estação de Greenwhich, serão usadas no novo instrumento que está sendo projetado para alcançar um nível de precisão melhor que 100 *mas*, para uma observação isolada. Com a automação das observações poderão ser registrados cerca de 50 trânsitos solares por sessão (Leste ou Oeste).

Terá óptica Newtoniana, com distância focal mantida em 3m e abertura efetiva de 10cm. Isto possibilitará, usando uma câmera CCD comercial de 512x512 pixels (12μ), alcançar um bom compromisso entre a resolução (0.5"/pixel) e a porção da imagem do Sol observada (15%).

Seu prisma não utilizará molas e sua estabilidade será alcançada através de balanceamento. Além disso o sistema aproveitará o eixo maior da pupila de entrada obtendo um aumento de resolução.

O instrumento utilizará um único espelho, o que irá minimizar os defeitos ópticos. No foco deste espelho será montada a câmara CCD. A passagem da luz no interior do instrumento será feita ao longo de um caminho vertical, atenuando os efeitos da turbulência através das camadas estratificadas de ar. As imagens poderão ser obtidas de modo direto ou através de um obturador. No primeiro caso, imagens duplas do limbo solar são registradas em cada *frame*. No segundo, imagens melhor definidas são obtidas em diferentes *frames*, muito próximas ao momento do trânsito.

Além de observações no contínuo (5480Å), um conjunto de filtros possibilitará observações em diferentes comprimentos de onda de banda passante estreita, possivelmente CaII (3934Å) e $H_\alpha$ (6562Å). Esta configuração permitirá a repetição de observações em comprimentos de onda diferentes.

Para enfrentar os possíveis problemas com a focagem e nitidez da imagem uma possibilidade, sob investigação no momento, é definir um modelo padrão (forma e comprimento) para a função de luminosidade do limbo e ajustar o foco instrumental por este modelo, compensando as mudanças nas condições ópticas, instrumental e atmosférica. Recursos para determinação da correção de *Flat Field* (tela branca e lâmpadas) também estão incluídos no projeto da cúpula.

Pretende-se que o método observacional e o tratamento dos dados sejam totalmente compatíveis com aqueles dos grupos da França (CERGA), Argélia e Turquia. Isto fornecerá meios de comparação direta dos resultados e completará os períodos sem observação. O projeto também visa a comparação com os resultados do satélite PICARD.

O monitoramento do diâmetro solar deverá se estender por um período compatível com os 11 anos do ciclo solar.

# Bibliografia

# Anexo I

Tabelas com os resultados completos de cada um dos parâmetros testados separadamente, com os seus valores brutos e normalizados pela média e desvios padrão.

Três testes foram realizados na forma de equações com as seguintes condições:

Teste a: Um único semidiâmetro médio e um valor único para o coeficiente paramétrico, tanto para as observações a Leste quanto para a Oeste:

$$SD = SD_{obs} - \left(k_{par} \cdot \text{parâmetro}\right)$$

Teste b: Um único semidiâmetro médio, mas dois coeficientes paramétricos independentes para as observações Leste e Oeste:

$$SD = SD_{obs} - \left(k_{par}^{Leste} \cdot \text{parâmetro}^{Leste}\right) - \left(k_{par}^{Oeste} \cdot \text{parâmetro}^{Oeste}\right)$$

Teste c: Semidiâmetros e coeficientes paramétricos independentes para as observações a Leste e a Oeste:

$$\begin{cases} SD^{Leste} = SD_{obs}^{Leste} - \left(k_{par}^{Leste} \cdot \text{parâmetro}^{Leste}\right) \\ SD^{Oeste} = SD_{obs}^{Oeste} - \left(k_{par}^{Oeste} \cdot \text{parâmetro}^{Oeste}\right) \end{cases}$$

# Simbologia:

| | |
|---|---|
| $\sigma$ | Dispersão da série |
| <sd> | Semidiâmetro médio da série |
| coef. | Valor do coeficiente |
| z | Distância zenital |
| tgz | Tangente trigonométrica da distância zenital |
| $\Delta$z | Evolução da distância zenital ao longo da sessão |
| A | Azimute |
| $\Delta$A | Evolução do azimute ao longo da sessão |
| hl | Latitude heliográfica |
| $\Delta$hl | Evolução da latitude heliográfica ao longo da sessão |
| T | Temperatura interpolada para o instante médio da medida |
| $\Delta$T | Evolução da Temperatura interpolada ao longo da sessão |
| p | Pressão atmosférica |
| $\Delta$p | Evolução da pressão atmosférica ao longo da sessão |
| FF | Fator de Fried |
| bdm | Largura média do bordo da imagem direta |
| $\Delta$bdm | Evolução da largura média do bordo da imagem direta ao longo da sessão |
| brm | Largura média do bordo da imagem refletida |
| $\Delta$brm | Evolução da largura média do bordo da imagem refletida ao longo da sessão |
| sigpd | Sigma do ajuste da parábola para o bordo da imagem direta |
| sigpr | Sigma do ajuste da parábola para o bordo da imagem refletida |

# Anexo I

| Parâmetros | | z | tgz | Δz | A | ΔA | hl | Δhl | T | ΔT |
|---|---|---|---|---|---|---|---|---|---|---|
| Valor bruto | σ | 0,552 | 0,552 | 0,553 | 0,553 | 0,552 | 0,553 | 0,552 | 0,551 | 0,552 |
| | <sd> | 958,861±0,033 | 958,927±0,023 | 959,111±0,006 | 959,013±0,031 | 959,109±0,006 | 959,096±0,011 | 959,110±0,006 | 959,376±0,038 | 959,080±0,008 |
| | coef. | 0,006±0,001 | 0,189±0,023 | -0,098±0,034 | ≈0 | 0,108±0,029 | ≈0 | 0,139±0,034 | -0,009±0,001 | 0,133±0,024 |
| Valor normalizado | σ | 0,552 | 0,552 | 0,553 | 0,553 | 0,552 | 0,553 | 0,553959 | **0,551** | **0,552** |
| | <sd> | 959,108±0,006 | 959,107±0,006 | 959,109±0,006 | 959,109±0,006 | 959,109±0,006 | 959,109±0,006 | 959,109±0,006 | **959,109±0,006** | **959,108±0,006** |
| | coef. | 0,044±0,006 | 0,048±0,006 | -0,017±0,006 | 0,018±0,006 | 0,022±0,006 | 0,008±0,006 | 0,023±0,006 | **-0,041±0,006** | **0,031±0,006** |

| Parâmetros | | p | Δp | FF | bdm | Δbdm | brm | Δbrm | sigpd | sigpr |
|---|---|---|---|---|---|---|---|---|---|---|
| Valor bruto | σ | 0,552 | 0,553 | 0,552 | 0,553 | 0,553 | 0,552 | 0,553 | 0,551 | 0,552 |
| | <sd> | 956,227±1,356 | 959,114±0,006 | 959,023±0,044 | 959,077±0,027 | 959,117±0,007 | 959,058±0,025 | 959,103±0,007 | 959,177±0,011 | 959,177±0,012 |
| | coef. | 0,004±0,002 | 1,074±0,445 | 0,002±0,001 | 0,006±0,004 | 0,560±0,232 | 0,007±0,004 | -0,324±0,204 | -0,037±0,006 | -0,045±0,007 |
| Valor normalizado | σ | 0,552 | 0,553 | **0,553** | 0,553 | 0,553 | 0,552 | 0,553 | 0,551 | **0,552** |
| | <sd> | 959,109±0,006 | 959,109±0,006 | **959,109±0,006** | 959,109±0,006 | 959,109±0,006 | 959,109±0,006 | 959,109±0,006 | 959,109±0,006 | **959,109±0,006** |
| | coef. | 0,012±0,006 | 0,015±0,006 | **0,011±0,006** | 0,007±0,006 | 0,014±0,006 | 0,012±0,006 | -0,009±0,005 | -0,040±0,006 | **-0,037±0,006** |

Tabelas dos coeficientes encontrados para o teste (a): $SD = SD_{obs} - \left(k_{par} \cdot \text{parâmetro}\right)$

| Parâmetros | | z | tgz | Δz | A | ΔA | hl | Δhl | T | ΔT |
|---|---|---|---|---|---|---|---|---|---|---|
| Valor bruto | σ | 0,550 | 0,550 | 0,553 | 0,553 | 0,552 | 0,551 | 0,553 | 0,550 | 0,552 |
| | <sd> | 958,830±0,033 | 958,907±0,022 | 959,140±0,012 | 959,020±0,031 | 959,088±0,013 | 959,093±0,011 | 959,089±0,011 | 959,408±0,046 | 959,074±0,008 |
| | Coef. (L) | 0,007±0,001 | 0,251±0,025 | 0,067±0,067 | ≈0 | 0,223±0,067 | ≈0 | 0,275±0,070 | -0,010±0,002 | 0,133±0,024 |
| | Coef. (O) | 0,006±0,001 | 0,180±0,023 | -0,331±0,086 | ≈0 | -0,012±0,071 | ≈0 | -0,015±0,077 | -0,010±0,001 | 0,215±0,062 |
| Valor normalizado | σ | 0,551 | 0,551 | 0,553 | 0,553 | 0,553 | 0,551 | 0,553 | 0,550 | 0,553 |
| | <sd> | 959,113±0,006 | 959,113±0,006 | 959,141±0,012 | 959,108±0,006 | 959,089±0,013 | 959,110±0,006 | 959,089±0,011 | 959,101±0,007 | 959,115±0,007 |
| | Coef. (L) | 0,076±0,008 | 0,081±0,007 | 0,011±0,011 | 0,029±0,009 | 0,045±0,014 | 0,054±0,009 | 0,046±0,012 | -0,058±0,010 | 0,022±0,007 |
| | Coef. (O) | 0,016±0,008 | 0,020±0.008 | -0,064±0,017 | 0,010±0,008 | -0,003±0,014 | -0,027±0,008 | -0,003±0,013 | -0,029±0,010 | 0,051±0,015 |

| Parâmetros | | p | Δp | FF | bdm | Δbdm | brm | Δbrm | sigpd | sigpr |
|---|---|---|---|---|---|---|---|---|---|---|
| Valor bruto | σ | 0,553 | 0,553 | 0,553 | 0,552 | 0,552 | 0,553 | 0,553 | 0,551 | 0,552 |
| | <sd> | 956,803±1,368 | 959,113±0,006 | 959,023±0,044 | 959,054±0,028 | 959,119±0,007 | 959,056±0,025 | 959,103±0,007 | 959,185±0,012 | 959,177±0,012 |
| | Coef. (L) | 0,003±0,002 | 1,369±1,419 | 0,003±0,001 | 0,015±0,005 | -0,652±0,342 | 0,012±0,004 | -0,670±0,286 | -0,034±0,005 | -0,041±0,008 |
| | Coef. (O) | 0,003±0,002 | 1,033±0,483 | 0,002±0,001 | 0,005±0,005 | 1,372±0,284 | 0,005±0,004 | -0,093±0,253 | -0,050±0,007 | -0,047±0,007 |
| Valor normalizado | σ | 0,553 | 0,553 | 0,553 | 0,552 | 0,553 | 0,553 | 0,553 | 0,551 | 0,552 |
| | <sd> | 959,110±0,006 | 959,105±0,007 | 959,109±0,006 | 959,113±0,006 | 959,112±0,006 | 959,108±0,006 | 959,108±0,006 | 959,117±0,006 | 959,108±0,006 |
| | Coef. (L) | 0,0003±0,008 | 0,032±0,015 | 0,006±0,007 | 0,076±0,010 | -0,010±0,009 | 0,047±0,009 | -0,007±0,009 | -0,061±0,008 | -0,060±0,009 |
| | Coef. (O) | 0,023±0,008 | 0,010±0,007 | 0,023±0,010 | -0,021±0,007 | 0,033±0,008 | -0,012±0,008 | -0,010±0,008 | -0,006±0,009 | -0,018±0,008 |

Tabelas dos coeficientes encontrados para o teste (b): $SD = SD_{obs} - \left(k_{par}^{Leste} \cdot \text{parâmetro}^{Leste}\right) - \left(k_{par}^{Oeste} \cdot \text{parâmetro}^{Oeste}\right)$.

| **Parâmetros** | | **z** | **tgz** | **Δz** | **A** | **ΔA** | **hl** | **Δhl** | **T** | **ΔT** |
|---|---|---|---|---|---|---|---|---|---|---|
| Valor bruto | σ | 0,550 | 0,548 | 0,553 | 0,552 | 0,552 | 0,551 | 0,552 | 0,550 | 0,552 |
| | <sd> (L) | 958,690±0,046 | 958,823±0,032 | 959,134±0,013 | 959,987±0,045 | 959,075±0,021 | 959,046±0,015 | 959,084±0,016 | 959,510±0,065 | 959,098±0,013 |
| | <sd> (O) | 958,961±0,047 | 958,994±0,032 | 959,167±0,027 | 959,047±0,043 | 959,096±0,016 | 959,132±0,015 | 959,093±0,015 | 959,322±0,064 | 959,063±0,011 |
| | Coef. (L) | 0,011±0,001 | 0,340±0,034 | 0,042±0,071 | ≈ 0 | 0,278±0,096 | 0,002±0,0003 | 0,299±0,088 | -0,014±0,002 | 0,086±0,029 |
| | Coef. (O) | 0,003±0,001 | 0,099±0,031 | -0,496±0,196 | ≈ 0 | 0,025±0,084 | -0,001±0,0003 | 0,009±0,094 | -0,007±0,002 | 0,265±0,070 |
| Valor normalizado | σ | 0,550 | 0,548 | 0,553 | 0,552 | 0,552 | 0,551 | 0,552 | 0,550 | 0,553 |
| | <sd> (L) | 959,144±0,009 | 959,146±0,009 | 959,135±0,014 | 959,127±0,008 | 959,075±0,020 | 959,130±0,008 | 959,084±0,016 | 959,090±0,011 | 959,117±0,009 |
| | <sd> (O) | 959,088±0,008 | 959,088±0,008 | 959,159±0,024 | 959,092±0,008 | 959,096±0,016 | 959,093±0,008 | 959,093±0,015 | 959,109±0,009 | 959,111±0,011 |
| | Coef. (L) | 0,082±0,008 | 0,086±0,009 | 0,007±0,012 | 0,026±0,008 | 0.057±0,020 | 0,054±0,008 | 0,050±0,015 | -0,065±0,011 | 0,021±0,007 |
| | Coef. (O) | 0,023±0,008 | 0,025±0,008 | -0,085±0,029 | 0,009±0,008 | 0,005±0,017 | -0,026±0,008 | 0,001±0,016 | -0,032±0,009 | 0,047±0,017 |

Tabelas dos coeficientes encontrados para o teste (c):. $\begin{cases} SD^{Leste} = SD^{Leste}_{obs} - \left(k^{Leste}_{par} \cdot \text{parâmetro}^{Leste}\right) \\ SD^{Oeste} = SD^{Oeste}_{obs} - \left(k^{Oeste}_{par} \cdot \text{parâmetro}^{Oeste}\right) \end{cases}$

| **Parâmetros** | | **p** | **Δp** | **FF** | **bdm** | **Δbdm** | **brm** | **Δbrm** | **sigpd** | **sigpr** |
|---|---|---|---|---|---|---|---|---|---|---|
| Valor bruto | σ | 0,553 | 0,553 | 0,553 | 0,550 | 0,552 | 0,552 | 0,553 | 0,550 | 0,551 |
| | <sd> (L) | 959,732±2,055 | 959,127±0,009 | 959,085±0,054 | 958,776±0,048 | 959,122±0,009 | 958,934±0,039 | 959,123±0,010 | 959,272±0,019 | 959,239±0,018 |
| | <sd> (O) | 954,322±1,833 | 959,097±0,009 | 958,906±0,077 | 959,184±0,033 | 959,114±0,009 | 959,142±0,033 | 959,083±0,010 | 959,118±0,016 | 959,125±0,016 |
| | Coef. (L) | -0,001±0,003 | 0,753±1,443 | 0,001±0,001 | 0,062±0,008 | -0,596±0,357 | 0,029±0,005 | -0,359±0,302 | -0,065±0,007 | -0,072±0,010 |
| | Coef. (O) | 0,006±0,002 | 0,551±0,526 | 0,005±0,001 | -0,015±0,005 | 1,281±0,311 | -0,007±0,005 | -0,414±0,278 | -0,017±0,009 | -0,021±0,009 |
| Valor normalizado | σ | 0,553 | 0,553 | 0,553 | 0,550 | 0,552 | 0,552 | 0,552 | 0,550 | 0,551 |
| | <sd> (L) | 959,129±0,009 | 959,122±0,011 | 959,128±0,008 | 959,137±0,008 | 959,130±0,009 | 959,128±0,008 | 959,129±0,009 | 959,151±0,009 | 959,128±0,008 |
| | <sd> (O) | 959,094±0,008 | 959,095±0,008 | 959,092±0,008 | 959,094±0,008 | 959,096±0,008 | 959,092±0,008 | 959,091±0,008 | 959,087±0,008 | 959,092±0,008 |
| | Coef. (L) | -0,003±0,009 | 0,014±0,018 | 0,006±0,007 | 0,080±0,010 | -0,013±0,009 | 0,045±0,009 | -0,009±0,009 | -0,070±0,008 | -0,060±0,009 |
| | Coef. (O) | 0,020±0,008 | 0,008±0,007 | 0,025±0,010 | -0,020±0,007 | 0,030±0,008 | -0,012±0,008 | -0,011±0,008 | -0,018±0,010 | -0,018±0,008 |

Tabelas dos coeficientes encontrados para o teste (c):. $\begin{cases} SD^{Leste} = SD_{obs}^{Leste} - \left(k_{par}^{Leste} \cdot parâmetro^{Leste}\right) \\ SD^{Oeste} = SD_{obs}^{Oeste} - \left(k_{par}^{Oeste} \cdot parâmetro^{Oeste}\right) \end{cases}$

# Anexo II

Tabelas com a contagens do número de observações, por heliolatitude, para as passagens Leste e Oeste de toda a campanha.

Conforme explicado no capítulo 5, o método de observação impõe simetria Norte-Sul e Leste-Oeste, e assim todos as contagens referem-se ao primeiro quadrante.

# Anexo II

| Passagem Leste (0°< θ < 30°) | |
|---|---|
| *heliolatitude* | *contagens* |
| 0.5 | 0 |
| 1.5 | 0 |
| 2.5 | 34 |
| 3.5 | 175 |
| 4.5 | 133 |
| 5.5 | 118 |
| 6.5 | 115 |
| 7.5 | 105 |
| 8.5 | 75 |
| 9.5 | 67 |
| 10.5 | 69 |
| 11.5 | 86 |
| 12.5 | 66 |
| 13.5 | 54 |
| 14.5 | 59 |
| 15.5 | 70 |
| 16.5 | 65 |
| 17.5 | 52 |
| 18.5 | 47 |
| 19.5 | 50 |
| 20.5 | 55 |
| 21.5 | 62 |
| 22.5 | 58 |
| 23.5 | 61 |
| 24.5 | 49 |
| 25.5 | 47 |
| 26.5 | 44 |
| 27.5 | 51 |
| 28.5 | 43 |
| 29.5 | 35 |

| Passagem Leste (30°< θ < 60°) | |
|---|---|
| *heliolatitude* | *contagens* |
| 30.5 | 37 |
| 31.5 | 25 |
| 32.5 | 28 |
| 33.5 | 66 |
| 34.5 | 39 |
| 35.5 | 31 |
| 36.5 | 22 |
| 37.5 | 29 |
| 38.5 | 39 |
| 39.5 | 37 |
| 40.5 | 26 |
| 41.5 | 25 |
| 42.5 | 30 |
| 43.5 | 37 |
| 44.5 | 27 |
| 45.5 | 30 |
| 46.5 | 26 |
| 47.5 | 39 |
| 48.5 | 40 |
| 49.5 | 35 |
| 50.5 | 46 |
| 51.5 | 53 |
| 52.5 | 45 |
| 53.5 | 47 |
| 54.5 | 41 |
| 55.5 | 44 |
| 56.5 | 43 |
| 57.5 | 43 |
| 58.5 | 54 |
| 59.5 | 53 |

| Passagem Leste (60°< θ < 90°) | |
|---|---|
| *heliolatitude* | *contagens* |
| 60.5 | 58 |
| 61.5 | 67 |
| 62.5 | 74 |
| 63.5 | 90 |
| 64.5 | 91 |
| 65.5 | 78 |
| 66.5 | 78 |
| 67.5 | 74 |
| 68.5 | 60 |
| 69.5 | 61 |
| 70.5 | 66 |
| 71.5 | 54 |
| 72.5 | 53 |
| 73.5 | 54 |
| 74.5 | 45 |
| 75.5 | 41 |
| 76.5 | 34 |
| 77.5 | 26 |
| 78.5 | 18 |
| 79.5 | 19 |
| 80.5 | 12 |
| 81.5 | 7 |
| 82.5 | 2 |
| 83.5 | 2 |
| 84.5 | 0 |
| 85.5 | 0 |
| 86.5 | 0 |
| 87.5 | 0 |
| 88.5 | 0 |
| 89.5 | 0 |

Contagens do número de observações do diâmetro solar, por heliolatitude, para as sessões a Leste.

| Passagem Oeste (0°< θ < 30°) | |
|---|---|
| ***heliolatitude*** | ***contagens*** |
| 0.5 | 70 |
| 1.5 | 84 |
| 2.5 | 95 |
| 3.5 | 114 |
| 4.5 | 121 |
| 5.5 | 130 |
| 6.5 | 160 |
| 7.5 | 151 |
| 8.5 | 62 |
| 9.5 | 46 |
| 10.5 | 56 |
| 11.5 | 52 |
| 12.5 | 35 |
| 13.5 | 30 |
| 14.5 | 24 |
| 15.5 | 31 |
| 16.5 | 45 |
| 17.5 | 57 |
| 18.5 | 26 |
| 19.5 | 26 |
| 20.5 | 32 |
| 21.5 | 48 |
| 22.5 | 70 |
| 23.5 | 47 |
| 24.5 | 37 |
| 25.5 | 41 |
| 26.5 | 46 |
| 27.5 | 53 |
| 28.5 | 62 |
| 29.5 | 80 |

| Passagem Oeste (30°< θ < 60°) | |
|---|---|
| ***heliolatitude*** | ***contagens*** |
| 30.5 | 47 |
| 31.5 | 36 |
| 32.5 | 44 |
| 33.5 | 37 |
| 34.5 | 40 |
| 35.5 | 34 |
| 36.5 | 49 |
| 37.5 | 45 |
| 38.5 | 39 |
| 39.5 | 45 |
| 40.5 | 33 |
| 41.5 | 43 |
| 42.5 | 44 |
| 43.5 | 34 |
| 44.5 | 47 |
| 45.5 | 50 |
| 46.5 | 47 |
| 47.5 | 45 |
| 48.5 | 51 |
| 49.5 | 50 |
| 50.5 | 50 |
| 51.5 | 53 |
| 52.5 | 65 |
| 53.5 | 50 |
| 54.5 | 45 |
| 55.5 | 54 |
| 56.5 | 60 |
| 57.5 | 65 |
| 58.5 | 63 |
| 59.5 | 73 |

| Passagem Oeste (60°< θ < 90°) | |
|---|---|
| ***heliolatitude*** | ***contagens*** |
| 60.5 | 65 |
| 61.5 | 70 |
| 62.5 | 69 |
| 63.5 | 85 |
| 64.5 | 88 |
| 65.5 | 97 |
| 66.5 | 91 |
| 67.5 | 102 |
| 68.5 | 104 |
| 69.5 | 89 |
| 70.5 | 99 |
| 71.5 | 79 |
| 72.5 | 84 |
| 73.5 | 67 |
| 74.5 | 59 |
| 75.5 | 48 |
| 76.5 | 46 |
| 77.5 | 32 |
| 78.5 | 26 |
| 79.5 | 23 |
| 80.5 | 19 |
| 81.5 | 12 |
| 82.5 | 14 |
| 83.5 | 8 |
| 84.5 | 9 |
| 85.5 | 6 |
| 86.5 | 3 |
| 87.5 | 3 |
| 88.5 | 0 |
| 89.5 | 0 |

Contagens do número de observações do diâmetro solar, por heliolatitude, para as sessões a Oeste.